\documentclass[
	aps, twocolumn, amsmath, amssymb,
	prl, superscriptaddress, floatfix]{revtex4}
\usepackage{graphicx}		
\usepackage{dcolumn}		
\usepackage{bm}				
\usepackage{amssymb}
\usepackage[hypertexnames=false]{hyperref}
\usepackage{multirow}
\usepackage{color}
\usepackage[normalem]{ulem}
\usepackage{mathtools}
\usepackage{tikz}
\usepackage{cancel}
\usetikzlibrary{arrows.meta}

\newcommand{\de}{\mathrm{d}}

\newcommand{\cc}[1]{\textcolor{magenta}{\textit{ }}}

\newcommand{\comments}[1]{}   

\begin{document}


\title{SWAP algorithm for lattice spin models}

\date{\today}

\author{Greivin Alfaro Miranda}
\affiliation{Sorbonne Universit\'e, Laboratoire de Physique Th\'eorique et Hautes Energies, CNRS UMR 7589,
    4 Place Jussieu, 75252 Paris Cedex 05, France}

\author{Leticia F. Cugliandolo}
\affiliation{Sorbonne Universit\'e, Laboratoire de Physique Th\'eorique et Hautes Energies, CNRS UMR 7589,
    4 Place Jussieu, 75252 Paris Cedex 05, France}
\affiliation{Institut Universitaire de France, 1 rue Descartes, 75005 Paris France}

\author{Marco Tarzia}
\affiliation{Sorbonne Universit\'e, Laboratoire de Physique Th\'eorique de la  Mati\`ere Condens\'ee, CNRS UMR 7600,
    4 Place Jussieu, 75252 Paris Cedex 05, France}
\affiliation{Institut Universitaire de France, 1 rue Descartes, 75005 Paris France}

\begin{abstract} 
We adapted the SWAP molecular dynamics algorithm for use in lattice Ising spin models. We dressed the spins
with a randomly distributed length and we alternated long-range spin exchanges  
with conventional single spin flip Monte Carlo updates, both accepted with a stochastic rule which respects detailed balance.
We show that this algorithm, when applied to the bidimensional Edwards-Anderson model,   
speeds up significantly the relaxation at low temperatures and
manages to find ground states with high efficiency and little computational cost. The exploration of spin models should help in understanding
why SWAP accelerates the evolution of particle systems  and shed light on relations 
between dynamics and free-energy landscapes. 
\end{abstract}

\maketitle
\vspace{0.25cm}


The slow relaxation of spin and structural glasses~\cite{Bouchaud98, BerthierBiroli, berthier2011dynamical, menon2012physics} poses a computational challenge because it is not feasible to investigate their progression towards equilibrium using standard numerical simulations. Efforts to overcome this slowdown have led to the development of specialized computers and numerical techniques.

Spin-glasses are interesting physical systems which also attract theoretical attention
because they map to a broad range of hard combinatorial optimization problems.
In the context of these frustrated magnets, methods such as parallel tempering~\cite{Earl05}, 
as well as the recently introduced replicated simulated annealing~\cite{Baldassi16, Ricci23}, enable the equilibration of larger sample sizes compared to conventional single spin flip Monte Carlo. The special purpose 
Janus machine has allowed to equilibrate spin-glass models of rather large sizes,  though only those with discrete variables and couplings~\cite{Janus09}. More recently, deep reinforcement learning methods have been  explored 
to find the ground states of finite dimensional spin-glasses~\cite{fan2023searching, Boettcher2023}. 

In the context of structural glasses, Berthier {\it et al.}~\cite{berthier16,Ninarello17,berthier_efficient_2019} made significant advancements over previous non-local particle exchange methods~\cite{berthier_modern_2022},
achieving accelerated equil\-i\-bration of several glass-forming liquids. 
Their breakthrough was first realized  in polydisperse mixtures which still exhibit the characteristics of glass\--forming liquids, and then generalized to many other models~\cite{Ninarello17}
The method incorporated standard Molecular Dynamics, periodically alternated with a step in which two randomly selected particles with typically different size are exchanged using a Monte Carlo rule satisfying detailed balance. This technique, referred to as SWAP, has proven successful in equilibrating particle systems of unprecedented size, all the way down to the glass transition temperature~\cite{Scalliet22}.

This spectacular effect was  interpreted as direct evidence against a static, cooperative explanation
	of the glass transition such as the one offered by the random first-order transition theory (RFOT)~\cite{Wyart17}. 	
	Yet this claim was contested in~\cite{Berthier19} where  the efficiency of SWAP was explained in 
	terms of its ability  to avoid  the  slowdown caused by numerous metastable local minima in the (free-)energy 
	landscape:
	the non-local particle exchanges thus might be overcoming barriers quickly, in contrast to a strictly local dynamics that would 
necessitate cooperative rearrangements to do it. Despite the distinct (mean-field) free-energy 
landscapes of structural and spin glasses,  they both harbor a multitude of metastable states. This 
suggests the potential success of a similar algorithm in accelerating the dynamics of spin glasses, models in 
which dynamics facilitation~\cite{chandler2010dynamics} is absent by construction.

In this Letter, we adapt the SWAP method for application to finite-dimensional spin models. In this reshaping we introduce an auxiliary model in which we assign a length to the spin variables, akin to the role of particle diameters in the original SWAP implementation. 
The exchange of constituents, in our case, the spins, 
will effectively mitigate the local energy barriers created by the quenched randomness, allowing us to explore configurations that would otherwise remain inaccessible. 

We have chosen to focus on two-dimensional (2D) problems for which we know the equilibrium phases
and ground states. In the Supplemental Material (SM) we 
validate the method with a study of  a clean and unfrustrated spin system 
with a finite-temperature second-order transition from a paramagnetic to a ferromagnetic phase. 
The core of this work is the study of the 2D Edwards-Anderson (EA) model, a random magnet with
spin-glass properties only at zero temperature, but exceedingly long physical relaxation at low temperatures. 
This 2D problem is not just a theoretical construct: thin film spin-glass materials have regained 
experimental interest in  recent years~\cite{Guchhait14}.
Moreover, the ground state configurations can be identified exactly with 
special algorithms~\cite{hartmann_ground_2011,Weigel18}. This gives us a knowledge against which we can 
confront the performance of our algorithm. 

Concretely, we modify the Ising models by introducing associated 
``$\Delta$-models" in which the lengths of the spins are 
drawn initially from a pre-determined probability distribution. 
We then perform a single-spin-flip evolution alternating at random time-steps with a non-local exchange of spins, 
the swaps.
The distribution of the spin-lengths remains unchanged  
 but local energy barriers are uplifted so that the acceptance of new configurations is propitiated. 
In the $\Delta$-model with uniform ferromagnetic couplings, SWAP does not improve over the standard single-spin-flip evolution
of the parent Ising model (see the SM). In contrast, in the frustrated case, 
SWAP manages to take the $\Delta$-model to equilibrium 
faster than usual implementations of parallel tempering do for the EA model at
low temperatures. 
After instantaneous quenches, we find the ground state of the $\Delta$-Model with $\Delta=2$ around $2$ decades faster
than with parallel tempering Monte Carlo~\cite{Roma09, Wang2015}.
This is further improved by a temperature annealing, that enables us to find at least $99\%$ of the ground states for 
$\Delta \geq {0.5}$.
\comments{
\textcolor{blue}{*** I have used the following two references to compare the algorithms w.r.t. the ground state \cite{Roma09} and \cite{Wang2015}, the main idea is that Romà et al found the heuristic formula for Parallel tempering
\begin{equation}
	t_{PT} = \frac{1}{m} \left(\frac{\mathcal{P}_0}{1-\mathcal{P}_0} \right)^{\frac{L^d}{q}} \exp(bL^c - a) 
\end{equation}	
where usual values, for the most efficient parameters presented in the article for the 2d case, are $m = 5$, $a = 5.86$, $b = 3.35$, $c = 0.5$, $d = 0.2$ and $q = 2$. If we perform exactly the same form of the fit we obtain
\begin{equation}
	t_{\rm swap} =  \left(\frac{\mathcal{P}_0}{A-\mathcal{P}_0} \right)^{\frac{1}{q}} \exp(bL^c - a) 
\end{equation} with $A = 3.19$, $a = -3.07$, $b = 1.80$, $c = 0.5$, $d = 0.0$ and $q = 2.42$. Where $A$ comes from the fact that our sigmoid fit now has an amplitude different from 1 i.e.
\begin{equation}
	\mathcal{P}_0 (\tau) = A \frac{e^{q \tau}}{e^{q \tau}+1}
\end{equation} However the proposed fit by Romà is not convenient for us, we find a better collapse for \begin{equation}
t^\star(L) = \alpha L^\gamma 
\end{equation}, so basically setting $a = 0$ and letting $c$ vary. We obtain \begin{equation}
t = \alpha L^\gamma \left(\frac{\mathcal{P}_0}{1-\mathcal{P}_0} \right)^{\frac{1}{q}}
\end{equation}, here $A = 1$ as in the first case, and $q = 2.50(1)$, $\alpha = 1.25(1)$, $\alpha = 3.73(2)$} 
}

\begin{figure}[b!]
	\vspace{-0.75cm}
    \centering
    	\includegraphics[width=0.95\linewidth]{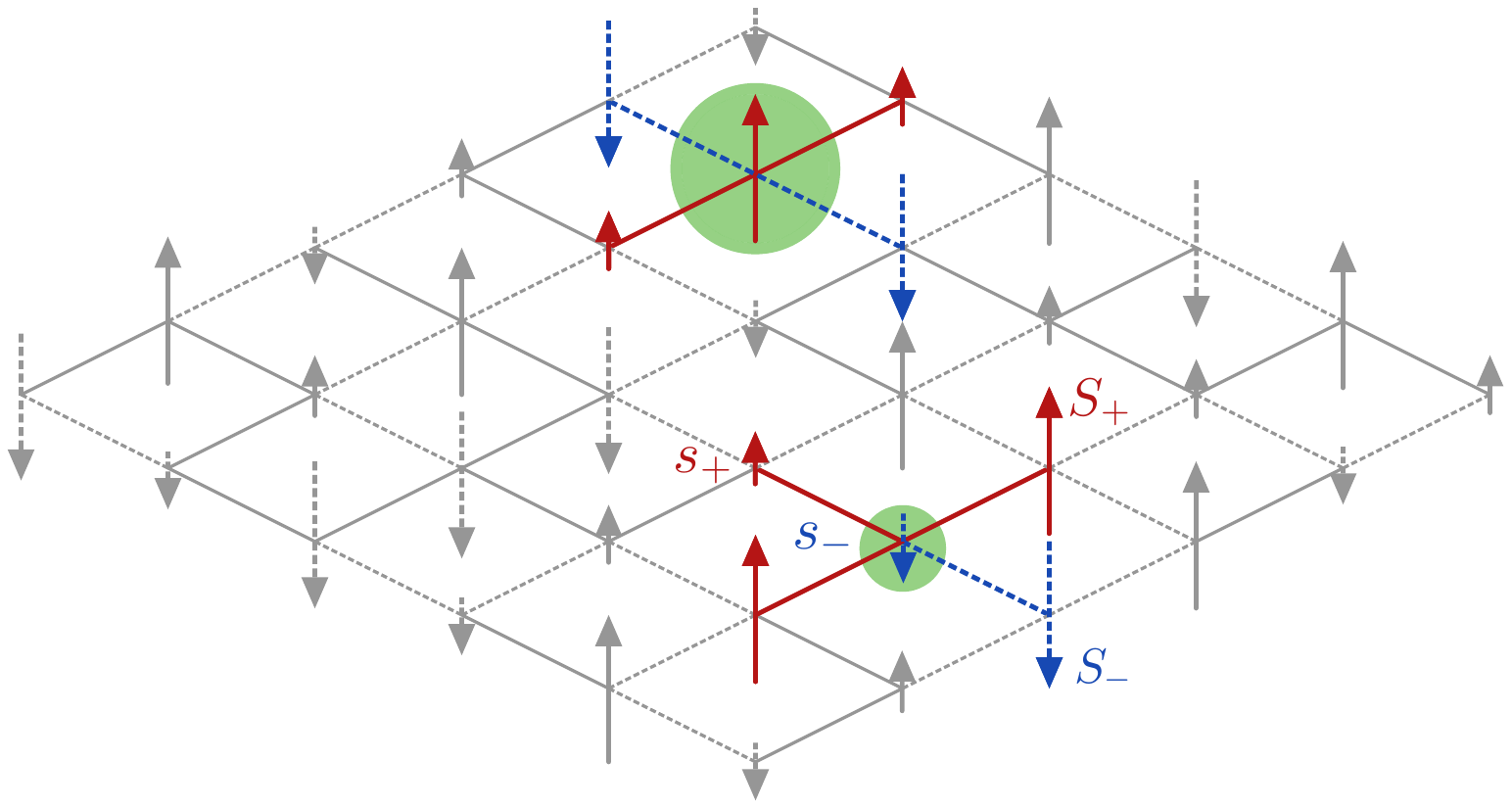}
	\vspace{-0.75cm}
	\caption{
	Sketch of a spin configuration of the modified 2DEA model. 
	Two spins are singled out for analysis (surrounded by green bubbles). 
	The neighboring up and down spins are colored red and blue, respectively.
	The solid (red) and dashed (blue) links represent $J_{ij}>0$ and $J_{ij}<0$, respectively.
	 The length of the arrows are proportional to the length of the spins, that is, the local $\tau_i$ values. They are 
	here chosen to take only two values, for simplicity. 
	}
	\label{fig:PlaquetteSWAP}
\end{figure}

We start from a finite-dimensional Ising Model (IM), 
\begin{equation}
\mathcal{H} = - \sum_{\langle ij \rangle } J_{ij} \sigma_i \sigma_j 
\; , 
\;\; \ \sigma_i = \pm 1
\; , 
\;\;
i = 1, \dots, N
\label{eq:IsingHamiltonian}
\end{equation}
where $N=L^{\mbox{\footnotesize 2}}$, 
 $\langle ij \rangle$ 
 indicates a sum over nearest neighbors on a two dimensional square lattice with linear size 
$L$ (each pair added once) and periodic boundary conditions. 
In a clean ferromagnetic model the coupling strengths are all equal $J_{ij}=J>0$. 
In the EA model, they are drawn from a symmetric probability 
distribution centered at zero, and with variance $J^2$. 
The latter has spin-glass ground states with a 2-fold degeneracy due to the global spin-reversal symmetry. For bimodal couplings this degeneracy increases drastically with a gap to the lowest excitations. For Gaussian distributed 
couplings there are only two ground states but the spectrum of excitations grows continuously. At non-vanishing temperatures, the model, with either choice of couplings, behaves paramagnetically.
The Gaussian system undergoes domain growth at $T=0$~\cite{kisker_off-equilibrium_1996}.
In the bimodal case, clusters of spins maintain their relative orientation in all ground states and form 
a backbone on which coarsening takes place while all other spins behave paramagnetically~\cite{Roma06}.
Single spin flip dynamics freezes at $T \lesssim$ 
0.2~(see Fig.~SM20 \cite{supp}).

We will consider a variation of model (\ref{eq:IsingHamiltonian}) in which the spins have different amplitudes~\cite{Krasnytska20} or lengths, interacting through unitary bimodal bonds, $J_{ij}$, so that $p(\{J_{ij}\}) = \frac{1}{2} \delta(J_{ij}-1)+ \frac{1}{2} \delta(J_{ij}+1)$, and 
\begin{equation}
\label{eqn:definition_softspins}
\mathcal{H} = - \sum_{\langle ij \rangle } J_{ij} s_i s_j \; , 
\qquad\quad
s_i = \sigma_i \tau_i \; .
\end{equation} 
The $\sigma_i$s are Ising variables 
$\sigma_i = \pm 1$
and the $\tau_i$ are independently and initially drawn
from a normalized box distribution, $p_\tau(\tau_i)$, {\it i.e.}
\begin{eqnarray}
\label{eqn:definition_lengths}
\tau_i \in [1-\Delta/2, 1+\Delta/2 ]    
\;, \qquad 0 \leq \Delta \leq 2 
\; . 
\label{eq:pdf-taui}
\end{eqnarray}
The average over the $\tau_i$ distribution is 
denoted $[...]$. The parameter $\Delta$ controls the spin length variation: their mean and  
variance are $[\tau_i] =1$ and $[\tau_i^2] - [\tau_i]^2 = \Delta^2/12 $. 
 $\Delta \leq 2$ ensures  that $\tau_i \geq 0$, and the Ising case is recovered for 
$\Delta =0$. 
The space-varying $\tau_i$  induce random interactions between the Ising spins even in the unfrustrated case since
(\ref{eqn:definition_softspins}) can be recast as 
\begin{equation}
\mathcal{H} = 
	- \sum_{\langle ij \rangle }\mathcal{J}_{ij} \sigma_i \sigma_j
	\; ,
	\qquad\qquad \mathcal{J}_{ij} = J_{ij} \tau_i \tau_j
	\; . 
	\end{equation}
The new continuously varying exchanges $\mathcal{J}_{ij}$  
follow a symmetric distribution function with a gap that closes for $\Delta \to 2$, see
Fig.~SM16, with local spatial correlations {\it via} the $\tau_i$ 
which can be further enhanced by SWAP, see Fig.~\ref{fig:Pvde}(c).

We proved that, in mean-field, the only effect of $\Delta$ 
is to modify $T_c$ (the calculations for the fully-connected and Bethe lattice $\Delta$-models are in the SM~\cite{mezard_bethe_2001}). Additionally, 
the real-space RG of the $\Delta$-model on the hierarchical lattice yields the same fixed point as the one
of the EA~\cite{Southern1977, Drossel2000, Drossel2001}. These results indicate that assigning a length to the spins does not change the physics of the problem.

At each Monte Carlo (MC) sweep, the nature of the microscopic moves is dictated as follows: 
\begin{eqnarray}
\left\{
\begin{array}{ll}
\textrm{with~}p_{\rm swap} & \mapsto N \, \mbox{(non-local) exchange attempts} \ \ \  
\nonumber
\\
& \qquad (\sigma_i, \tau_i) \leftrightarrow (\sigma_j, \tau_j) 
\; , 
\label{eq:SWAP-def}
\\
[6pt]
\textrm{with}~1-p_{\rm swap} \!\! & \mapsto N \, \mbox{single spin flip attempts} 
\nonumber
\\
& \qquad \sigma_i \to -\sigma_i
\; . 
\end{array}
\right.
\end{eqnarray} 
The microscopic moves are accepted with the Metropolis acceptance probability
$	p_{\rm acc} = \min(1, e^{-\beta \Delta E}) $. 
The inverse temperature is $\beta=1/(k_BT)$ and $\Delta E$ is the energy variation due to the $i$-th spin flip or the spin exchange between the $i$-th and $j$-th spins, chosen with the constraint of being more than one lattice spacing apart, making the exchanges strictly non-local. In later versions of the algorithm, we found that the overall number of accepted moves could be increased by considering pure length exchanges, $\tau_i \leftrightarrow \tau_j$, in the SWAP step. The following results hold for either option, although the latter is recommended.  
For $p_{\rm swap} = 0$, all sweeps consist solely of single spin flips, thus the $\tau_i$ variables remain quenched throughout the entire MC evolution, exactly like the couplings $J_{ij}$. Conversely, for $p_{\rm swap} = 1$, all attempted moves are spin-exchanges, allowing the length variables to fluctuate. Any intermediate value of $p_{\rm swap}$ yields an evolution such that a fraction $1-p_{\rm swap}$ of the total number of sweeps keeps $\tau_i$ unchanged, while for the remaining $p_{\rm swap}$ the $\tau_i$ are annealed. We thus refer to the $\tau_i$ as
\textit{partially} annealed. Note that the total set of possible lengths $\{ \tau_i\}$ remains fixed during a run of the algorithm.
In the following, $\langle ... \rangle$ represents an average over thermal MC noise 
and initial conditions of the  $\{\sigma_i\}$
and  $[...]$ stands for the average over different realizations of the quenched couplings $\{ J_{ij} \}$ and of the partially annealed length variables~$\{ \tau_i \}$.

\begin{figure}[b!]
\includegraphics[width=0.87\linewidth]{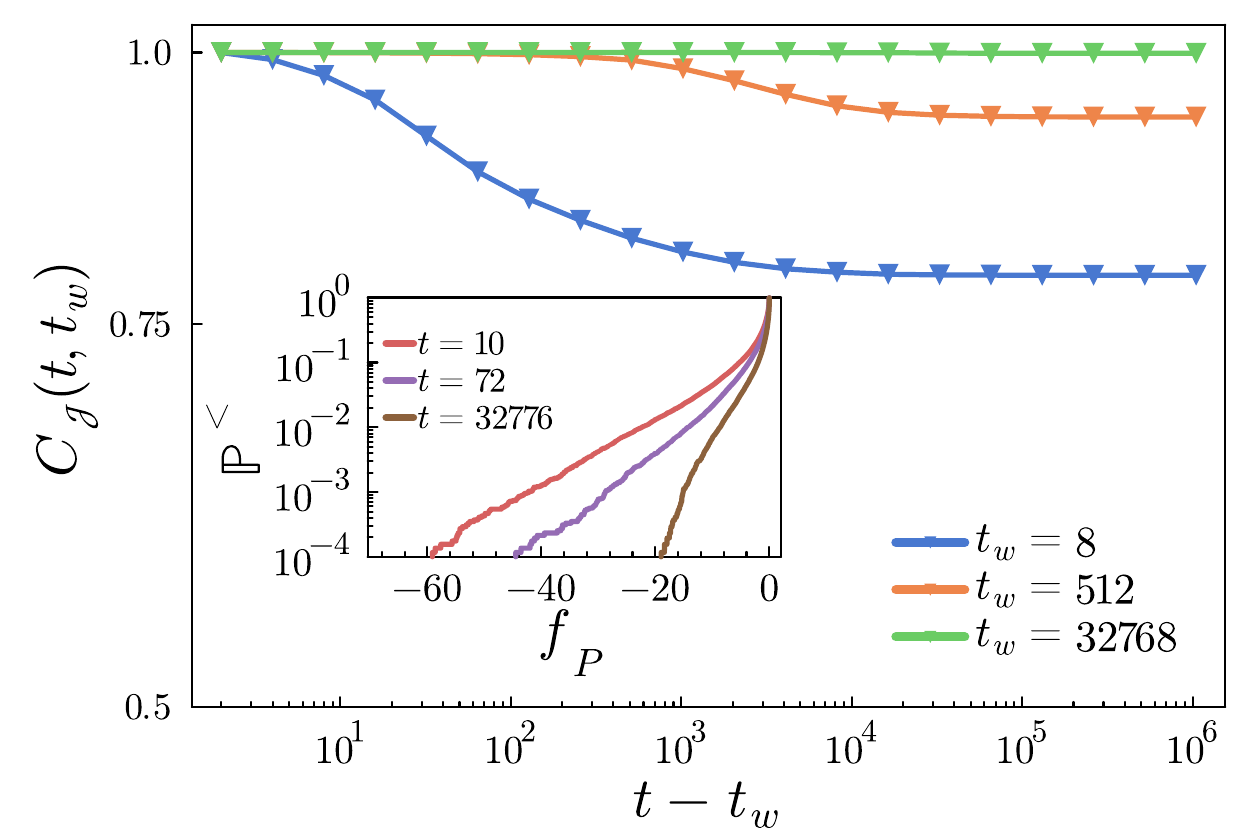}
\vspace{-0.2cm}
\caption{
The two-time correlations of the couplings ${\mathcal J}_{ij}$ in the  
$\Delta$-model with $\Delta = 1.5$, $L=32$ quenched to $T=0$ evolved with SWAP ($p_{\rm swap} = 0.1$). 
The waiting times $t_w$ are given in the key. 
Inset: the cumulative probability of local frustrations $f_P$ at three times after the quench.
}
\label{fig:Jcorrelation-main} 
\end{figure}

The mechanism whereby the swaps accelerate the dynamics is exemplified 
in Fig.~\ref{fig:PlaquetteSWAP}.
For simplicity, in the sketch 
we use two spin lengths only, a large one (long arrows and $S_\pm$),
 and a smaller one (short arrows and $s_\pm$), and  bimodal 
 $J_{ij}=\pm J$ interactions. 
 The energy barrier to  flip  the upper-left highlighted spin is $\Delta E = 
	4J (S^2 + S s) > 0$ (as $S > 0$ and $s > 0$); therefore, this spin is blocked at low temperatures.
Instead, the energy variation after an exchange of the two highlighted spins is 
	 $\Delta E = J (s^2-  S^2) < 0$ (as $s < S$). This non-local move will take place since it is  energetically 
	 favorable, and it may thus help unblocking the upper-left spin and its surroundings. This mechanism
	 will be confirmed by the analysis below.

Frustrated plaquettes remain frustrated since neither the signs of $J_{ij}$ nor ${\mathcal J}_{ij}$ change.
However, the diffusion of the $\tau_i$ can affect the magnitude of the local frustration, quantified
by  $f_P \equiv \prod_{\langle ij\rangle \in P} {\mathcal J}_{ij}(t)$, where the product runs over the links of a plaquette $P$.
The cumulative probability, defined as $\mathbb{P}^{<}(f_P < x) = \int^x_{-\infty} p_{f_P}(y) dy$ for negative $f_P$ i.e. frustrated plaquettes, is plotted for 3 times reached with SWAP after a $T=0$ quench in the inset of Fig.~\ref{fig:Jcorrelation-main}. SWAP reduces the magnitude of the frustration, as the probability of finding large negative values for $f_P$ decreases with time, 
up until a constant functional form is reached. The two-time local correlation, 
$		
C_\mathcal{J}(t, t_w) = \sum_{i, j} [ \langle  \mathcal{J}_{ij}(t) \mathcal{J}_{ij}(t_w) \rangle]/\sum_{i, j} [\langle \mathcal{J}^2_{ij}(t_w) \rangle] 
$,
	displayed in the main part of Fig.~\ref{fig:Jcorrelation-main}, 
becomes stationary and $\lim_{t\gg 1} \lim_{t_w\gg 1} C_{\mathcal J}(t,t_w) =1$. This can also be confirmed by tracking the energy density as a function of time, where a stationary plateau is reached at the same values of $t-t_w$.
Effectively quenched configurations of the effective couplings ${\mathcal J}^*_{ij} = {\mathcal J}_{ij}(t_{\rm max})$
are reached in each run after $\sim 10^5$ 
sweeps in a system with $L=32$. Importantly enough, not all quenches 
lead to the same final  ${\mathcal J}^*_{ij}$ configuration.  
This is observed from the calculation of the correlation of the $\tau_i$ sampled in
different runs of the dynamics (according to the definition in Eq.~(SM30), results not shown).

\begin{figure}[t!]
	\centering
	\includegraphics[width=0.85\linewidth]{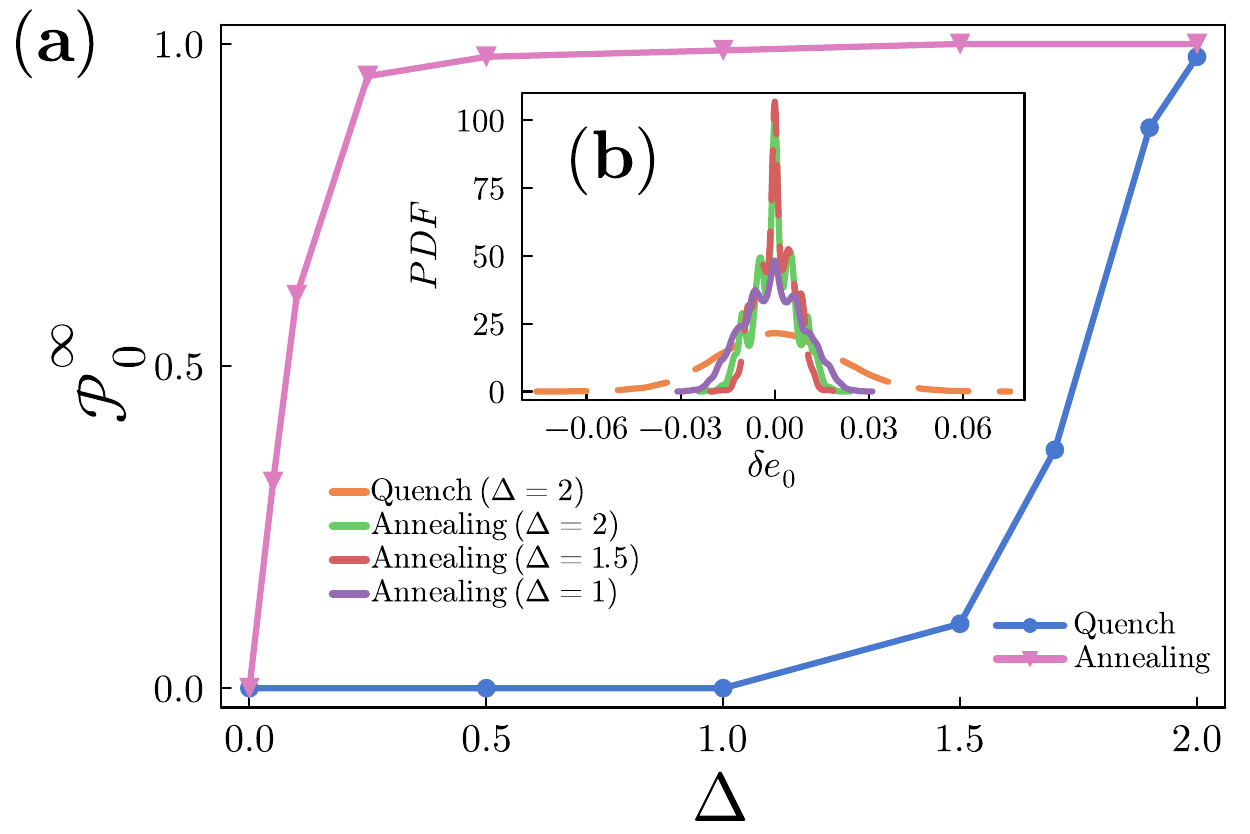}
	\\
	\centering
	\includegraphics[width=1\linewidth]{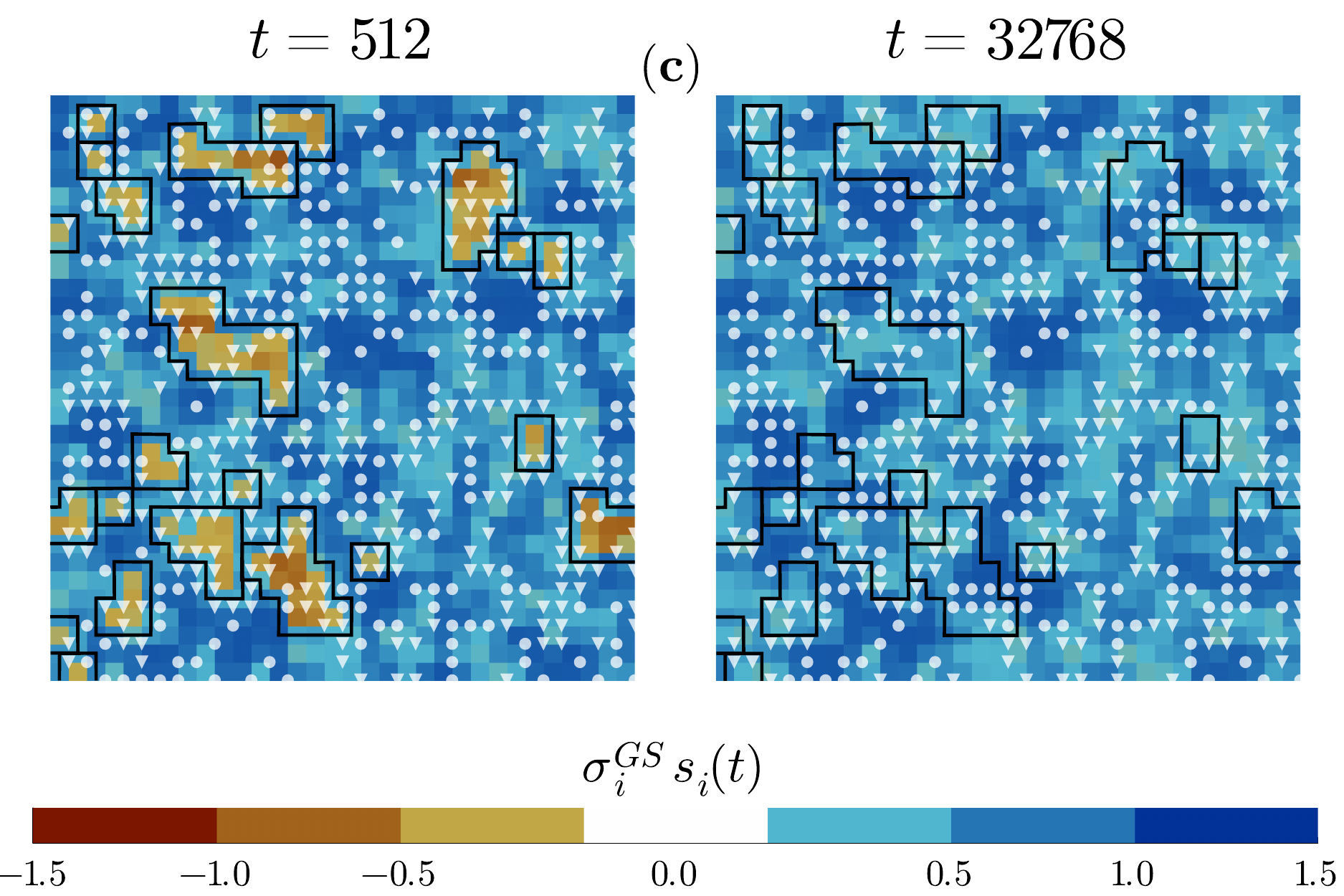}
	\caption{
	(a) Asymptotic probability of reaching a ground state after a $T=0$ quench and a 
	quadratic annealing, starting from $T_0 = 1.0$ during $t_f \approx 10^7$ MCs.
	 $L=32$.  Inset (b)
 	Probability distributions of the ground state energy density differences found after $T=0$ quenches and 
	annealing protocols, Eq.~(\ref{eq:annealing_protocol-def}), of models with different $\Delta$.  
	In all cases, the $J_{ij}$ and initial 
	lengths $\{\tau_i(t = 0)\}$ are the same, and the data are sampled over $10^3$ initial Ising spin conditions $\{\sigma_i(t = 0) =\pm 1\}$. 
	(c) The overlap of an early (left panel) and final (right panel) 
	$s_i$ configuration with the ground state of the model with couplings ${\mathcal J}^*_{ij}
	={\mathcal J}_{ij}(t_{\rm max})
	$, for a quench to $T=0$ with $\Delta = 1.5$. The light bullets and triangles are located
	at frustrated plaquettes with local frustration $f_P$ being greater or smaller than one-half in magnitude, respectively.
	}
	\label{fig:Pvde}
\end{figure}

We now investigate the efficiency of the individual SWAP runs to reach the ground state
of the 2DEA model with interactions $\mathcal J_{ij}^*$.
With this aim we stored the couplings 
${\mathcal J}_{ij}(t) $ and the Ising spins  $\sigma_i(t)$. 
Concomitantly, we used the facility in Bonn~\cite{CJMM22} to find the 
unique (apart from global spin reversal) ground state $ \sigma_i^{\rm gs}$ of a 2DEA 
model with the ${\mathcal J}^*_{ij}$ interactions.
Then, we calculated the overlap and  the probability of reaching the ground state~\cite{Roma09}
\begin{equation}
q(t) = \frac{1}{N} \sum_{i=1}^N \sigma_i^{\rm gs} \sigma_i(t)
\; , 
\ \ \ 
{\mathcal P}_0(t) = \frac{1}{N_{\rm r}} \sum_{\alpha = 1}^{N_{\rm r}} \; \delta_{|q_\alpha(t)|, 1}
\; .
\end{equation}
The index $\alpha$ runs over the simulation runs and $N_r$ is its total number, 
that is, the total number of MC simulations which sample different realizations of the couplings $J_{ij}$, 
the set of lengths $\{\tau_i\}$ and the initial conditions $\{\sigma_i(t=0)\}$. In the simulations shown this number 
was always around $100$.
 Figure~SM18 demonstrates that, for an instantaneous quench to $T = 0$, ${\mathcal P}^\infty_0  \sim 0.5$ for $\Delta =1$, 
 ${\mathcal P}^\infty_0  \sim 0.75$ for $\Delta =1.5$ and 
 ${\mathcal P}^\infty_0$ saturates to one for $\Delta =2$, 
 last blue bullet in Fig.~\ref{fig:Pvde}. 
 Optimization with respect to $p_{\rm swap}$
shows no strong dependence on $p_{\rm swap}$ for values in between 0.05 and  0.5, say. We 
 use $p_{\rm swap}=0.1$ henceforth. However, an inconvenience resides in the fact that even in the cases in which the ground states are reached,
 for the same realization of the  $J_{ij}$ 
and initial conditions of the $\tau_{i}$, 
different thermal evolutions produce different final configurations with a broad 
energy density distribution, inset in Fig.~\ref{fig:Pvde}.
Therefore, although being ground states of the 2DEA model with couplings ${\mathcal J}_{ij}^*$, 
these configurations originate from 
metastable states of the $\tau_i$ variables. We are interested in finding a way of closing this gap, and thus finding a global minimum of the full model in Eq.~(\ref{eqn:definition_softspins}).
\comments{ *** to the SM ****
displays ${\mathcal P}_0(t)$ for three values of $\Delta$ and $p_{\rm swap} = 0.5$. In all cases, after a fast increase
ending at $t \lesssim 10^3$ MCs,   ${\mathcal P}_0(t)$ saturates to a ${\mathcal P}_0^\infty$
which increases with $\Delta$ and gets very close to 1 for $\Delta =2$. Further optimization
of the algorithm achieved by gauging  $p_{\rm swap}$ is studied in
the lower inset which shows the dependence of  ${\mathcal P}_0^\infty$ on $p_{\rm swap}$ in the $\Delta$-model with $\Delta = 2$.
For intermediate values, $0.1 \lesssim p_{\rm swap} \lesssim 0.9$, ${\mathcal P}_0^\infty$ remains approximately 
constant apart from numerical noise, and it decays to zero at the two extremes of 
either non-local spin exchanges ($p_{\rm swap} \to 1$) or pure single-spin-flips ($p_{\rm swap} \to 0$), 
Finally, we checked the dependence on system size in the lower panel using $\Delta =2$ and $p_{\rm swap} =0.1$. 
The curves are similar and the percentage of ground states found is independent of $L$. The upper inset
shows the scaling of ${\mathcal P}_0$ against $t/t^\star(L)$ with $t^\star(L) \sim 1.25 \,  L^{3.75}$. 
*** end comments to SM ***}

 In order to optimize both the  $\tau_i$ and  $\sigma_i$, 
we adopt the thermal protocol~\cite{rubin_dual_2017}
\begin{equation}
	T(t) = T_0 \left(1 - t/t_f\right)^a \; ,
	\label{eq:annealing_protocol-def}
\end{equation}
with $T_0 = 1.0$, $t_f$ the total number of MC-sweeps, and $a=1$ (linear) or $a=2$ (quadratic). 
SWAP is now able to find ground states $100 \%$ of the runs for almost all $\Delta$ values, pink triangles in Fig.~\ref{fig:Pvde}. 
The spread of ground state energies (Gaussian distributed) 
narrows considerably with respect to the one  of $T=0$ quenches, inset in Fig.~\ref{fig:Pvde}
(although it does not disappear completely).  The configurations reached are ground states of models 
with only slightly different $\mathcal J_{ij}^*$. This is confirmed by the 
evolution of the self correlation of the length variables in two
different runs, $\tau_i^{(1)}$ and $\tau_i^{(2)}$: after an initial decorrelation which increases with $\Delta$, 
these variables progressively correlate again to reduce frustration until becoming almost 
identical when $T\to 0$ at the final annealing time
(Fig.~SM23). The snapshots shown in Fig. 3(c) confirm the initial hypothesis illustrated in Fig.~1 and reveal more detailed mechanisms. By analyzing the overlap of an evolving configuration with its respective expected ground state, a quench at $T = 0$ displays clear coarsening behavior. Warm colors (enclosed with black lines) denote droplets, regions where spins are unaligned with the expected ground state, while the cool colors indicate domains where spins are properly aligned. We tag the frustrated plaquettes ($f_P < 0$) at their centers with circles (for $|f_P| > 0.5$) and triangles (for $|f_P| \leq 0.5$) depending on the magnitude of the frustration strength.
At $t = 512$, the frustration decreases in several of the enclosed droplets (more triangles accumulate around the droplets), favoring spin flips that will destroy the droplets and align their spins with the ground-state orientations. After this process, when the ground state has been reached at $t = 32768$, the newly ground-state-aligned spins can acquire a larger magnitude (lowering the energy further) explaining the proliferation of circles in the regions previously occupied by the droplets.

\begin{figure}[h!]
\includegraphics[width=0.85\linewidth]{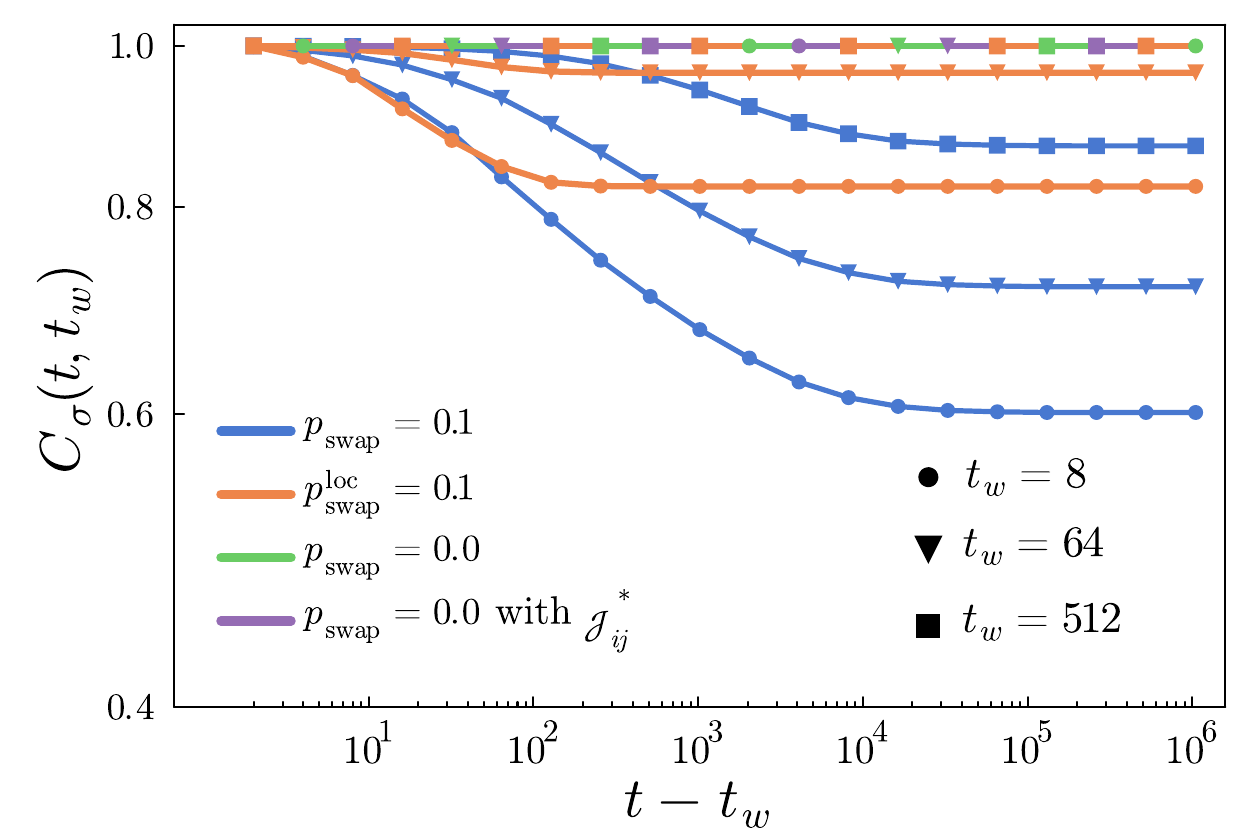}
\caption{
The two-time Ising spin self-correlation at three waiting-times. 
Data for four kinds of $T=0$ evolution of a system with $L=32$ and $\Delta = 1.5$ starting from 
random initial conditions:  \textbf{(i)} SWAP with non-local moves, \textbf{(iv)} local spin exchanges and spin flips, and solely single spin flips of the $\Delta$ model with \textbf{(ii)} random ${\mathcal J}_{ij}$ 
and \textbf{(iii)} optimized ${\mathcal J}^*_{ij}$ couplings. }
\label{fig:Cttw}
\end{figure}

Our SWAP has two ingredients: annealing of disorder because of the $\tau_i$ exchanges 
(absent in the original SWAP method used for interacting particle systems) 
and non-local moves (the essential feature of the original SWAP). To determine if the annealing of the $\tau_i$ generates a less frustrated landscape, for which the ground state can be found more easily, we compared the outcome of \textbf{(i)} our SWAP algorithm to the following cases:
\begin{enumerate}
    \item[\textbf{(ii)}] Purely single spin flip dynamics (i.e. $p_{\textrm{swap}} = 0$) with randomly generated ${\mathcal J}_{ij}$ bonds.
    \item[\textbf{(iii)}] Purely single spin flip dynamics with the optimized ${\mathcal J}_{ij}^*$, found with the quadratic temperature annealing using the strictly non-local SWAP method (with $p_{\textrm{swap}} = 0.1$). In this case the $\tau_i$ variables are quenched but organized in the pattern produced by the SWAP implementation.
\end{enumerate}
Furthermore, we wanted to distinguish the effect of the non-local moves and we also  compared to:
\begin{enumerate}
    \item[\textbf{(iv)}]
     An evolution with only local exchanges, in which we restrict the spin exchanges to be just between  nearest neighbors. As with the non-local SWAP implementation the $\tau_i$'s are still partially annealed along the evolution but only through local moves.
\end{enumerate}
We then calculate the self-correlation $C_\sigma(t,t_w) = N^{-1} \sum_{i=1}^N [\langle \sigma_i(t) \sigma_i(t_w)\rangle]$
after a $T=0$ quench of the $L = 32$ system at three waiting times. The results are plotted in Fig.~\ref{fig:Cttw}, they show that single spin flips, for both ${\mathcal J}_{ij}$ and ${\mathcal J}^*_{ij}$, fail to decorrelate the configurations, as their curves lie in the plateau of $C(t,t_w)=1$ for all waiting times. The ${\mathcal J}^*_{ij}$ are not special in this respect. Besides, while local exchanges are able to decorrelate configurations, they also reach the $C(t,t_w)=1$ plateau at the last waiting time considered ($t_w = 512$), and the associated configuration of the $\sigma_i$ variables is not the ground state of the converged $\mathcal J_{ij}^*$, unlike for the SWAP. Thus, non-local SWAP is more efficient in advancing the evolution than just local exchanges.

\begin{figure}[t!]
\includegraphics[width=0.85\linewidth]{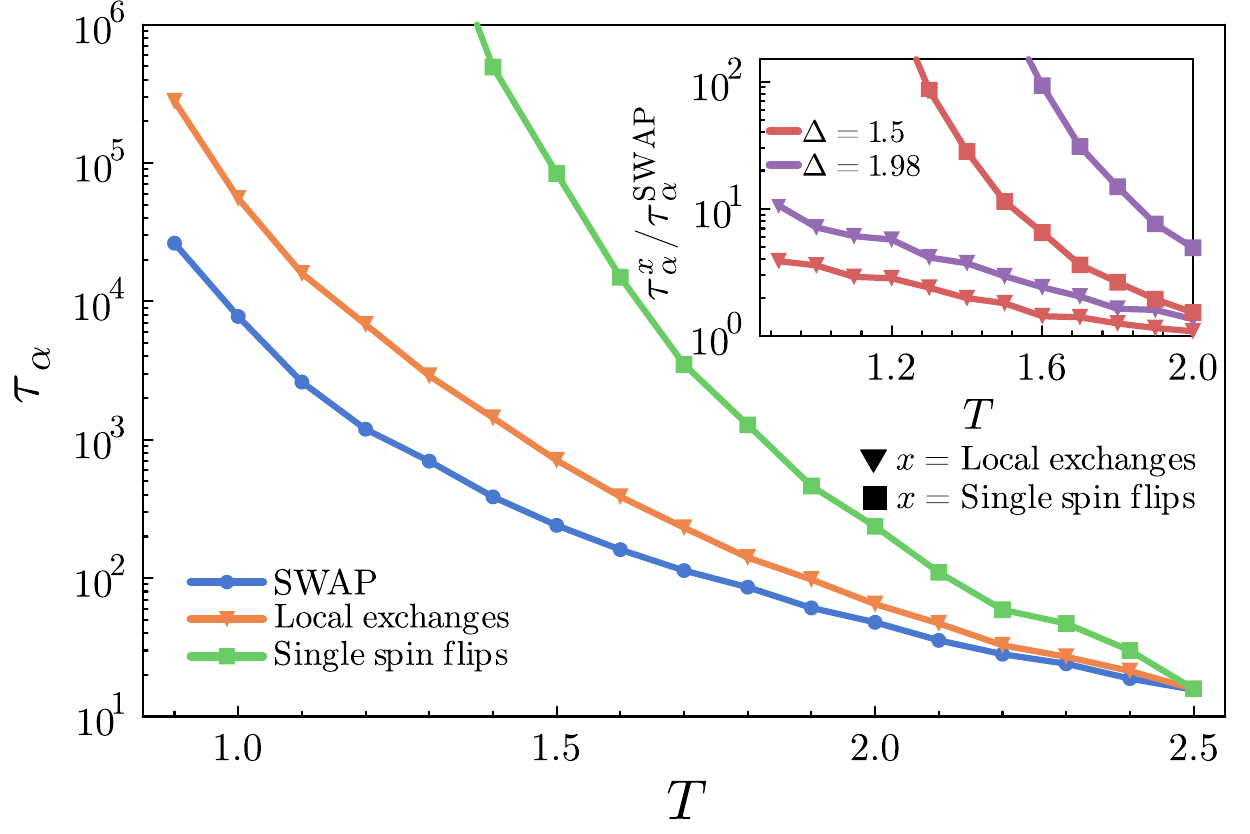}
\vspace{-0.2cm}
\caption{
Characteristic relaxation time $\tau_\alpha$ extracted from the decay of the $\sigma$
self-correlation after quenches to the target temperatures, evolved with \textbf{(i)} SWAP (i.e. non-local moves), \textbf{(iv)} local spin exchanges, both with the same $p_{\rm swap} =0.1$, and \textbf{(iii)} pure single-spin-flip dynamics ($p_{\rm swap} =0$) with optimized bonds ($\mathcal{J}^*_{ij}$), for a model with $\Delta = 1.5$. In the inset, comparison of the two relaxation times, for local exchanges and pure single-spin-flips with respect to SWAP, as a function of temperature for two $\Delta$-models.}
\label{fig:relax-time}
\end{figure}

Finally, we measured the characteristic relaxation time, $\tau_\alpha$, of the algorithms at finite temperature. We have excluded algorithm \textbf{(ii)}, the case of pure single spin flips with randomly generated $\mathcal{J}_{ij}$ bonds, because our previous analysis concluded that its dynamics are similar to those with optimized bonds, $\mathcal J^*$. This time the optimized couplings are taken from the values produced at the end of the non-local SWAP quench when equilibrium is reached, for each respective temperature. We define this $\tau_\alpha$ as the time for which the self-correlation has become age independent (i.e. $C(t,t_w) = \hat{C}(t-t_w)$) and has decayed to $20 \%$ of its original value (i.e.~$\hat{C}(\tau_\alpha) = 0.2~\hat{C}(0)$). As can be seen in Fig.~\ref{fig:relax-time}, the partial annealing of the disorder plays a prominent role in accelerating the dynamics with respect to the single spin flip case with optimized couplings. The addition of non-local moves accelerates the dynamics even further, gaining one decade at $T = 0.9$ with respect to the local exchanges. The three dynamics are indistinguishable at higher temperatures (here, around $T \approx 2.5$).

Let us summarize our results. We adapted the SWAP algorithm to act on finite dimensional spin models. We showed that 
it accelerates the evolution of a 2D disordered model at very low temperatures.
The method
allowed us to sample ground states of an Ising spin-glass with little numerical effort. Other parameters
to optimize, which we have not explored in depth yet, are $P_\tau(\tau_i)$ 
and  the annealing scheme. 
The knowledge gained from the  2D  models studied here 
will serve us to extend this study to the more interesting 3D  
cases with a finite temperature phase transition.

In the disordered spin model, the sluggish dynamics at low temperatures is due to the intricate nature of the (free-)energy landscape. Hence, the efficiency of the SWAP algorithm in accelerating the dynamics can only be attributed to the 
partial smoothing of the landscape, which mitigates higher barriers as needed. This is further evidenced by the lack of acceleration in the non-disordered Ising case, characterized by a ``simple'' landscape. However, the acceleration achieved in our spin model is modest compared to the one accomplished 
in structural glasses, suggesting that facilitation~\cite{chandler2010dynamics} 
may also be at work in the latter case, consistent with previous studies concerned with kinetically constrained models \cite{Gutirrez2019}. Besides, we observe correlated patterns 
of spin lengths, $\tau_i$,  at low temperatures, hinting at similar phenomena in structural glasses, which calls for further investigation.

\noindent
   {\it Acknowledgments.}
    LFC acknowledges  financial support from ANR-19-CE30-0014 and ANR-20- CE30-0031. We thank L. Berthier, G. Biroli, J. Kurchan, E. Marinari, F. Ricci-Tersenghi, F. Rom\'a, J. J. Ruiz-Lorenzo 
    and F. Zamponi  for discusions and suggestions.

\bibliographystyle{apsrev4-1}
\bibliography{References}




\begin{center}
    {\bf SUPPLEMENTAL MATERIAL}
\end{center}




\vspace{0.25cm}
\tableofcontents

\setcounter{equation}{0}
\renewcommand*{\theequation}{SM\arabic{equation}}

\setcounter{figure}{0}
\renewcommand*{\thefigure}{SM\arabic{figure}}

\vspace{1cm}

\twocolumngrid

\section{A. The ferromagnetic Model}
\label{sec:FM}

In this Section we test the performance of the SWAP method when applied to a clean ferromagnetic (FM) model. Concretely, 
we compare the rate of approach to equilibrium of the SWAP algorithm applied to a $\Delta$-model built upon the 
standard 2DIM, to the one of single spin flips applied to the unfrustrated Ising case. In 
both cases we work  below their finite temperature critical points.

We start by generalizing the Hamiltonian of the 2DIM
\begin{equation}
\mathcal{H} = - J \sum_{\langle ij \rangle } s_i s_j = - J \sum_{\langle ij \rangle }  \tau_i \sigma_i \tau_j \sigma_j
\; .
\label{eqn:IsingHamiltonian}
\end{equation}
$J>0$ and, in the rest of this Section we rescale the interactions 
so as to set $J=1$. Concretely, we work with a square lattice with periodic boundary conditions. We use the summation convention in Eq.~(\ref{eqn:IsingHamiltonian}) 
such that the critical temperature of the  Ising model is $T^{\rm IM}_c= 2.27$.

A ferromagnetic model with spins with variable size has been 
considered in~\cite{Krasnytska20,Krasnytska21,Dudka23}. One of the motivation was
to describe ``structurally-disordered magnets'' with 
two (or more) chemically different magnetic components. Specificaly, the critical properties of 
model with variable Ising spin lengths
drawn from a bimodal distribution function, was studied in detail in~\cite{Dudka23}, with special emphasis on 
its critical properties. We will comment on their findings when discussing the critical properties of our model.

At the initial time of the simulation we need to choose the orientation of the Ising spins $\sigma_i$ and the lengths $\tau_i$  of the 
spins $s_i$. For the former we consider two cases, $\sigma_i=\pm 1$ with probability a half, mimicking an infinite temperature
initial state, or $\sigma_i = 1$ for all $i$ representing a zero temperature one. For the latter,  we draw the $\tau_i$ 
independently from the box distribution Eq. (3).

As explained in the main text, we alternate conventional Monte Carlo (MC) updates 
and spin exchanges with probability $p$.
For this model, the energy variation employed in the acceptance probability 
$p_{\rm acc} = \min(1, e^{-\beta \Delta E}) $, is  
\begin{equation}    
	\Delta E =  
	\begin{dcases}
		2 s_i \sum_{j \in \partial i} s_j & \mbox{$\sigma_i \rightarrow -\sigma_i$} \\ 
		\left(s_i - s_j \right) \left( \sum_{k \in \partial i} s_k - \sum_{k \in \partial j} s_k \right) &  \mbox{$s_i \leftrightarrow s_j$} \\ 
	\end{dcases}
\nonumber
\end{equation}
The first line corresponds to the flipping of the $i$-th site and the second one to the exchange between 
the spins on the $i$-th and $j$-th sites, respectively, which are more than one lattice spacing apart.
The symbol $\partial i$ represents the neighbors of the $i$th spin, i.e. the four nearest neighbors on the 
square lattice.
Unless $p_{\rm swap} = 1$, 
the algorithm produces non-conserved order parameter kinetics. 

\subsection{A.1 An equivalent Random Bond Ising Model}
\label{subsec:equivalent-RBIM}

After the introduction of $s_i = \tau_i \sigma_i$,  random interactions between the Ising spins $\sigma_i$ emerge. 
It is straightforward to see that by separating the Ising degrees of freedom and the length-ones, 
the product of the latter plays the role of random couplings in the original Ising Hamiltonian~(\ref{eqn:IsingHamiltonian}).
A  Random Bond Ising Model (RBIM) is then recovered
\begin{equation}
\mathcal{H} = 
 -
   \sum_{\langle ij \rangle } {\mathcal J}_{ij} \sigma_i \sigma_j 
   \qquad \mbox{with} \qquad 
   {\mathcal J}_{ij}=J \tau_i \tau_j  \; ,
\label{eqn:RBIMHamiltonian}
\end{equation}
that is, a specific and structured distribution of the couplings ${\mathcal J}_{ij}$ induced by the one of the $\tau_i$ lengths. 
This distribution is neither  the box nor the bimodal one usually considered in the literature.
Being the variables $\tau_i$ and $\tau_j$ i.i.d.
for $i\neq j$, then $p({\mathcal J}_{ij}) = p_\tau(\tau_i) p_\tau(\tau_j)$, so that the probability distribution function (pdf) of the 
new couplings can be found by means of a Mellin transform, yielding
\begin{equation}    
 p({\mathcal J}_{ij}) = 
    \begin{dcases}
        \, \frac{1}{\Delta^2} \log\left(\frac{{\mathcal J}_{ij}}{u_-^2}\right) \quad & u_-^2 \leq {\mathcal J}_{ij} \leq u_-u_+ \\
       \,  \frac{1}{\Delta^2} \log\left(\frac{u_+^2}{{\mathcal J}_{ij}}\right) \quad &  u_-u_+ \leq {\mathcal J}_{ij} \leq u_+^2 \\
        \, 0 \quad &  \textrm{elsewhere} \\
    \end{dcases}
    \nonumber
\end{equation}
with $u_- = 1 - \Delta/2$ and $u_+ = 1 + \Delta/2$. 

\begin{figure}[h!]
    \centering
    \includegraphics[width=0.87\linewidth]{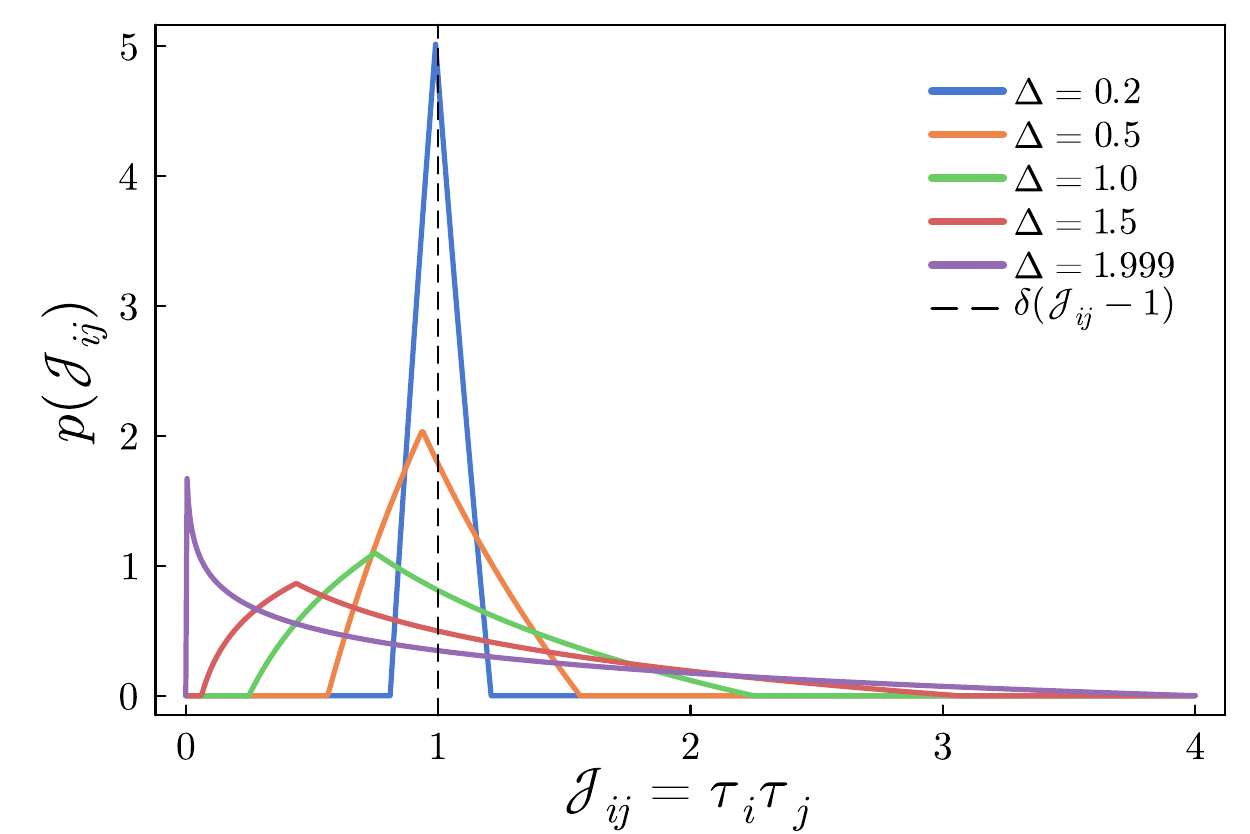}
    \caption{The  probability distribution function of the coupling strengths ${\mathcal J}_{ij} = \tau_i \tau_j$
    ($J=1$)
    arising from the product of the spin-length variables. Several values of the length-controlling parameter $\Delta$ given in
    the key are considered.
    }
    \label{fig:CouplingsDistro}
\end{figure}

The mean and variance  are
\begin{equation}
\label{eq:meanvarferro}
[{\mathcal J}_{ij}] =1 
\; ,
\qquad
[{\mathcal J}^2_{ij}] - [{\mathcal J}_{ij}]^2= 
\frac{\Delta^2}{144}
(\Delta^2 +24)
\; .
\end{equation}
A plot of $ p({\mathcal J}_{ij})$ for several values of the length controlling-parameter $\Delta$
is shown in Fig.~\ref{fig:CouplingsDistro}, where it is clear that as we take $\Delta \rightarrow 0$ the pdf 
tends to a Dirac-delta distribution centered at 1 as it is expected in the hard-spin limit. These new distributions maintain ${\mathcal J}_{ij} \geq 0$
for all $\Delta$, 
so frustration is avoided. 

However, notice that the ${\mathcal J}_{ij}$ are not i.i.d. variables, as for two bonds with a common spin, $k$,  the 
exchanges are correlated
\begin{figure}[h!]
	\begin{tikzpicture}
	\coordinate (i) at (-2,0);
	\coordinate (k) at (0,0);
	\coordinate (j) at (2,0);
	
	\node[fill=black, circle, inner sep=2pt, label=below:$i$] at (i) {};
	\node[fill=black, circle, inner sep=2pt, label=below:$k$] at (k) {};
	\node[fill=black, circle, inner sep=2pt, label=below:$j$] at (j) {};
	
	\draw (i) -- node[below] {${\mathcal J}_{ik}$} (k) -- node[below] {${\mathcal J}_{kj}$} (j);
\end{tikzpicture}
\end{figure}
\vspace{-0.5cm}
\begin{eqnarray}
	&& \;  [{\mathcal J}_{ik} {\mathcal J}_{kj}] =  [\tau_{i} \tau^2_{k} \tau_{j}] = [\tau_{i}] [\tau^2_{k}] [\tau_{j}] = 1 + \frac{\Delta^2}{12}
	\nonumber\\
	&& \qquad\qquad \! \neq   \; [{\mathcal J}_{ik}] [{\mathcal J}_{kj}] = [\tau_i] [\tau_k]^2 [\tau_j] =1
	\; . 
	\label{eq:correlationofJij}
\end{eqnarray}
Therefore, the joint pdf of all couplings ${\mathcal J}_{ij}$ is not just the product of the
individual pdfs $p({\mathcal J}_{ij})$.

We will characterize the dynamics using 
both interpretations: firstly, as a ferromagnetic model with soft-spins and secondly, as a RBIM with the 
above kind of bonds. 

\subsection{A.2 Equilibrium properties} 

In order to appropriately describe the coarsening dynamics we need to first locate the equilibrium critical temperature. 
In particular, we have to establish its dependence on the parameter $\Delta$ and also characterize the equilibrium properties in
the spontaneous symmetry broken phase and close to the critical point. Finally, we have to prove that the equilibrium 
properties do not depend on the microscopic dynamic rules. 

\begin{figure}[h!]
	\centering
	\includegraphics[width = 0.87\linewidth]{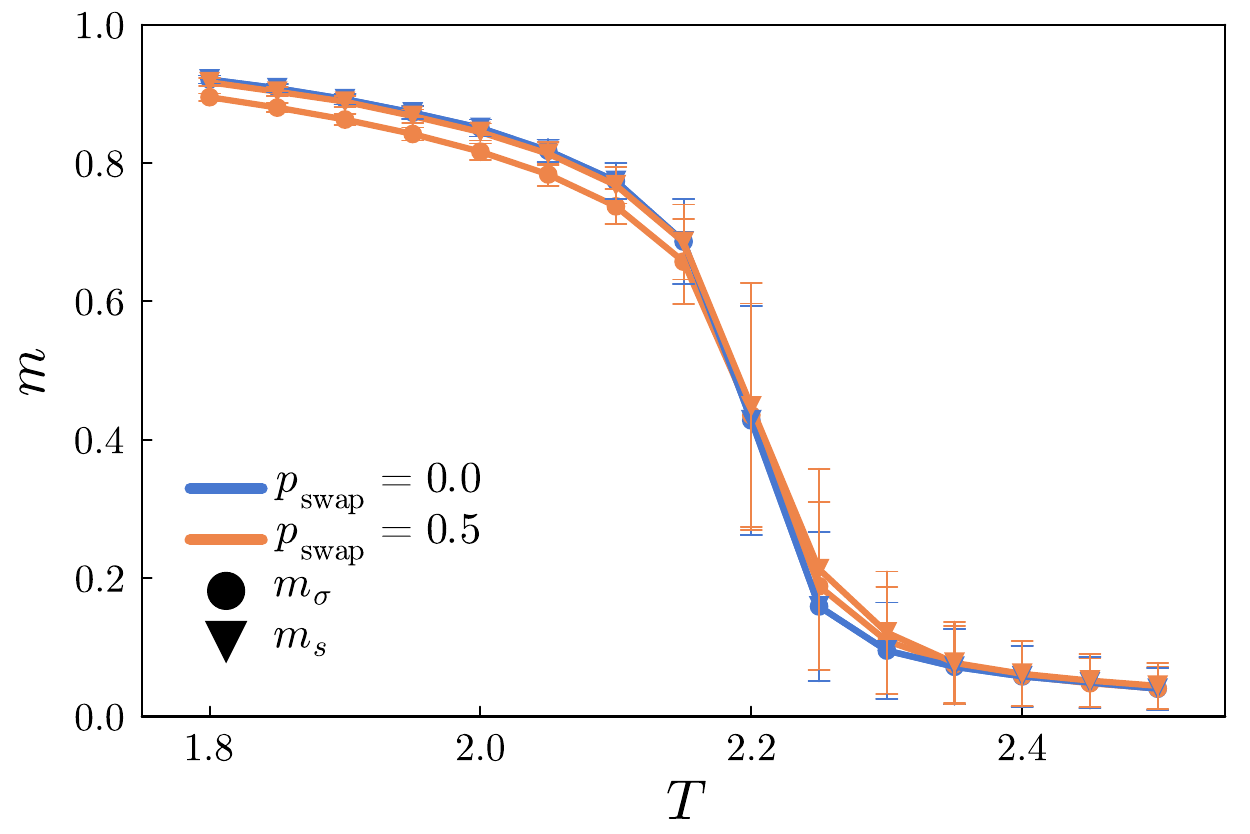}
	\caption{The temperature dependence of both equilibrium magnetization densities (\ref{eq:ms-def}) and (\ref{eq:msigma-def}) in the FM model with $L=160$ and $\Delta = 1$. 
	A hundred zero temperature, completely ordered, initial conditions with different choices of the $\tau_i$ variables were evolved with both single-spin flip dynamics and the SWAP method with $p_{\rm swap} = 0.5$
		during  $\mathcal{T} = 2^{16}$ MC-sweeps. After this time, considered to be sufficient for equilibration,  
		we sampled the magnetization densities at $10^4$ MC-sweeps.
The small difference between the $\sigma$ and $s$ magnetization densities 
drops at high temperatures, as expected for the convergence of $m_\sigma = m_s$ to zero in the paramagnetic phase.
Here and in all other plots the errorbars are estimated from the standard deviations. 
						}
\label{fig:msvTstdvswap.pdf}
\end{figure}

Since the equilibrium configurations below the critical point should be magnetized, 
for this study we initiate all simulations in $\sigma$-ordered configurations, $\sigma_i=1$ for all $i$. 
In this way we force a positive magnetization at low temperatures. The length variables $\tau_i$ 
are drawn from the box distribution initially. In this way we sample different positively magnetized initial 
configurations of the $s_i$ soft spins.

\subsubsection{A.2.1 The averaged magnetizations}

Two magnetization densities can be defined, 
\begin{eqnarray}
m_s &=& \frac{1}{N} \; \sum_{i=1}^{N} \, [\langle s_i\rangle]
\label{eq:ms-def}
\; , 
\\
m_\sigma&=& \frac{1}{N} \sum_{i=1}^{N} \, [\langle \sigma_i\rangle]
\label{eq:msigma-def}
\; , 
\end{eqnarray}
where the angular brackets denote average over thermal noises and the 
square brackets average over the distribution of the $\tau_i$s. If 
one assumes that, in equilibrium, the average $[\langle \tau_i \sigma_i\rangle]$ factorizes as
$[ \tau_i ] [\langle \sigma_i \rangle]$, and 
using the fact that  $[\tau_i] = 1$ (for $L\to\infty$), then
\begin{equation}
m_s = \frac{1}{N} \sum_i \, [ \tau_i ] [\langle \sigma_i \rangle] = m_\sigma
\quad \forall \, T, \Delta
\; .
\label{eq:msequivmsigma}
\end{equation}
This hypothesis is put to the test in Fig.~\ref{fig:msvTstdvswap.pdf} 
where we plot both magnetization densities for an equilibrated  $L = 160$ lattice and a not too strong disorder, 
$\Delta = 1$. The data show a very small systematic deviation, with $m_\sigma \leq m_s$, 
suggesting that there might be a weak correlation between the $\tau_i$ and $\sigma_i$ variables along the evolution, disappearing at 
high temperatures when the paramagnetic disordered phase prevails with $m_s=m_\sigma=0$. 
Indeed, the difference is due to the fact that thermal fluctuations 
tend to favor the reversal of shorter spins in the soft-spin case, since these flips cost less energy 
than the ones of longer spins, and hence $m_s$ is slightly higher than $m_\sigma$ in the 
ordered phase. (The same features are visible within the domains in the out of equilibrium 
snapshots in Fig.~\ref{fig:DomainsD2}.)

In Fig.~\ref{fig:msvTstdvswap.pdf} we  show equilibrium data obtained with the two microscopic 
dynamics under study, pure single spin-flip and SWAP.
The static equilibrium properties are blind to 
the choices $p_{\rm swap} = 0.5$ and $p_{\rm swap} = 0$. The equivalence between the datasets built with the two dynamic
rules can be verified for other values of $p_{\rm swap}$ in the range $0 \leq p_{\rm swap} < 1$
and other values of $\Delta$.

\begin{figure}[h!]
	\centering
	\includegraphics[width = \linewidth]{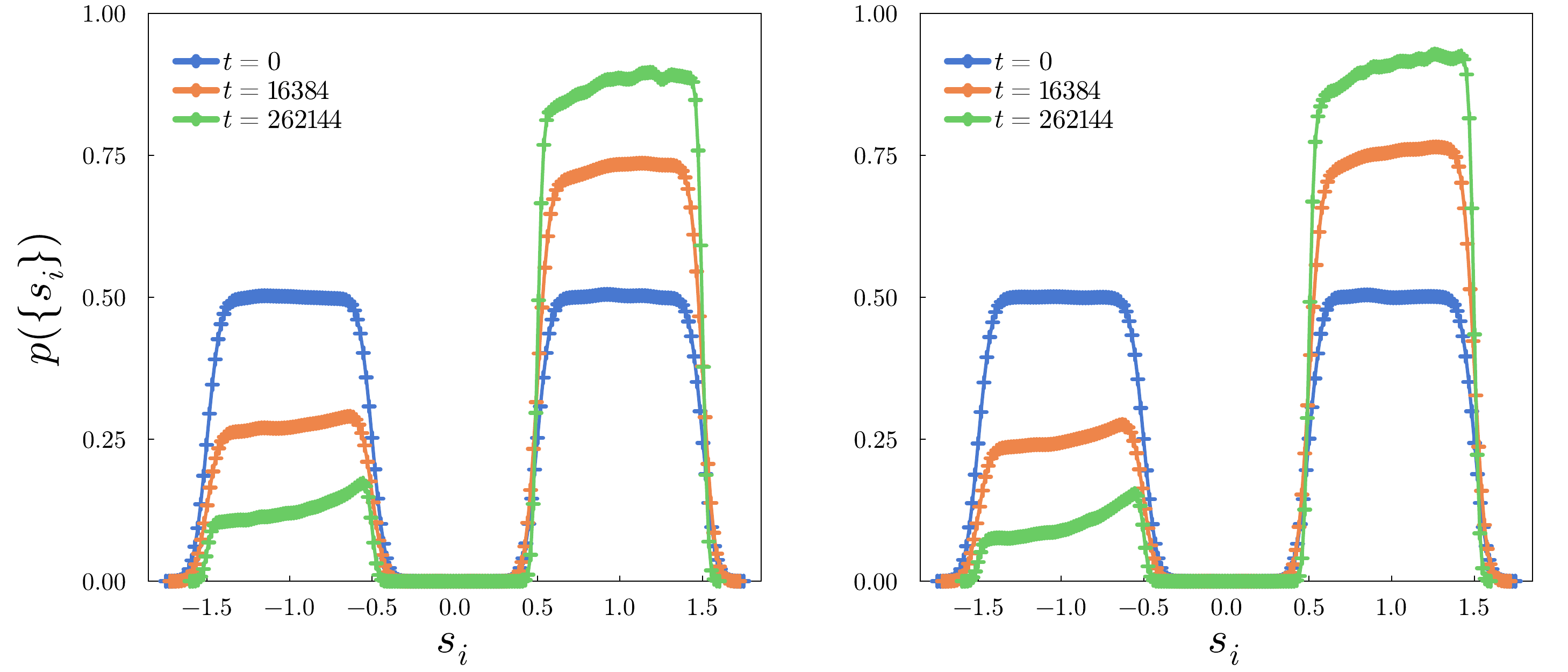}
	\caption{Probability distribution of the spins $s_i$ 
	after a quench to $T = 0.77 \, T_c$ from an infinite temperature initial condition,  with (left) single-spin-flip kinetics and (right) SWAP dynamics ($p_{\rm swap} = 0.5$). Data were sampled using 
	$50$ realizations of the $\tau_i$s, 
	with $L = 128$ and $\Delta = 1$.}
	\label{fig:DistributionSoftSpins_Ferro}
\end{figure}

The distribution of the spin $s_i$ values at three times, $t=0$ and two subsequent ones, for evolutions that led to a positive magnetized domain are shown in Fig.~\ref{fig:DistributionSoftSpins_Ferro}. The $t=0$ data are the distribution of the spin-lengths Eq.~(3), multiplied by $\pm 1$. 
The evolution drives the system to positive magnetization and the weight of the pdf progressively moves to 
the positive support. Note that the two peaks are not symmetric around their midpoints for $t>0$.

\begin{figure}[h!]
	\centering
	\includegraphics[width=0.87\linewidth]{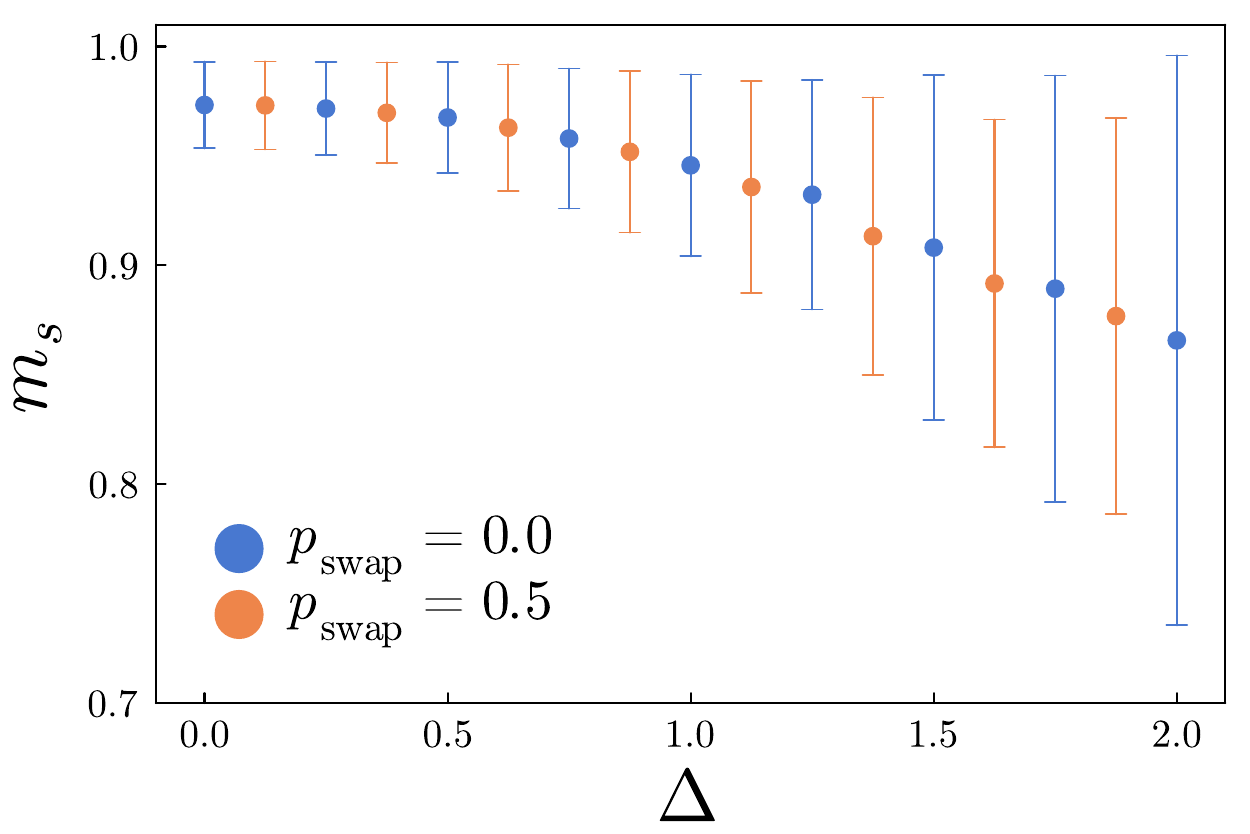}
	\caption{The $\Delta$ dependence of the spin $s$ magnetization density 
		for $L = 16$ using both kinds of dynamics. The temperature is set to $T = 1.67$ in all cases which corresponds 
		to 		$0.74 < T/T_c < 0.84$ depending on the $\Delta$ considered. The equilibration time was set to $2^{18}$ MC-steps
		and data points and error bars were calculated using  initial states with $\sigma_i=1$ and 800 choices of the~$\tau_i$.
	}
	\label{fig:mvDelta2}
\end{figure}

The relaxation time depends on the size of the system, temperature and the length-controlling parameter (or disorder-width), $t_{\rm eq}(L;  T,\Delta)$.
Therefore, by increasing $L$ to reduce finite size effects we are in turn increasing the relaxation time, 
and this  makes equilibrium harder to access. In order to test the equilibrium properties at large values of $\Delta$ and, in particular, 
the fact that they do not depend on the microscopic dynamics, it is convenient to use small $L$. In Fig.~\ref{fig:mvDelta2}
we compare the $\Delta$ dependence of the equilibrium magnetization obtained with single spin flip and SWAP dynamics. 
Just a few averaged values show a deviation, with the SWAP ones being slightly below the single spin flip ones. This 
difference is not systematic and in any case very weak so we do not consider it relevant.

\subsubsection{A.2.2 The phase transition}

 \begin{figure}[b!]
    \centering
     \includegraphics[width = 0.87\linewidth]{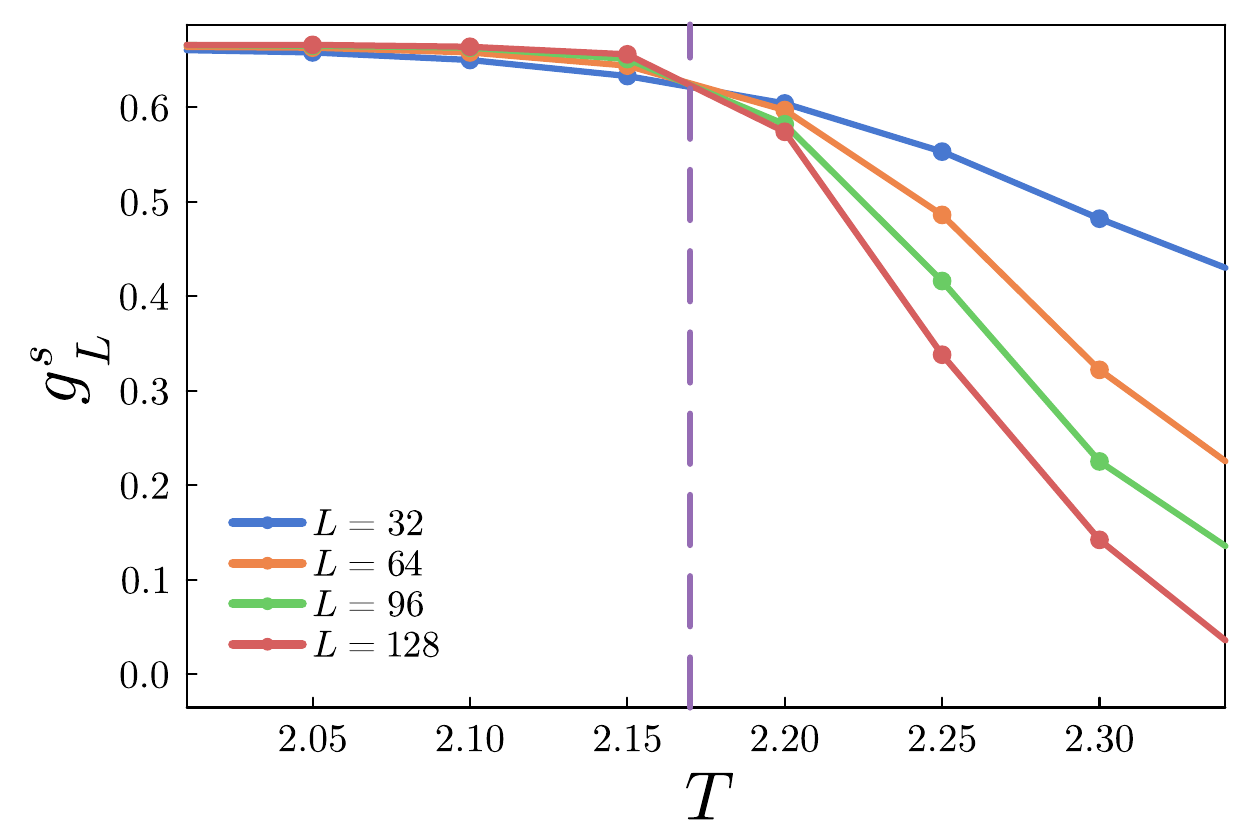}
    \caption{Binder cumulant's temperature dependence for the equilibrium spin $s$ magnetization defined in Eq.~(\ref{eq:binder-def}) obtained 
    with single spin flip dynamics and $\Delta = 1$. The results are equivalent for equilibrium data obtained with 
    the  SWAP evolution with $p_{\rm swap} = 0.5$.   
    }
    \label{fig:BCDelta1}
\end{figure}

\begin{figure}[h!]
\vspace{0.25cm}
	\centering
	\includegraphics[width=0.87\linewidth]{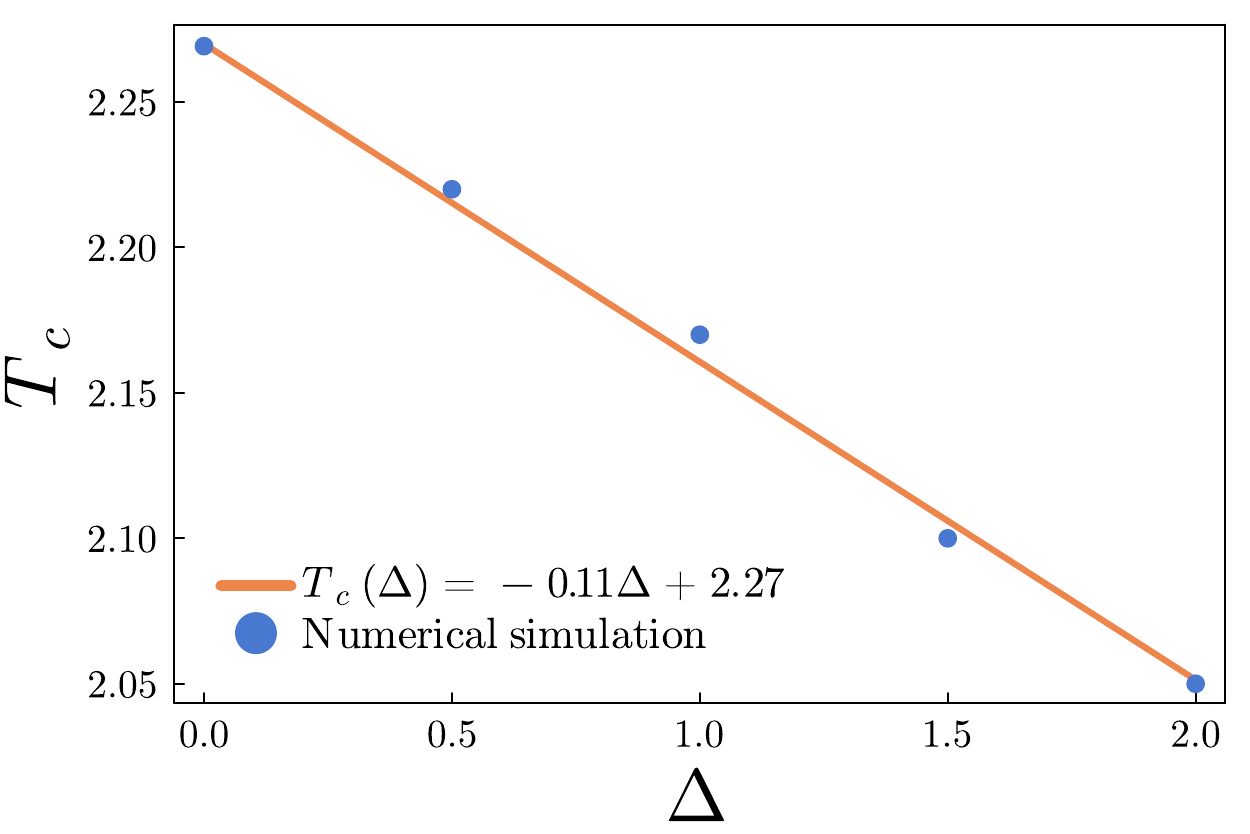}
	\caption{The critical temperature $T_c$  estimated with MC simulation (datapoints) and a linear fit (curve)
		against $\Delta$. The 2DIM critical value is found at $\Delta = 0$. The variation 
		of $T_c$ with $\Delta$ is quite weak.
	}
	\label{fig:phasediagram2}
\end{figure}

A second order ferromagnetic-paramagnetic transition separates magnetized and paramagnetic phases at a $\Delta$ dependent $T_c$.
The Binder Cumulant, defined as\begin{equation}
	g^s_L = 1 -  \frac{\left[ \left\langle \left( \sum^{N}_{i = 1} s_i \right)^4 \right\rangle \right]}{3\left [\left\langle \left( \sum^{N}_{i = 1} s_i \right)^2 \right\rangle \right]^2}
	\; , 
	\label{eq:binder-def}
\end{equation}
allows one to pin-down the critical point. It locates the critical temperature where
 the $g_L^s$ data for different $L$  cross, as displayed in Fig.~\ref{fig:BCDelta1}. We 
 find  $T_c(\Delta = 0) \sim 2.27$ in good agreement with $T^{\rm IM}_c$. Then, 
 $T_c(\Delta = 1) = 2.17$, a slightly lower value than $T^{\rm IM}_c$.  
 In Fig.~\ref{fig:phasediagram2} we plot the full $\Delta$ dependence of the critical 
 temperatures estimated from the crossing of the Binder parameter. The continuous line is a linear fit which represents the
 data rather accurately. The range of variation of $T_c$, $2.05 - 2.27$,  
 with the parameter $\Delta \in [0, 2]$ is of the same order as the one found in other works for the conventional RBIM with a box distribution of couplings and $[J_{ij}] =1$~\cite{henkel_superuniversality_2008}. The analysis 
 of the Binder cumulant of the magnetization $m_\sigma$ and the two equilibrium magnetizations obtained with the SWAP method
 yield a $T_c(\Delta)$ which is equivalent to this one within error bars. 
 
 We have also performed a high temperature
 series expansion analysis (not shown) 
 that tends to confirm that $T_c(\Delta)$ decreases with increasing $\Delta$.

\begin{figure}[h!]
   \vspace{0.25cm}
        \centering
        \includegraphics[width=0.87\linewidth]{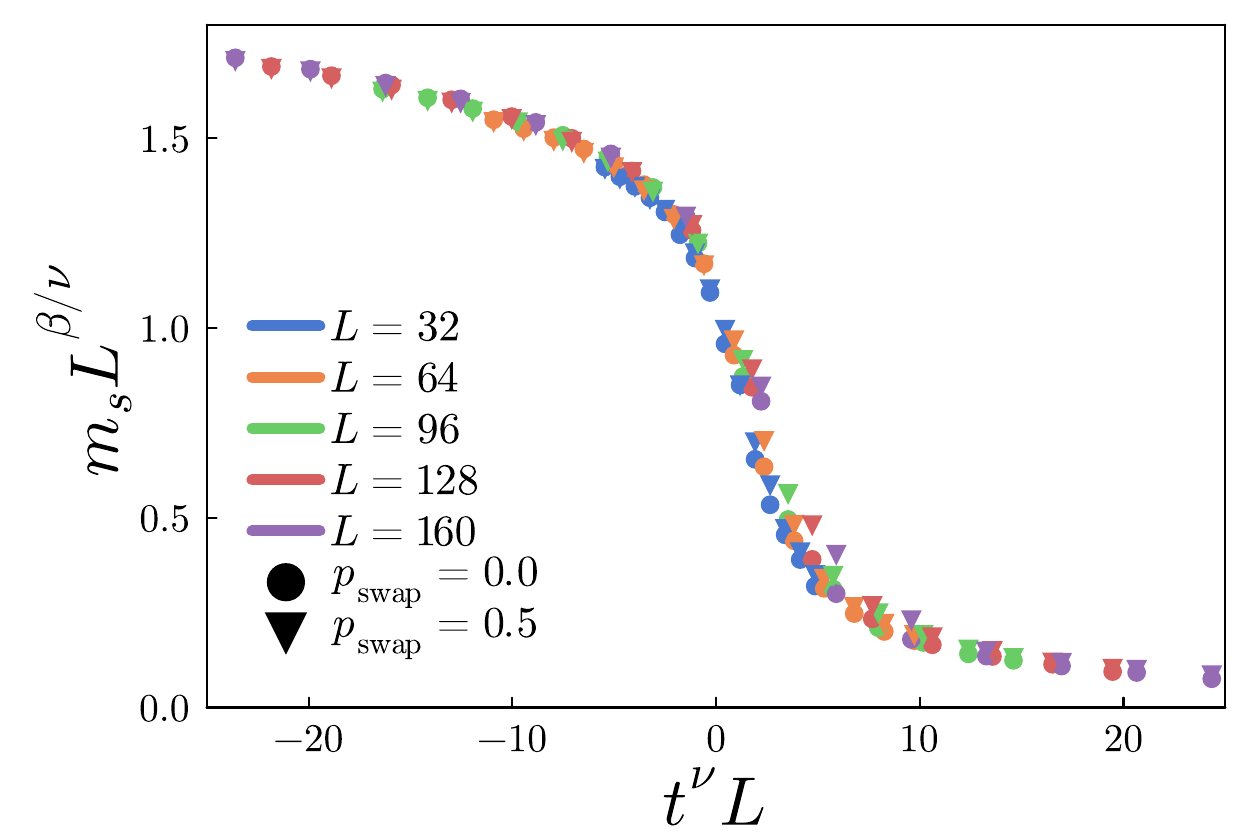}
        \caption{Critical scaling of the equilibrium magnetization density $m_s$, with $t \equiv (T-T_c)/T_c$ the reduced temperature.
        $\Delta = 1$. We fixed  the critical exponents to the values of the clean ferromagnetic 2DIM, 
        $\beta = 1/8$ and $\nu = 1$, and we used the estimated value of $T_c$ for  $\Delta=1$.
        }
        \label{fig:criticalexpDelta1}
\end{figure}

With the critical temperature and its dependency on $\Delta$ assessed, we proceed to probe the equilibrium behavior 
around criticality at $T_c(\Delta)$, and compare it to that of the 2D pure ferromagnetic Ising universality class, for which 
the magnetization and correlation length critical exponents are
\begin{equation}
	\beta = 0.125
	\; , \qquad \qquad \nu = 1 \; ,
	\label{eq:Ising_criticalexp-def}
\end{equation} 
respectively.
In Fig.~\ref{fig:criticalexpDelta1}  the equilibrium magnetizations $m_s$ of the model with $\Delta=1$ and different system 
sizes are scaled using the Ising values~(\ref{eq:Ising_criticalexp-def}). The 
data fall on a single master curve, confirming that the critical properties of the 2DIM are preserved in the $\Delta$-model. 

In~\cite{Dudka23} the critical properties of this very same model with a bimodal distribution of 
lengths was studied analytically and numerically. It was shown in this paper that, apart from a special 
length distribution, in all other cases the criticality is  the one of the 2D dilute Ising model. 
According to the Harris criterium, the 2D dilute Ising model is marginal 
and the critical exponents  (apart from logs) are the same as the ones of the 2DIM.

\subsection{A.3 Dynamical properties}

Having  checked that the equilibrium properties of the $\Delta$-model do not deviate significantly from the ones of the clean ferromagnetic 
Ising universality class,
we proceed to study the domain growth. We start the dynamics from an infinite temperature initial configuration and we perform an instantaneous sub-critical quench. Studying the coarsening phenomena in this set-up will provide us with a direct survey of the microscopic evolution, hence letting us contrast the efficiency of the SWAP method to the one of the standard single-spin-flip kinetics.

\subsubsection{A.3.1 Instantaneous configurations}

We first present snapshots of the system domains along the Monte Carlo evolution. As with the magnetization densities,  there 
are two ways to analyze the domains: 1) to consider the $s_i$ spins as a whole, keeping track of both their sign and length, 
or 2) to  isolate the Ising dependence $\sigma_i$. Clearly, the former carries more information than the latter.

\begin{figure*}
	\centering
	\includegraphics[height = 12cm]{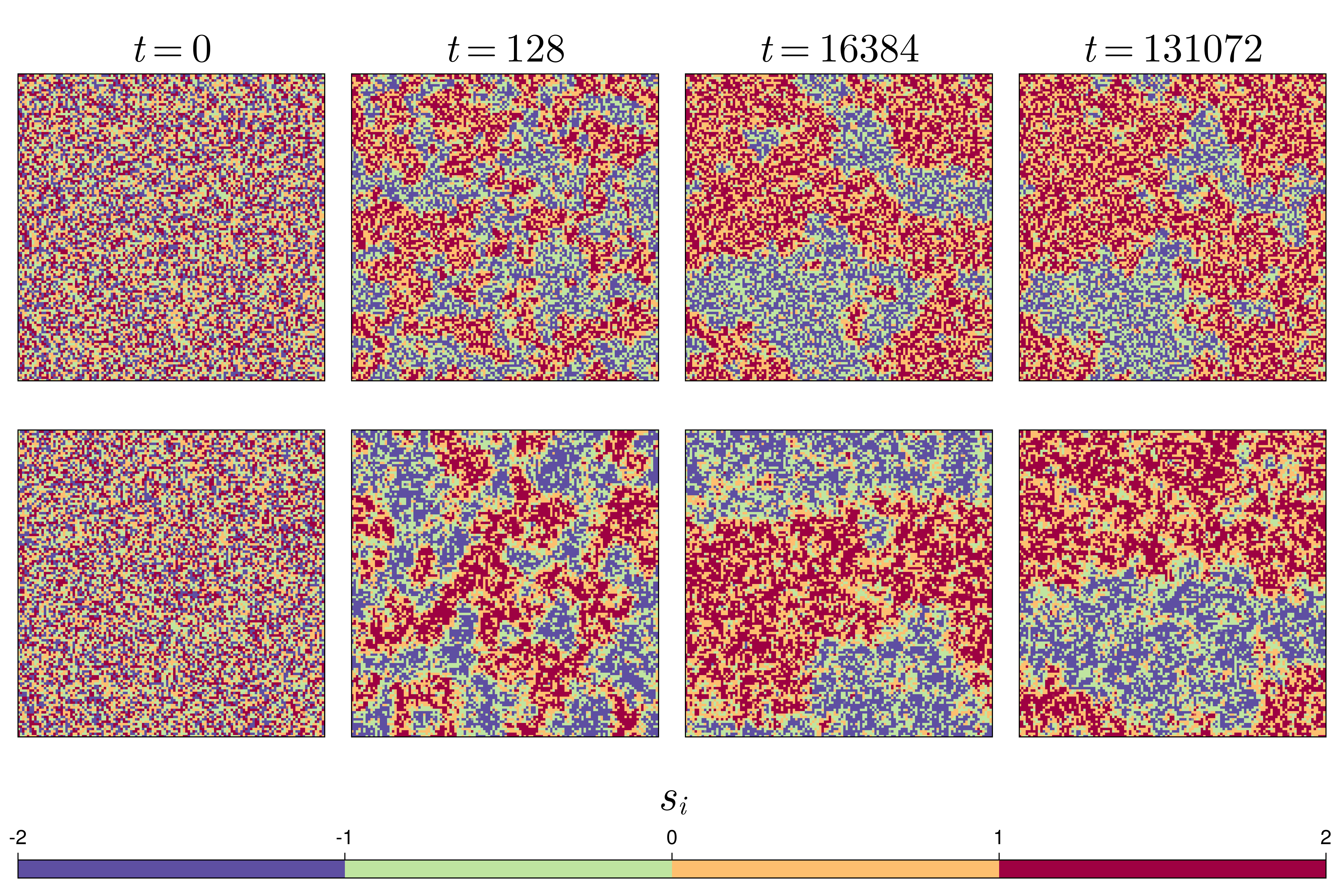}
	\caption{Instantaneous snapshots of the $\Delta$-model with $\Delta = 2$ for quench at $T = 0.8\, T_c$. First row: single-spin-flip dynamics. 
	Second row: SWAP dynamics.
	The color scale binned as indicated in the bar shows the lengths of the local spins. Note that the sizes of the domains with the same orientational order look very similar but the darkness within them different. 
	With SWAP longer spins can get together more easily.
	}
	\label{fig:DomainsD2}
\end{figure*}

Snapshots of the $s_i$ spins obtained with single spin flip and $p_{\rm swap} = 0.5$ SWAP dynamics, are 
displayed in Fig.~\ref{fig:DomainsD2} top and bottom, respectively. As we
lost the binary description of the spin variables, we add a color heat map to distinguish between the different local spin lengths. 

As in any domain growth process the size of the domains increases with time.
However, here, there are fluctuations within the domains with the same orientation but smaller absolute value than the rest
	(green within blue, and yellow within red). The interfaces are typically constituted by a bilayer with short spin length (green and yellow) 
	located in between oriented regions with large spin length (blue and red).
The domains built with SWAP (lower row) allow the spins with longer length to get together more easily and hence 
appear with darker color  than the ones grown with single spin flip dynamics.

To quantify the rate at which domain growth occurs we need to measure the time-dependent typical domain size, $R(t)$, along the simulation. There are several numerical ways to get an indirect or direct measurement of this quantity. In the following we extract it from the inverse domain perimeter density and the space-time correlation. 

\subsubsection{A.3.2 The energy density}

The inverse domain perimeter density, 
\begin{equation}
	R(t) = - \frac{e_{eq}}{e(t)-e_{\rm eq}} = - \frac{e_{eq}}{\delta e(t)}
	\; , 
	\label{eq:inverseperimdensity-def}
\end{equation}
gives a first estimate of the growing length.
The denominator $\delta e(t)$ is the distance between the time-dependent averaged energy density and the equilibrium value, 
$e_{eq} = \lim_{t \rightarrow \infty} e(t)$. 
For the clean Ising model, $R(t) \sim t^{1/z_d}$ with $z_d$ the dynamical exponent.  
This in turn fixes a decay exponent for the energy density difference that would go as $\delta e \sim t^{-1/z_d}$. 
$z_d = 2$ for non-conserved order parameter dynamics while $z_{d} = 3$ for the locally  conserved order parameter ones~\cite{Bray94}. 
Comparing the dynamic exponents should be the simplest  way to see whether one algorithm drives the system towards 
equilibrium faster than the other. 

\begin{figure}[h!]
	\centering
	\includegraphics[width = \linewidth]{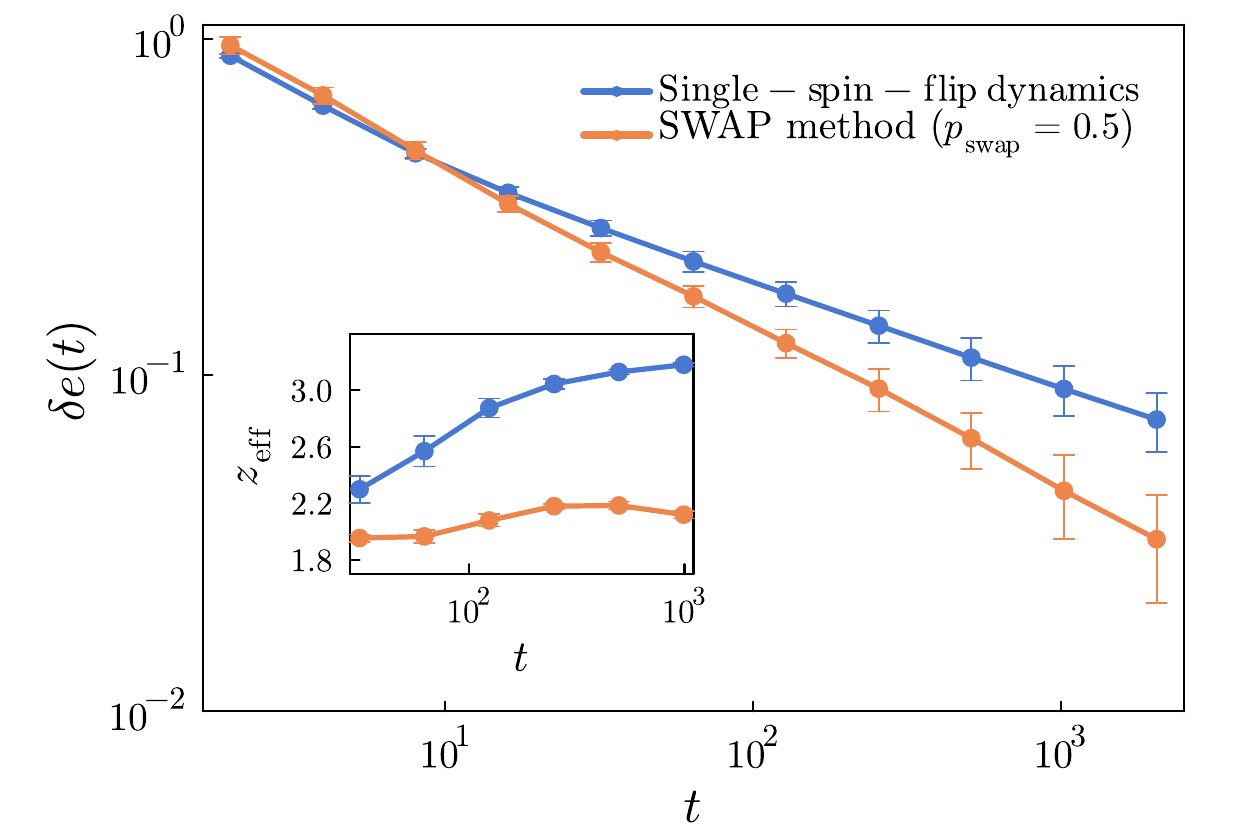}
	\caption{
	Averaged energy density relaxation of the  ferromagnetic $\Delta$-model in double logarithmic scale. 
	Disordered initial configurations prepared at $T_0 = \infty$ were evolved at $0.55 \, T_c$
		using single-spin flip and SWAP kinetics with $p_{\rm swap} = 0.5$. Data correspond to $\Delta = 1$ and $L = 128$, 
		and were averaged over $100$ different 
		initial conditions for both the $\sigma_i$s and  the $\tau_i$s. 
	A power law fit over moving time windows including three data points 
	yields the effective exponent $z_{\rm eff}$ reported in the inset. 
	}
	\label{fig:energydensity_witheffexp}
\end{figure}

We calculated the equilibrium energy density in the long-time limit ($t >2^{18}$ MC-sweeps)
of runs initiated in $\sigma$-ordered initial conditions and averaged 
over $100$ realizations of the $\tau_i$s. The time-dependent $e(t) = [\langle {\mathcal H}(t)\rangle]/N$ 
was computed, instead, after quenches from 
completely disordered initial conditions, with parameters such that the equilibrium state is ordered. 
The decay of the excess energy curves is shown in Fig.~\ref{fig:energydensity_witheffexp} 
in double logarithmic scale. A power law fit of the energy decays obtained with both methods leads to
a time-dependent effective exponent $z_{\rm eff}$ reported in the caption.

After a short transient in which the effective exponent of the single spin flip dynamics is close to $2$, the 
decay of the excess energy slows down and the effective exponent  increases
in time reaching a value close to 3 in the considered time window (blue data points in the main panel and the 
inset). The reason for this is that the 
variable length model has, from the point of view of the Ising spin variables 
$\sigma_i$, quenched random bonds. Therefore, the single spin flip dynamics feels these randomness
and its evolution is slowed down, with the effective exponent developing a dependence on the 
disordered strength, as pointed out in the literature~\cite{huse_pinning_1985,rivera_diluted_1990,Sicilia08,iguain_growing_2009,corberi_growth_2011,corberi_coarsening_2017, paul_domain_2004,henkel_superuniversality_2008}.

The SWAP method circumvents the slowing down introduced by the variable length of the spins. 
The dynamical exponent of single-spin-flip kinetics of the 2DIM~\cite{Bray94}, $z_d=2$, 
is recovered when one adds spin exchanges (orange data points in the main panel and the inset).

\subsubsection{A.3.3 The space-time correlation}
\label{sec:space-time}

We now focus on the one-time spatial correlation function which also carries information on the typical domain size along the evolution. 
As with the magnetization densities, we either measure the space-time correlation of the complete spin-variables,
\begin{eqnarray}
	\!\!\!\!\!\!\!\!\!\!\!\!\!\!\!\!\!\!\!\!\!
	N C_s(r,t) 
	&=& 
	\sum^{N}_{j=1}\sum^{N}_{k=1} [\langle s_{j}(t) s_{k}(t) \rangle] \Big{|}_{|\vec r_j -\vec r_k| =r}
	\nonumber\\
	&=&
	 \sum^{N}_{j=1}\sum^{N}_{k=1} [\langle \tau_{j} \sigma_{j}(t)  \tau_{k} \sigma_{k}(t) \rangle]\Big{|}_{|\vec r_j -\vec r_k| =r}
	 \label{eq:Cs-def}
	\end{eqnarray} 
where the double sum over indices $j$ and $k$ is restricted by the condition on the distance between the spins. 
We normalize by the number of terms considered.
Or else we measure the space-time  correlations of the Ising variables, in direct correspondence with the RBIM interpretation, 
\begin{equation}
	N C_\sigma(r,t) = \sum^{N}_{j=1} \sum^{N}_{k=1} [\langle \sigma_{j}(t)  \sigma_{k}(t) \rangle ] \Big{|}_{|\vec r_j -\vec r_k| =r}
	\; .
	\label{eq:CorrelationSigma}
\end{equation} 
In both cases, $\langle \dots \rangle$ is the average over $\sigma_i$ initial conditions and noise realizations of the 
dynamics on the one hand,  and over the 2D  equidistant lattice sites on the other. As stated in prior Sections, $[...]$ represents a \textit{disorder average} over several realizations of the lengths $\{\tau_i\}$. 
 We note that 
\begin{eqnarray}
C_s(r=0,t) &=& 
\frac{1}{N} \sum_j \, [\tau^2_j ] = 1+ \frac{ \Delta^2}{12}
\; , 
\\
C_\sigma(r=0,t) &=& 1
\; ,
\end{eqnarray} 
at all times $t$.

Dynamic scaling states that a {\it single} domain length $R(t)$ should scale all correlation functions. 
We now study separately the correlations of the $s_i$ and $\sigma_i$ spins to confirm that this
is indeed the case in this problem.

Not too close to the critical temperature, where the equilibrium magnetizations are not too low, 
one can estimate the typical domain size $R_{s, \sigma}(t)$  from the $r$ 
such that the correlations $C_{s, \sigma}(r,t)$ decay to, say, $1/e$ of their zero distance values.
A dynamic scaling regime, in which
\begin{equation}
C_{s,\sigma}(r,t) \sim f\left(\frac{r}{R_{s,\sigma}(t)}\right)
\end{equation}
is expected for $\xi_{\rm eq} \ll r \ll L$ with $\xi_{\rm eq}$ the equilibrium correlation length and $L$ the linear system size. Numerically, it is convenient to measure 
$r$  along any of the two axes of the 2D lattice with PBCs.

\begin{figure}
    \centering
    \includegraphics[width = \linewidth]{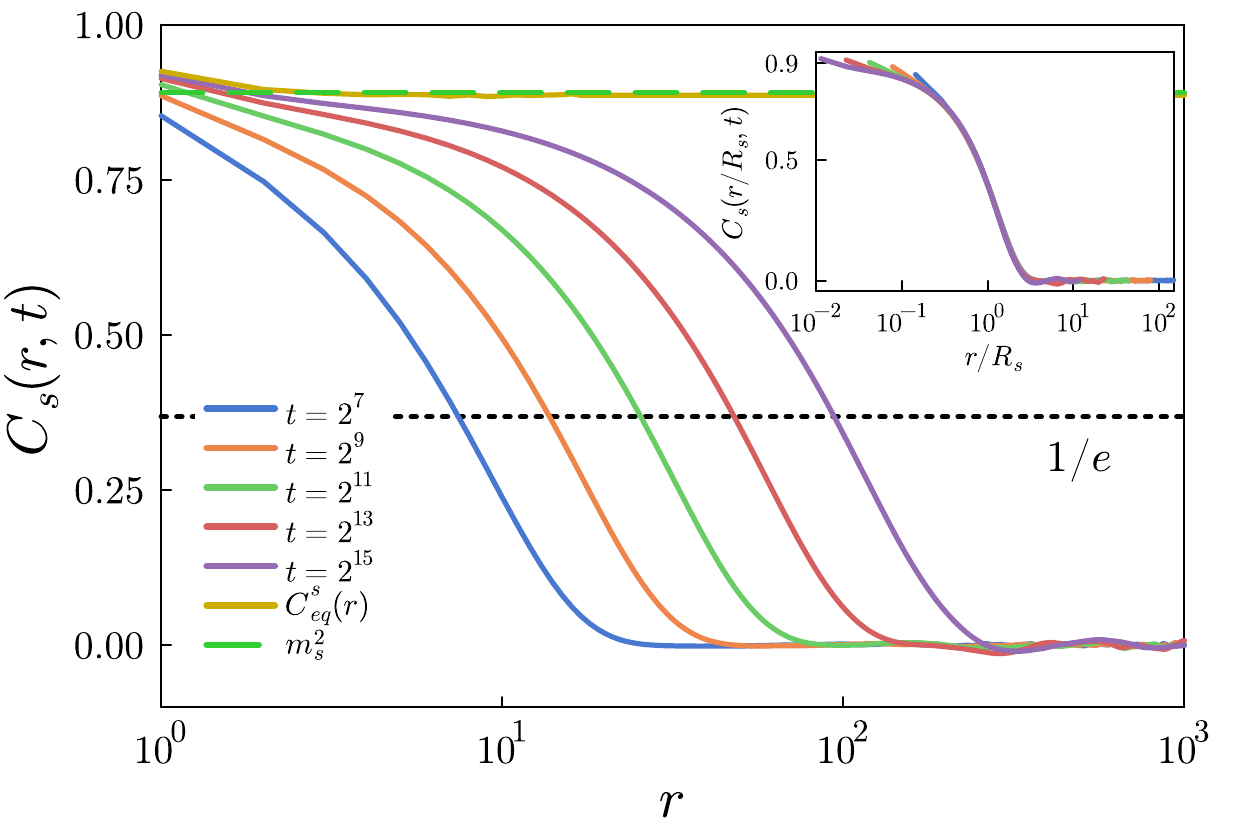}
    \caption{SWAP dynamics with $p_{\rm swap} = 0.5$ of
  an $L = 2048$ ferromagnetic 
    system with $\Delta = 1$, quenched from $T_0 = \infty$ to $T \approx 0.77~T_c$.
Main panel: space-time correlation of the $s$ spins at several times given in the key. 
The solid (gold) line is the equilibrium correlation while the dashed green line is $m^2_s$ in equilibrium.
    	Inset: test of dynamic scaling with the typical domain size $R_{s}(t)$ estimated from $C_{s}(R_s(t),t) = C_{s}(r=0,t)/e$, with $C_s(r=0,t)=1+\Delta^2/12$.    
	Averages are performed over $10$ runs.    }
    \label{fig:CorrelationDecayDelta1swap}
\end{figure}

The decay and scaling of the space-time correlation
$C_s(r,t)$ in a model with $\Delta=1$ are studied in Figs.~\ref{fig:CorrelationDecayDelta1swap}, 
using the SWAP method with $p_{\rm swap}=0.5$ and single spin-flips.
At different times, the curves differ, demonstrating, once again, the out of equilibrium character of the dynamics.

The (golden) curve, which approaches at far distances a finite constant,
corresponds to the correlation decay of an ordered initial configuration that we heated up to the desired temperature $T < T_c$
and evolves in equilibrium. At very large $r$, the two spins in Eqs.~(\ref{eq:Cs-def}) and (\ref{eq:CorrelationSigma}) are expected to 
become independent and the average factorize, $\langle s_j s_k \rangle \sim \langle s_j \rangle \langle s_k\rangle$, leading to
\begin{eqnarray}
	\lim_{r\to\infty} C^s_{\rm eq}(r) = [\langle s_i\rangle^2] = m_s^2 
	\; ,
	\\
	\lim_{r\to\infty} C^\sigma_{\rm eq}(r) = [\langle \sigma_i\rangle^2] = m_\sigma^2 
	\; .
\end{eqnarray} 
Moreover, in equilibrium the two magnetization densities are almost identical.
The first of these limits is verified numerically in Fig.~\ref{fig:CorrelationDecayDelta1swap}
and the other one as well.

\begin{figure}
    \centering
    \includegraphics[width=0.87\linewidth]{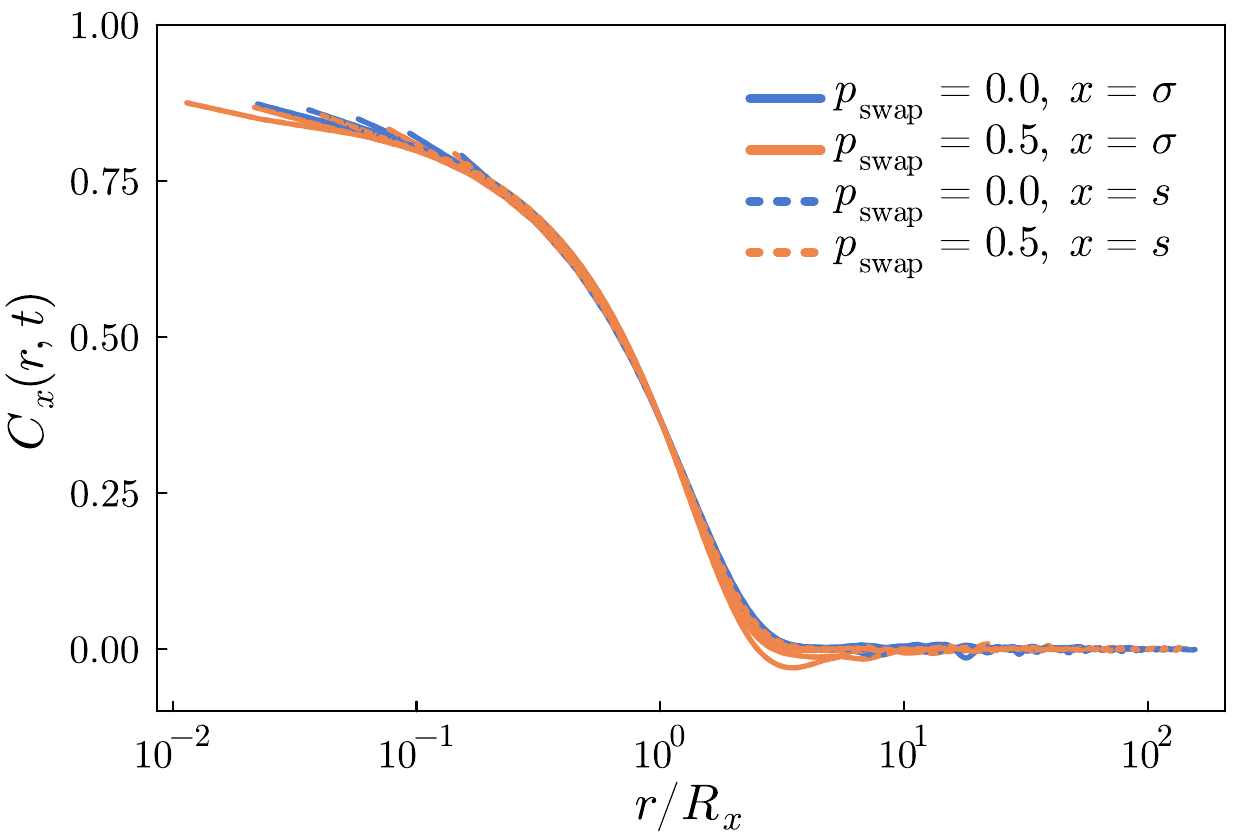}
    \caption{Scaling of the space-time correlations $C_s$ and $C_\sigma$
    for the single-spin flip ($p_{\rm swap} = 0$) and SWAP dynamics with $p_{\rm swap} = 0.5$. 
    The same master curve describes both sets of data.
    Same parameters as in Fig.~\ref{fig:CorrelationDecayDelta1swap}.
}
    \label{fig:Scaling}
\end{figure}

Dynamic scaling
 is checked in the inset of Fig.~\ref{fig:CorrelationDecayDelta1swap}
 and it works equally fine for the $\sigma$ correlations, see Fig.~\ref{fig:Scaling}.
  
 The dynamical scaling master curves, $f(x)$, for $C$ not too close to zero coincide, as can be seen in Fig.~\ref{fig:Scaling}.  
There are differences when the $C$s get close to zero, with oscillations for 
SWAP which are absent for single spin flip (somehow 
reminiscent of the oscillations also present when working with local-spin-exhanges as in the case of Kawasaki dynamics).

\subsubsection{A.3.4 The growing length for the soft spins}

\begin{figure}[h!]
    \centering
    \includegraphics[width=\linewidth]{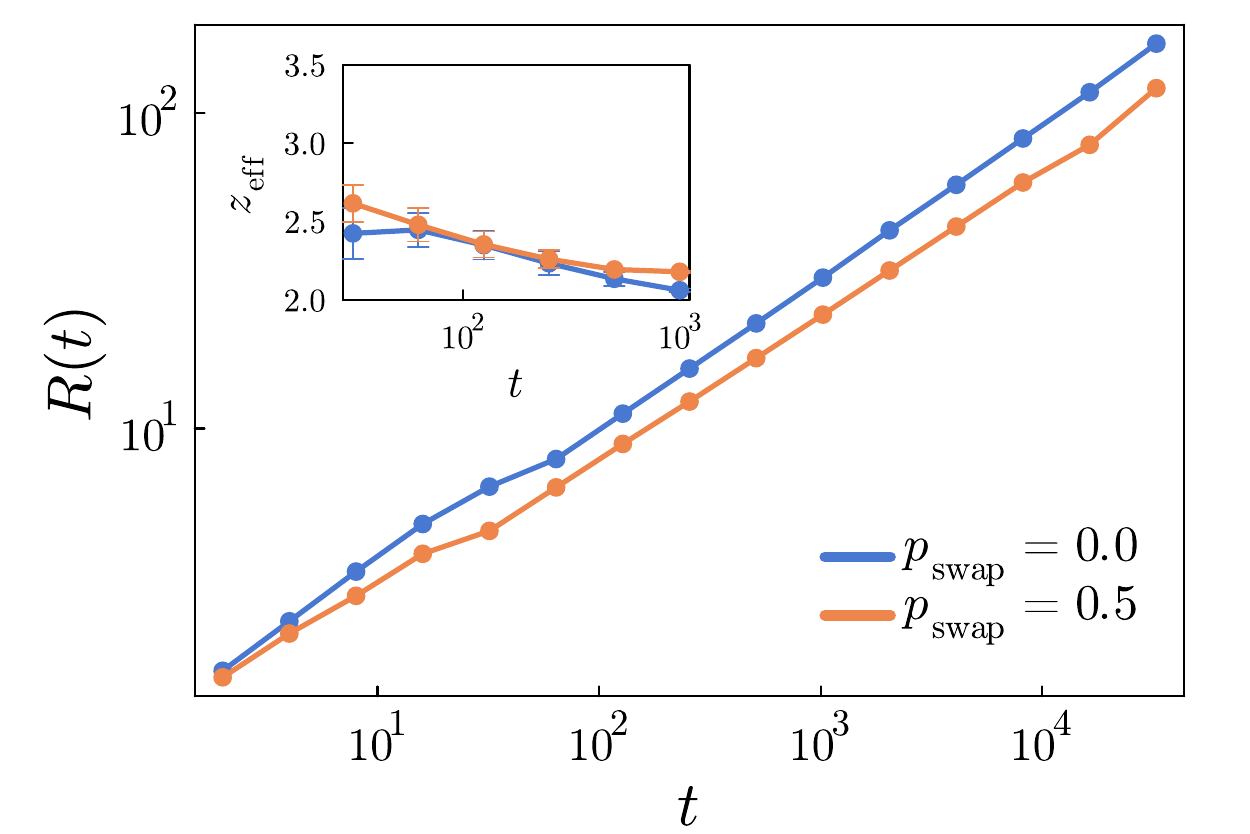}
    \caption{The growing length of the clean Ising model, $\Delta = 0$, with single spin flip and SWAP dynamics. 
     The inset shows the time dependence of the effective exponent $z_{\rm{eff}}$, 
     measured by performing a fit of the data at (moving) six consecutive times.}
    \label{fig:EffectiveExponentDelta0}
\end{figure}

\begin{figure}[h!]
    \centering
    \includegraphics[width=\linewidth]{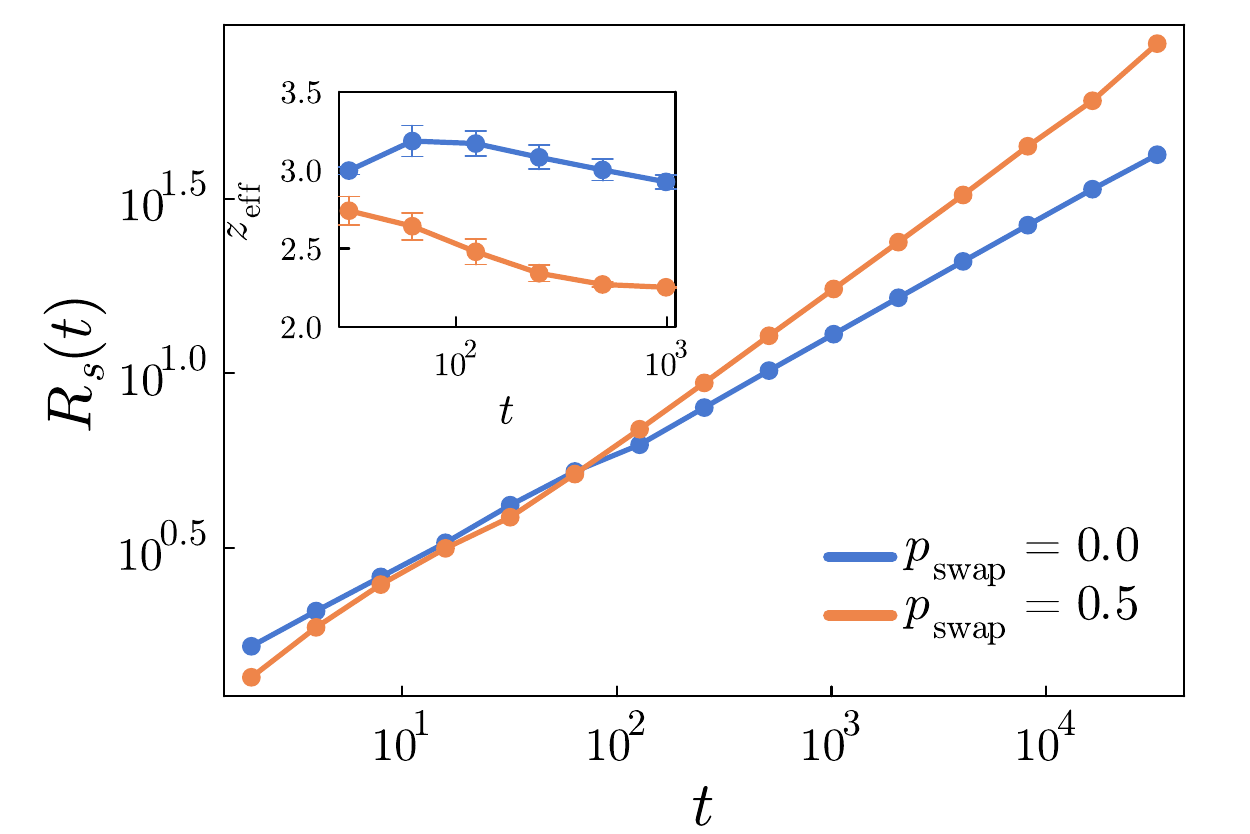}
    \caption{The growing length $R_s$ (estimated from the analysis of the $s$ spins space-time correlation) of the $\Delta = 1$-model, with single spin flip and SWAP dynamics. 
    Sub-critical quench to $T \approx 0.74~T_c$ and $L=2048$.
     In the inset, the time dependence of the effective exponent $z_{\rm{eff}}$, 
     measured by performing a fit of the data at (moving) six consecutive times. }
    \label{fig:EffectiveExponentDelta1}
\end{figure}

The growing length $R_s$ measured from the spins $s$ space-time correlations 
generated with the two updating rules are studied in Fig.~\ref{fig:EffectiveExponentDelta0} and Fig.~\ref{fig:EffectiveExponentDelta1}. In the former, the clean ferromagnetic
Ising model's $R_s = R$ is analyzed as a benchmark. In the latter, the $R_s$ of the soft spin model is studied.
As stated before, the typical domain sizes measured are fitted via a power law, with an effective exponent which depends on the width~$\Delta$, 
\begin{equation}
	R_s(t) = \lambda \, t^{z^{-1}_{\textrm{eff}} (\Delta)}
	\label{eq:Rtyp_def}
	\; .
\end{equation}
$\lambda$ is a non-universal parameter.
The exponent $z_{\textrm{eff}}$ is measured by taking averages over successive  time-windows along the domain growth, as we did in the previous Section when the inverse perimeter density was calculated.
The convergence of $z_{\rm eff}$ towards $z_d=2$ in the clean case is verified in Fig.~\ref{fig:EffectiveExponentDelta0} for single-spin-flip kinetics, while the SWAP method is unable to accelerate the dynamics, and produces a slightly slower convergence.

In Fig.~\ref{fig:EffectiveExponentDelta1} we study $z_{\rm{eff}}$ for the soft-spin model with $\Delta = 1$. 
When the system is evolved with single-spin-flip kinetics, 
we get  $z_{\rm eff} \sim 3$, and these update rules are not convenient.
However, when we implement the SWAP method, the effective exponent decreases to  $2.125$, 
a  value that is very close to the theoretical $z_d=2$ of the regular 2DIM with non-conserved order parameter dynamics. 

\subsubsection{A.3.5 The growing length for the Ising spins}

\begin{figure}
	\centering
	\includegraphics[width=\linewidth]{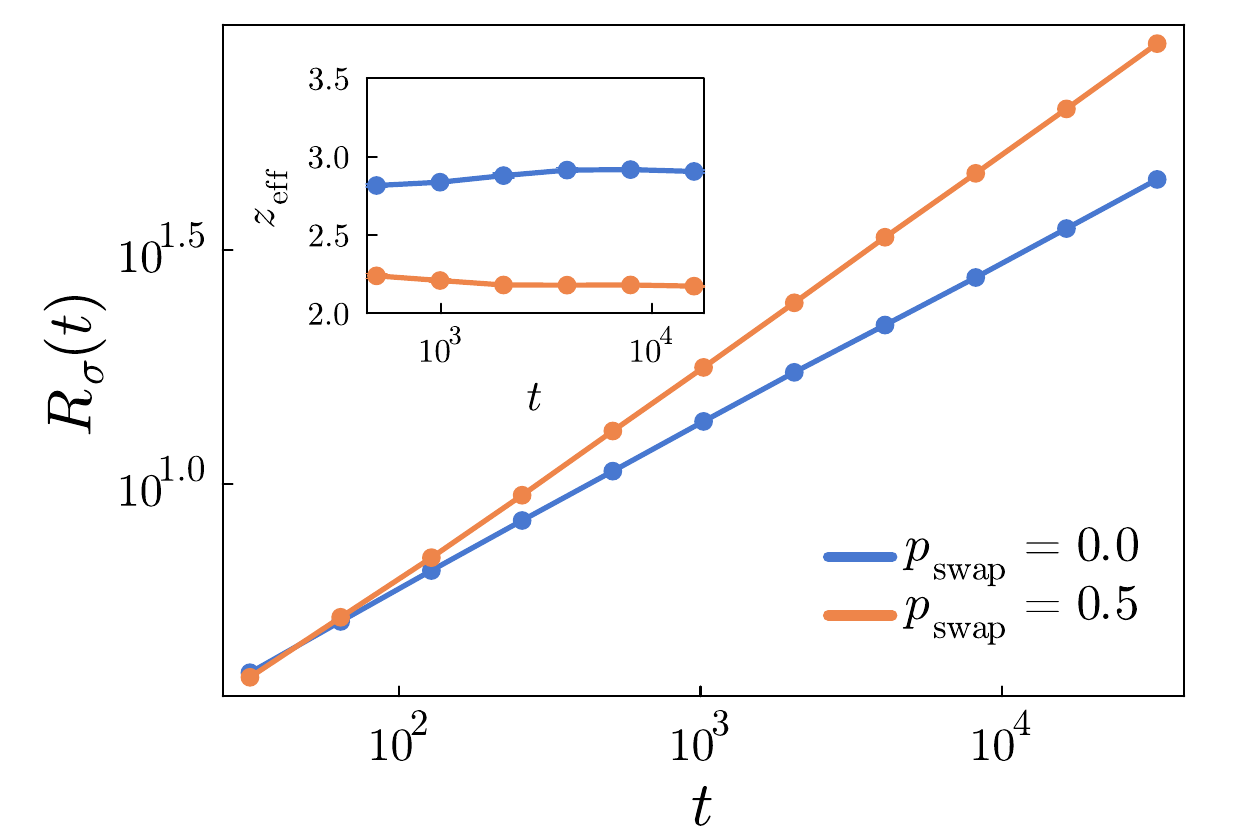}
	\caption{The growing length $R_\sigma$ (estimated from the analysis of the Ising $\sigma$ 
	spins space-time correlation).
	 Sub-critical quench to $T \approx 0.74~T_c$ and $L=2048$. The inset shows the effective exponent 
	 $z_{\rm{eff}}$ variation  in time, 
     measured by performing a fit of the data at (moving) six consecutive times. 
     }
	\label{fig:EffectiveExponentDelta1_Sigma}
\end{figure}

Now, we repeat the analysis above but this time measuring $R_{\sigma} (t)$. 
Tracking the domain growth of the Ising variables  in Fig.~\ref{fig:EffectiveExponentDelta1_Sigma}, 
similar conclusions are reached. The 
dynamic exponent remains close to $3$ for the single-spin-flip updates,
while  $z_{\rm eff}$ decreases when SWAP is performed, approaching  a value close to 
$2.25$ in the numerical interval explored, 
which is slightly larger than $z_d = 2$ for the 2DIM with 
non-conserved order parameter~\cite{Bray94}. 

\begin{figure}[h!]
	\centering
	\includegraphics[width=\linewidth]{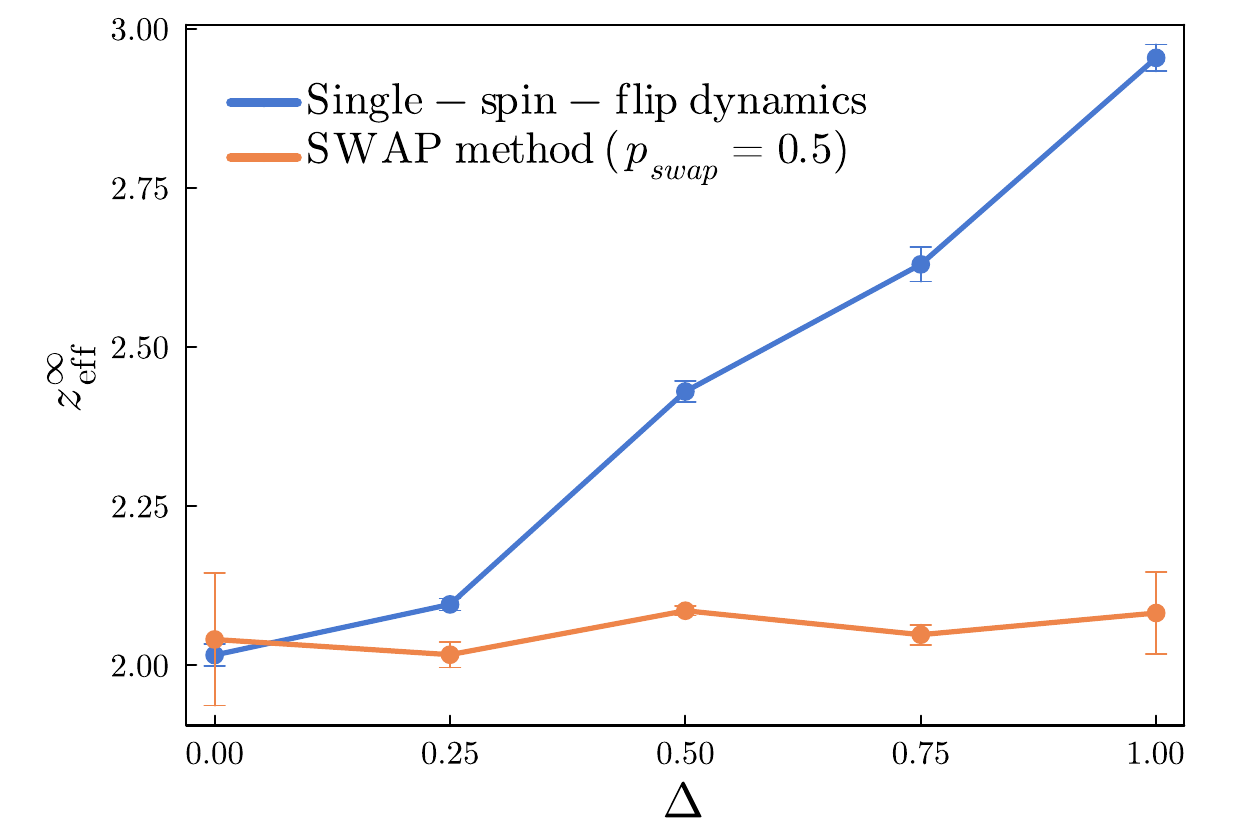}
	\caption{The asymptotic dynamical exponent, $z_{\rm eff}^\infty$, estimated from the effective exponent
		$z_{\rm eff}$ of the spins $s$ growing length, in  the latest time interval accessed by the simulation,
		against the disorder width $\Delta$. Sub-critical quench to $T \approx 0.74 \, T_c$, and $L = 2048$. 
	}
	\label{fig:zdvsDelta}
\end{figure}

Finally, we study the disorder dependence, $\Delta$, of the asymptotic value of  $z_{\rm{eff}}$, which we call 
$z_{\rm eff}^\infty$. We measure it in the last available time-interval. 
The results are plotted in Fig.~\ref{fig:zdvsDelta}. There is a large increase 
of $z_{\rm eff}^\infty$ with the width of the spin length distribution for the single spin flip 
dynamics, while there is none, apart from noise, in the SWAP simulations. 

\subsection{A.4 Conclusions}

With the SWAP dynamics one is able to obtain the equilibrium phase diagram of the $\Delta$-model, 
and characterize its second order phase transition and magnetized low temperature
phase. They do not vary considerably with respect to the ones of the standard 2DIM.

Concerning the out of equilibrium dynamics and further approach to equilibrium following a 
quench from the disordered phase, 
with the SWAP method one recovers the dynamic 
exponent of the original 2DIM, $z^\infty_{\rm eff} \sim 2$ for any $\Delta$. 
There is no improvement compared to the standard single spin flip 
dynamics of the original 2DIM in this unfrustrated case but, at least, there is no deterioration either.
Therefore, we have validated the use of the SWAP algorithm for this model, although we do not
improve with respect to the standard single spin flip MC using it. Instead, in cases with frustration, as we 
show in the main text and in the next section, the use of SWAP does accelerate the approach to 
equilibrium considerably.

\section{B The spin-glass Model}
\label{sec:SG}

In this Section we provide more details and additional results on the behavior of the 
by-pass $\Delta$-model that modifies the $\pm J$ 2DEA model. 
Concretely, in Eq.~(2) we used quenched random interactions  
drawn from a bimodal symmetric distribution
\begin{equation}
	p(J_{ij}) = \frac{1}{2}\delta(J_{ij} - J) + \frac{1}{2}\delta(J_{ij} + J)
	\; . 
	\label{eq:bimodaldist-def}
\end{equation} 
We set $J = 1$ in the numerical applications. In equilibrium this 2DEA 
model has no finite temperature phase transition:
it has spin-glass ground states and is paramagnetic at non-vanishing temperatures~\cite{bhatt_numerical_1988,wang_low-temperature_1988,hukushima_monte_1993,thomas_zero-_2011}.
Still, it presents a non-trivial slow 
relaxation towards the equilibrium paramagnetic state at low enough temperatures~\cite{Barahona82a,Barahona82b,takayama_monte_1983,huse_monte_1991,rieger_aging_1994,Roma06,hartmann_ground_2011,thomas_numerically_2013,xu_dynamic_2017,rubin_dual_2017,fernandez_out--equilibrium_2018,khoshbakht_domain-wall_2018,Weigel18,fernandez_dimensional_2019,fernandez_experiment-oriented_2019}.

The Hamiltonian of the frustrated $\Delta$-model  takes the form
\begin{equation}
	\mathcal{H} = 
	- \sum_{\langle ij \rangle }J_{ij} s_i s_j 
	= 
	-
	\sum_{\langle ij \rangle } J_{ij} \tau_{i} \tau_{j} \sigma_i \sigma_j 
	\; . 
	\label{eq:SpinGlassPreliminarHamiltonian-def}
\end{equation} 
For pure Ising spins, $\tau_i=1$ $\forall i$, it boils down to the 2DEA model with the bimodal 
couplings in Eq.~(\ref{eq:bimodaldist-def}). 

\subsection{B.1 An equivalent frustrated Ising model}
\label{subsec:equivalent-frustrated}

By following the same recipe used to analyze the ferromagnetic model, 
one can re-express the Hamiltonian with randomly chosen lengths $\tau_i$ 
as an EA Ising spin-glass 
\begin{equation}
	\mathcal{H} = 
	- \sum_{\langle ij \rangle }\mathcal{J}_{ij} \sigma_i \sigma_j \;,
	\label{eq:SpinGlassHamiltonian-def}
\end{equation} 
with an uncommon kind of couplings, defined as $\mathcal{J}_{ij} = J_{ij} \tau_i \tau_j$. 
The ${\mathcal J}_{ij}$
are correlated  through the site dependence of the $\{ \tau_i\}$, similarly to what happened with the ones of the ferromagnetic model, 
see Eq.~(\ref{eq:correlationofJij}).  

\begin{figure}[h!]
	\centering
	\includegraphics[width=0.87\linewidth]{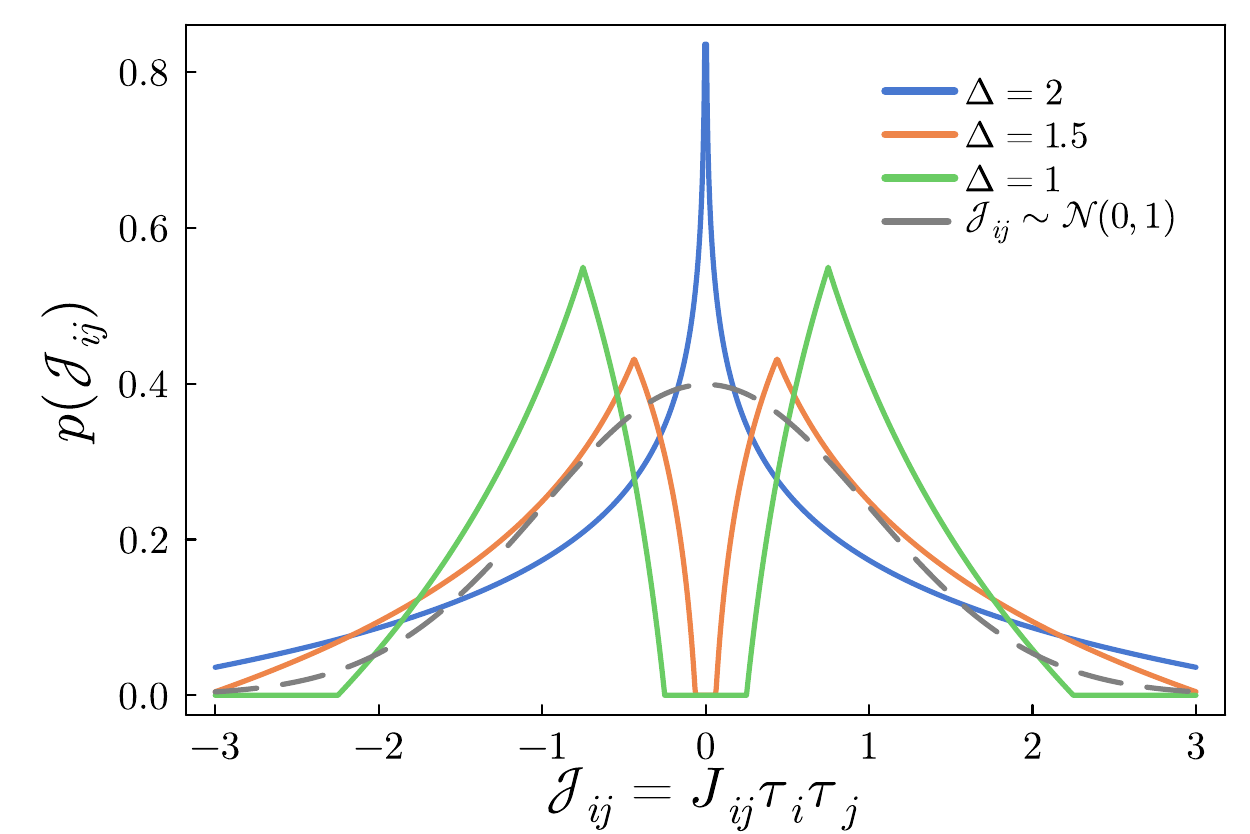}
	\caption{The  probability distribution function of the coupling strengths $\mathcal{J}_{ij} = J_{ij} \tau_i \tau_j$,
		arising from the product of the spin-lengths $\tau_i \tau_j$ and the quenched couplings $J_{ij}$ taking $\pm 1$ values with probability a half. 
		The dashed line represents a Gaussian distribution in normal form for comparison.
	}
	\label{fig:CouplingsDistro_EA}
\end{figure}

The mean and variance of the new effective couplings $\mathcal{J}_{ij}$ are
\begin{equation}
	\label{eq:meanvarsg}
	[{\mathcal J}_{ij}] = 0 
	\; ,
	\qquad
	[{\mathcal J}^2_{ij}] - [{\mathcal J}_{ij}]^2 = J^2
	\left( 1 + \frac{\Delta^2}{12} \right)^2
	\!\! .
\end{equation}

Moreover, there is a persistent quenched randomness, the $J_{ij}$, that is unaffected 
by the choice of the dynamics, unlike the $\{\tau_i\}$ that remain locally unmodified 
only when the dynamics do not involve spin exchanges. 

In Fig.~\ref{fig:CouplingsDistro_EA} we show the distribution of the couplings  ${\mathcal J}_{ij}$ induced by the one of the $\tau_i$ for three values of $\Delta$. The usual Gaussian distribution with zero mean and unit variance is also shown for reference. The distribution of the couplings  ${\mathcal J}_{ij}$ is just two delta peaks at $\pm J$ for $\Delta\to 0$  and it progressively shrinks the gap for increasing $\Delta$ until 
closing it completely when $\Delta=2$.

As explained in the main text, the frustration in the model is only determined by the $J_{ij}$ 
couplings. The introduction of the spin lengths and the insofar induced  ${\mathcal J}_{ij}$ 
do not remove it.

\subsection{B.2 The mean-field critical temperature}

\begin{figure}[t!]
\includegraphics[scale=0.4]{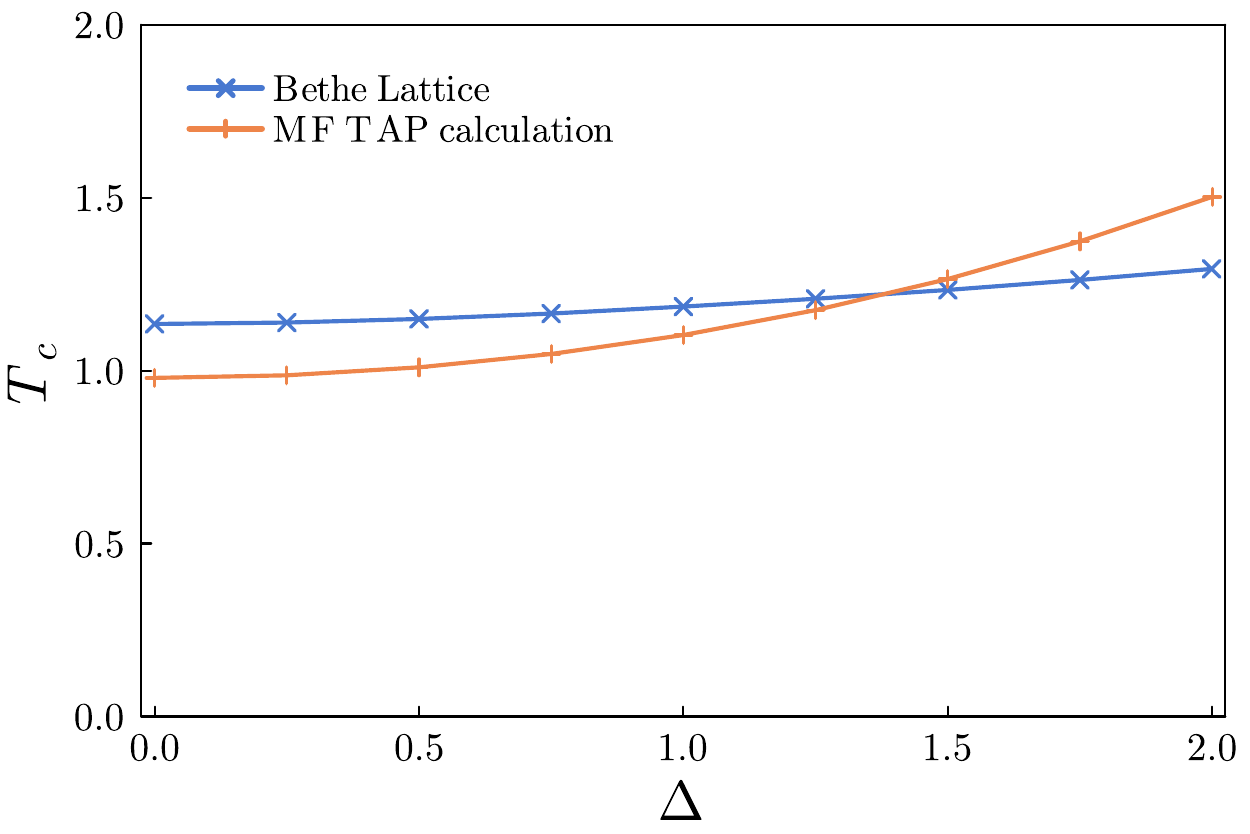}
\caption{The $\Delta$ dependence of the critical temperature in mean-field. Orange data: results for the 
fully connected model with the TAP approach. Blue data: results for the Bethe lattice model with connectivity $k = 2$.
The trend is the same and the range of variation of the critical temperature with $\Delta$ is very weak in 
both cases.}
\label{fig:criticalT-spinglass} 
\end{figure}

In mean-field (within the fully connected approximation) the critical temperature of an Ising model with {\it fixed} couplings ${\mathcal J}_{ij}$ (i.e. when both the $J_{ij}$'s and the $\tau_i$'s are quenched) in 
equilibrium at an inverse temperature $\beta$ can be estimated from the use of the TAP approach. After having 
conveniently scaled the ${\mathcal J}_{ij}$ with $N$ to ensure a good thermodynamic limit,  
one derives the following equations for the local magnetizations:
\begin{displaymath}
m_i = \tanh\left[ \beta \sum_{\partial i} {\mathcal J}_{ij} m_j + \beta h_i - \beta^2 \sum_{\partial i} {\mathcal J}^2_{ij} (1-m_j^2) m_i \right]
\end{displaymath}
The sums $\sum_{\partial i}$ indicate the spins $j$ connected to $i$.
Assuming a continuous phase transition and taking $h_i \sim 0$ as well, the local magnetization 
should be $m_i \sim 0$. If, moreover, one replaces $ {\mathcal J}^2_{ij}$ by  $[{\mathcal J}^2_{ij}]$
in the Onsager reaction term, 
\begin{eqnarray*} 
m_i &\sim& \beta \sum_{\partial i} {\mathcal J}_{ij} m_j + \beta h_i - \beta^2 \sum_{\partial i} [{\mathcal J}^2_{ij}] m_i 
\nonumber\\
&\sim& \beta \sum_{\partial i} {\mathcal J}_{ij} m_j + \beta h_i - \beta^2 J^2 \left(1+ \frac{\Delta^2}{12}\right)^2  m_i 
\; . 
\end{eqnarray*}
This equation can now be taken to the basis of eigenvectors of the matrix with elements ${\mathcal J}_{ij}$.
Calling $\vec v_\mu$ the eigenvector associated to the eigenvalue $\lambda_\mu$, 
and $m_\mu = \vec m \cdot \vec v_\mu$, 
\begin{eqnarray*} 
m_\mu &\sim& \beta \lambda_\mu m_\mu + \beta h_\mu - \beta^2J^2  \left(1+ \dfrac{\Delta^2}{12}\right)^2 m_\mu 
\; . 
\end{eqnarray*}
The linear susceptibilities are 
 \begin{eqnarray*} 
\chi_\mu = \left. \frac{\partial m_\mu}{\partial h_\mu} \right|_{\vec h  =0} 
&\sim& \frac{\beta }{1-\beta \lambda_\mu + \beta^2J^2 \left(1+ \dfrac{\Delta^2}{12}\right)^2 } 
\; . 
\end{eqnarray*}
The first susceptibility to diverge is the one associated to the largest eigenvalue $\lambda_{\rm max}$ 
and this arises at
\begin{equation}
\beta_c = \frac{\lambda_{\rm max} \pm \left[\lambda_{\rm max}^2 - 4 J^2 \left(1+ \dfrac{\Delta^2}{12}\right)^2 \right]^{1/2}}{2 J^2\left(1+ \dfrac{\Delta^2}{12}\right)^2}
\; . 
\end{equation}
In the Sherrington-Kirkpatrick model, the mean-field limit of the EA model, 
$\Delta \to 0$ and $\lambda_{\rm max} =2 J$. 
Then,  $\beta_c = J^{-1}$. If, in the $\Delta$-model, $\lambda_{\rm max} = 2 [{\mathcal J}^2_{ij}]^{1/2}$, which seems 
reasonable, then 
\begin{equation}
\beta_c \propto [{\mathcal J}^2_{ij}]^{-1/2}  \quad
\Rightarrow \quad
T_c = J \left(1+ \frac{\Delta^2}{12}\right)
\; . 
\end{equation}
In Fig.~\ref{fig:criticalT-spinglass} we plot, with orange data points, 
the critical temperature $T_c(\Delta)$
obtained from diagonalizing symmetric matrices with such elements and linear size $L=32$ ($J=1$).
The datapoints are consistent with the quadratic dependence
on $\Delta$
linking $T_c=J$ at $\Delta =0$ and $T_c=4/3\, J$ at $\Delta =2$.

An alternative way to estimate the $\Delta$ dependence of the critical temperature with a mean-field 
approach is to place the model on a Bethe lattice with connectivity $k$. By defining the cavity field acting on site $i$ produced by the effect of $k$ neighboring spins (in the absence of its $j$-th neighbor), we obtain the recursive equations   
\begin{align}
	h_{i \rightarrow j}(\tau_i)  =  \nonumber\\ 
	& \hspace{-1.25cm} \sum_{m \in \partial i/j} \frac{1}{\beta} \textrm{atanh} \left[ \tanh \left( \beta J_{im} \tau_i \tau_m \right) \tanh \left ( \beta h_{m \to i} (\tau_m) \right) \right ] \, . 
	\label{eq:local-fields}
\end{align} 
The sum excludes $m=j$. 
These equations admit $h_{i \to j} = 0$ as a solution on all sites, corresponding to the paramagnetic phase. 

Since we will be interested in the temperature regime in the vicinity of the critical point, $T \lesssim T_c$, in which the cavity fields are small, we can expand the 
right-hand-side of Eq.~(\ref{eq:local-fields}):
\begin{equation}
	h_{i \to j} (\tau_i) \simeq \sum_{m \in \partial i / j} \tanh \left( \beta J_{im} \tau_i \tau_m \right) h_{m \to i} (\tau_m)
	\; .
\end{equation}
Due to the randomness of the spin-amplitudes, 
these equations must be interpreted as a self-consistent integral equation for the probability distribution of the local cavity fields
\begin{align}
	P(h|\tau) = \nonumber \\ & \hspace{-1cm}\int \prod_{m=1}^k \left[ \sum_{J_m} \de \tau_m \de h_m \, p_{\tau}(\tau_m) \, P(h_m|\tau_m) \right ] \delta ( h - \tilde{h} ), \nonumber
\end{align}with $\tilde{h} = \sum_{m} \tanh \left( \beta J_{m} \tau \tau_m \right) h_{m }$. Using the integral representation of the $\delta$-function, one obtains the following equation for the Fourier transform of the probability distribution
\begin{equation}
	\hat{P}(q|\tau) = \left[ \int \sum_J  \de \tau^\prime p_{\tau}(\tau^\prime) \, \hat{P} \left( q \tanh \left( \beta J \tau \tau^\prime \right) |\tau^\prime \right) \right ]^k, \nonumber
\end{equation} 
Assuming that the fields follow a Gaussian distribution with variance, 
$\sigma^2_h = \overline{h(\tau)^2} - \overline{h(\tau)}^2$, we have
\begin{equation} \label{eq:self}
	\hat{P}(q|\tau) = 1 - \mathrm{i} q \overline{h(\tau)} - \frac{q^2}{2} \overline{h^2(\tau)} + \ldots \nonumber
\end{equation}
plugging this expression into the equation above, using the fact that $\overline{h(\tau)^n} \ll 1$ for $T \lesssim T_c$, and expanding up to second order one finds
\begin{subequations}
\begin{align}
	   & \overline{h(\tau)} \ =  k \int \sum_J \de \tau^\prime\  p_{\tau}(\tau^\prime) \, \tanh \left( \beta J \tau \tau^\prime \right) \overline{h(\tau^\prime)}, \nonumber \\
	 &  \overline{h^2(\tau)}   =  k \int \sum_J  \de \tau^\prime \ p_{\tau}(\tau^\prime) \, \tanh^2 \left( \beta J \tau \tau^\prime \right) \overline{h^2(\tau^\prime)} 
	 \nonumber\\
	 & \qquad\quad\
	 - \frac{k-1}{k} \overline{h(\tau)}^2. \nonumber
\end{align}
\end{subequations}
For $\overline{h(\tau)} = 0$ we are interested in the second equation, that defines a linear integral operator $f(\tau) = \int \de \tau^\prime \ \Gamma (\tau^\prime,\tau) f(\tau^\prime)$, with $f(\tau) = \overline{h^2(\tau)}$ and the (non-symmetric) kernel 
\begin{equation}
\Gamma(\tau^\prime,\tau) = k \sum_{J = \pm J} p_{\tau}(\tau^\prime) \tanh^2(\beta J \tau \tau^\prime)
\; . 
\end{equation} 
Therefore a solution of Eq.~\eqref{eq:self} with a non-vanishing function $\overline{h^2(\tau)}$ only exists if such integral operator has an eigenvector with eigenvalue $1$. For the specific case of the box distribution of width $\Delta$ we have diagonalized the integral operator numerically for $J = 1$ and $k=2$, for several values of $\beta$ and $\Delta$, on a grid of $2048 \times 2048$ intervals. The results for the critical temperature are reported in blue Fig.~\ref{fig:criticalT-spinglass}. Note that for $\Delta \to 0$ the critical temperature tends to the Bethe lattice value for  the Ising spin-glass, $\beta_c J = \textrm{atanh}(\sqrt{1/k}) \sim 0.88$, that is $T_c = 1.13 \, J$.
The increasing 
trend is the same as the one derived in the fully connected model with the TAP method.
The range of variation of the critical temperature with $\Delta$ is very weak in 
both cases.

 \subsection{B.3 The probability of reaching the ground state after a $T=0$ quench} 

We investigate the efficiency of SWAP to reach the ground states of the 2DEA model with the interactions $\mathcal J_{ij}^*$
obtained at the latest time reached after a zero temperature quench with SWAP. 

\begin{figure}[t!]
    \centering
    	\centering
	\includegraphics[width=0.87\linewidth]{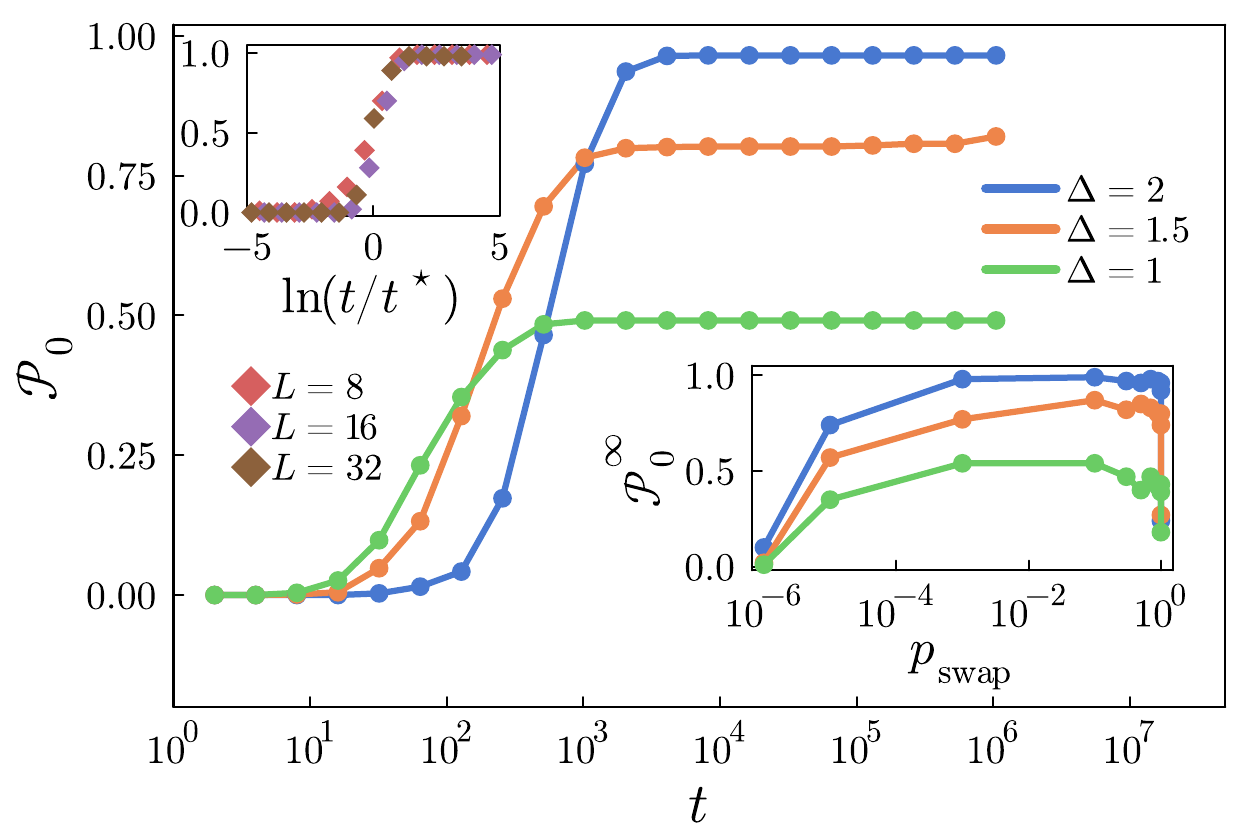}
	\caption{
	The probability of reaching the ground state of the 
	$\Delta$-model with ${\mathcal J}_{ij}^*= {\mathcal J}_{ij}(t_{\rm max})$, in systems with $L=8$ after
	zero temperature quenches. 
	Main panel: three $\Delta$ values given in the right key and $p_{\rm swap} = 0.5$. For $\Delta=2$, $96.5\% $ of the runs find a ground state.
	Lower inset: the asymptotic value against the parameter $p_{\rm swap}$ for $\Delta = 2$.
	Upper inset:  Scaling with $t^\star(L) = 1.25 \, L^{3.75}$ of the data for $\Delta=2$, $p_{\rm swap} = 0.1$ and three system sizes specified in the 
	left key.
	}
	\label{fig:P0}
\end{figure}

The main panel in Fig.~\ref{fig:P0} displays
${\mathcal P}_0(t)$ for three values of $\Delta$ and $p_{\rm swap} = 0.5$. In all cases, after a fast increase
ending at $t \lesssim 10^3$ MCs,   ${\mathcal P}_0(t)$ saturates to a ${\mathcal P}_0^\infty$
which increases with $\Delta$ and gets very close to 1 for $\Delta =2$. Further optimization
of the algorithm achieved by gauging  $p_{\rm swap}$ is studied in
the lower inset which shows the dependence of  ${\mathcal P}_0^\infty$ on $p_{\rm swap}$ in the $\Delta=2$-model.
For intermediate values, $0.1 \lesssim p_{\rm swap} \lesssim 0.9$, ${\mathcal P}_0^\infty$ remains approximately 
constant apart from numerical noise, and it decays to zero at the two extremes of 
either non-local spin exchanges ($p_{\rm swap} \to 1$) or pure single-spin-flips ($p_{\rm swap} \to 0$).
Finally, we checked the dependence on system size using $\Delta =2$ and $p_{\rm swap} =0.1$. 
The curves are similar and the percentage of ground states found is independent of $L$. The upper inset
shows the scaling of ${\mathcal P}_0$ against $t/t^\star(L)$ with $t^\star(L) \sim 1.25 \,  L^{3.75}$. 

\subsection{B.4 The overlap correlation and the spin-glass growing length} 

The overlap or four spin correlation is defined as
\begin{eqnarray}
	&& \!\!\!\!\!\!\!\! \!\!\!\!
	N C_4(r, t) = 
	\nonumber\\
	&&
	\!\!\!\!\!\!\!\!\!\!\!\!
	\sum^N_{\substack{i,j=1\\ |\vec r_i-\vec r_j| = r}} 
	 \left[ \left\langle \sigma^{(1)}_j(t) \sigma^{(2)}_j (t)\sigma^{(1)}_i(t) \sigma^{(2)}_i(t) \right \rangle\right]  
	\label{eq:CorrelationSG-def-SM}
\end{eqnarray} 
and the characteristic length over which it decays defines the spin-glass ordering length. It is here 
estimated from 
$ 
		R_\sigma(t) = 2 \int^\infty_{0} dr~C_4(r, t)
$,
as defined in~\cite{rieger_aging_1994}.

The Metropolis dynamics of the 2DEA at $T\simeq 0.5$ yield a super exponential decay, $C_4 \sim e^{-(r/R_\sigma(t))^\beta}$, with $\beta>1$. 
The spin-glass length scales as $R_\sigma(t)~\sim (t/\tau(T))^{1/z}$ 
with $z\sim 7$ and an Arrhenius characteristic time $\tau(T)$ at these relatively 
high temperatures~\cite{fernandez_out--equilibrium_2018}. 

\begin{figure}[h!]
	\centering
	\includegraphics[width=0.87\linewidth]{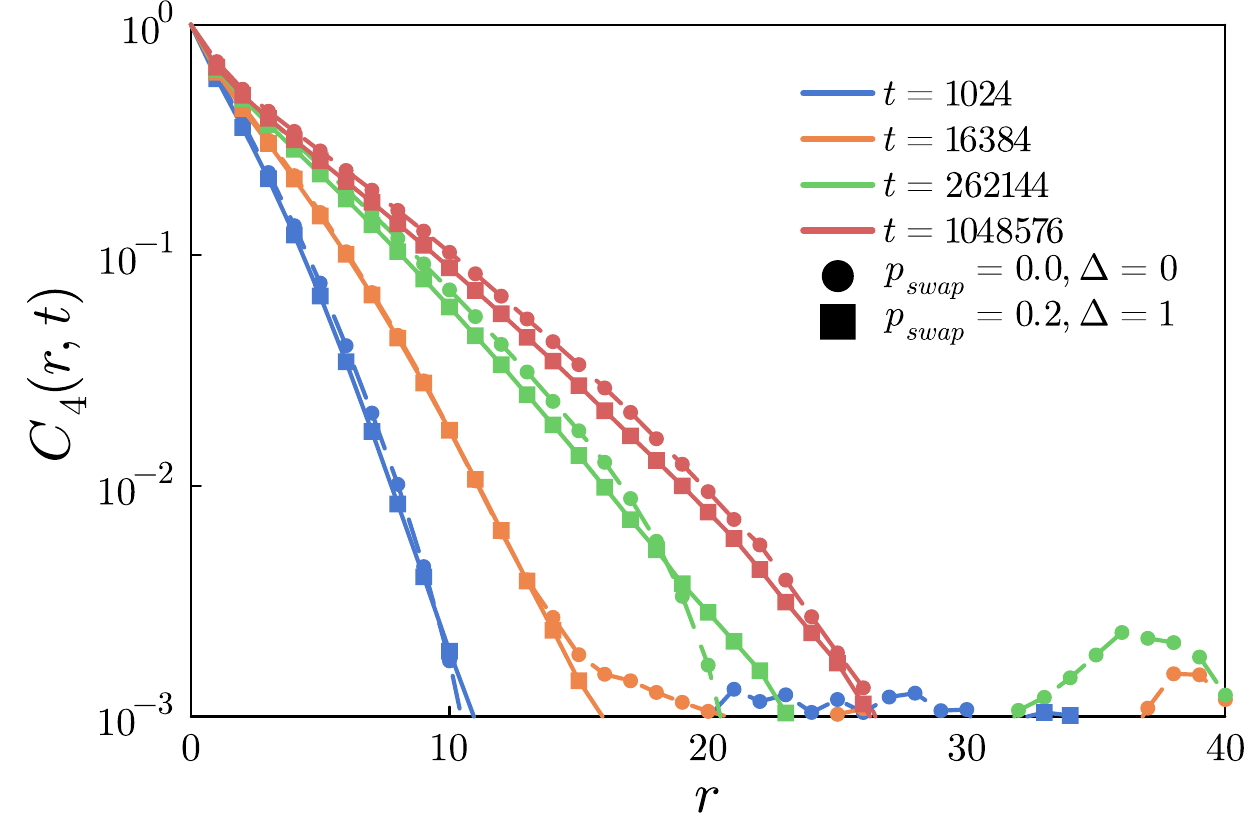}
	\includegraphics[width=0.87\linewidth]{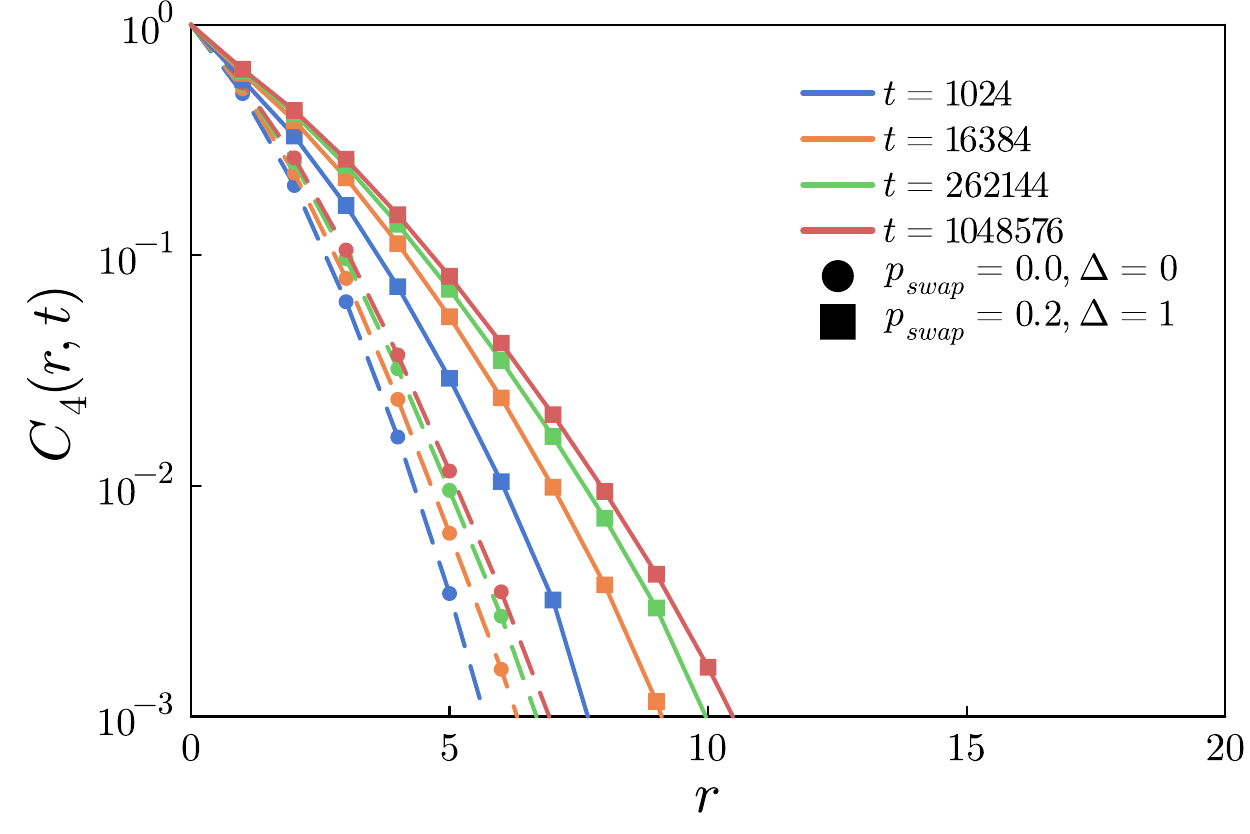}
	\caption{ 
	Overlap correlation $C_4(r,t)$, Eq.~(\ref{eq:CorrelationSG-def-SM}), versus the Cartesian distance $r$ for 
	several times given in the keys. $L = 512$ $\pm J$  2DEA ($\Delta=0$) model 
	evolved with single-spin and $\Delta=1$ model evolved with SWAP and $p_{\rm swap} = 0.2$. 
	Data for $T = 0.5$ (above) and $T=0.1$ (below). While single spin flips and SWAP yield equivalent results
	at high temperature, the latter builds longer correlations at low temperature. 
	}
	\label{fig:CorrelationSG-T05-T01}
\end{figure}

The spatial dependence of $C_4$, at different times, is studied in Fig.~\ref{fig:CorrelationSG-T05-T01} at $T=0.5$ and 
 $T=0.1$. At high temperatures SWAP on the $\Delta=1$ model 
	yields equivalent results to the single-spin flip evolution of the 
	original $\Delta=0$ model. At the longest time $t\sim 10^6$ considered the growing length is of the order of 20, 
	say, while the system's linear size is $L=512$.

	At low temperatures instead SWAP is much more efficient in building long-range
	correlations, even between copies that may have evolved towards different effective random couplings
	${\mathcal J}_{ij}$, than the single-spin flip evolution of the original $\Delta=0$ model. Still, the correlations 
	obtained with SWAP at the same time $t\sim 10^6$ decay faster than at higher $T$, reaching a shorter
	$R_\sigma(t)$.

\begin{figure}[h!]
	\centering
	\includegraphics[width=\linewidth]{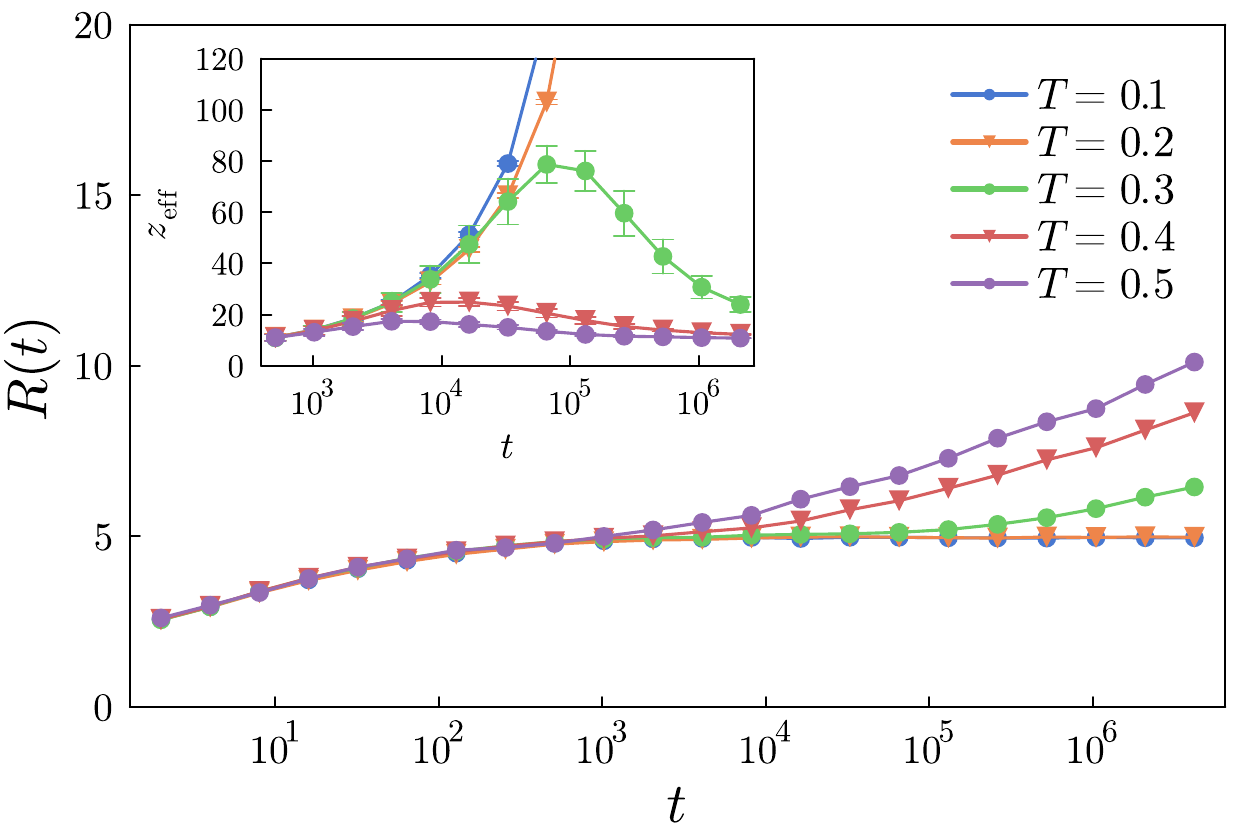}
	\caption{Typical growing length in the $\pm J$  2DEA model with $L=512$ evolved with single spin flip dynamics
	at several temperatures displayed in the key. 
	The inset shows the time dependence of the dynamical exponent $z_{\rm eff}$: it 
	converges to $\sim 8.5$ after the saturation of the growing length 
	at the value $R_p = 4.483 \pm 0.004$ is superseded at high-enough temperatures ($T > 0.2$).
}
	\label{fig:Rvt2dEA}
\end{figure}

For single-spin-flip dynamics, the growing length of the $\pm J$  2DEA Ising model, which we simply call
$R(t)$ in  Fig.~\ref{fig:Rvt2dEA},
freezes at low enough temperatures 
($T \leq 0.2$), saturating at a very short value~\cite{rieger_monte_1995}. The dynamic exponent, plotted
in the inset of the same figure, diverges. 
By increasing the temperature the plateau is surpassed and the evolution persists. The  latter growth can be 
described with the power-law $t^{1/z_{\rm eff}}$, as in Eq.~(\ref{eq:Rtyp_def}), and 
$z_{\rm eff}$ converges to a value close to 8.5 at $T$ larger than 0.3, say.

\begin{figure}[h!]
	\centering
	\includegraphics[width=\linewidth]{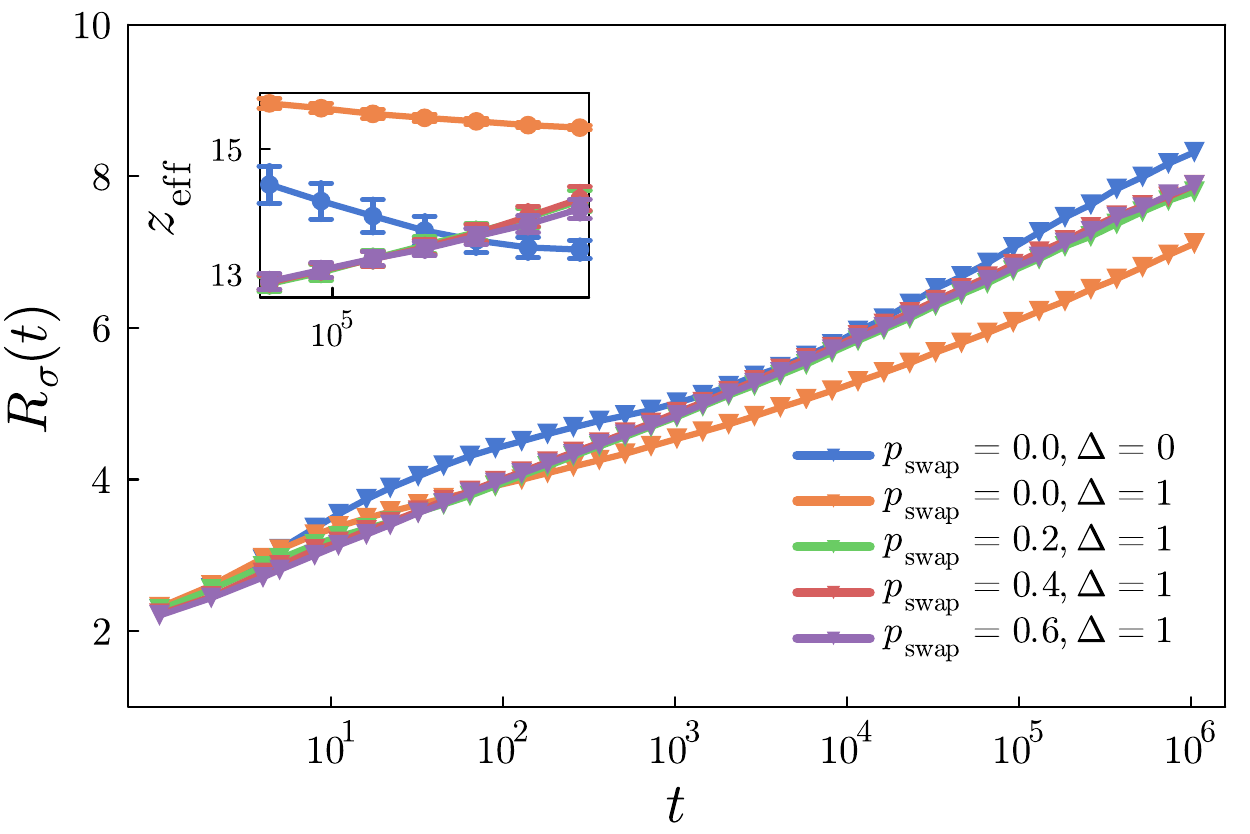}
	\caption{Typical growing length of the frustrated soft spin model with $L=512$ and
	$\Delta=1$, evolving with SWAP and single spin flip dynamics
	at a relatively high temperature, $T = 0.5$.
	The growth in the $\pm J$ 2DEA with single spin flip dynamics is plotted for comparison. 
	Several values of $p_{\rm swap}$ were used.
	The inset shows the time evolution of the dynamical exponent.
	In all cases a similar long-time limit, close to 8, is reached.   
	}
	\label{fig:Rvt2dEAandSWAPT05}
\end{figure}

Finally, in Fig.~\ref{fig:Rvt2dEAandSWAPT05} we compare the performance of SWAP with 
different values of the parameter $p_{\rm swap}$ to the one 
of the single spin flip updates when applied to the $\Delta=1$ model at high temperature. The 
data in the figure demonstrate that the two methods perform similarly.

\begin{figure}[h!]
	\centering
	\includegraphics[width=\linewidth]{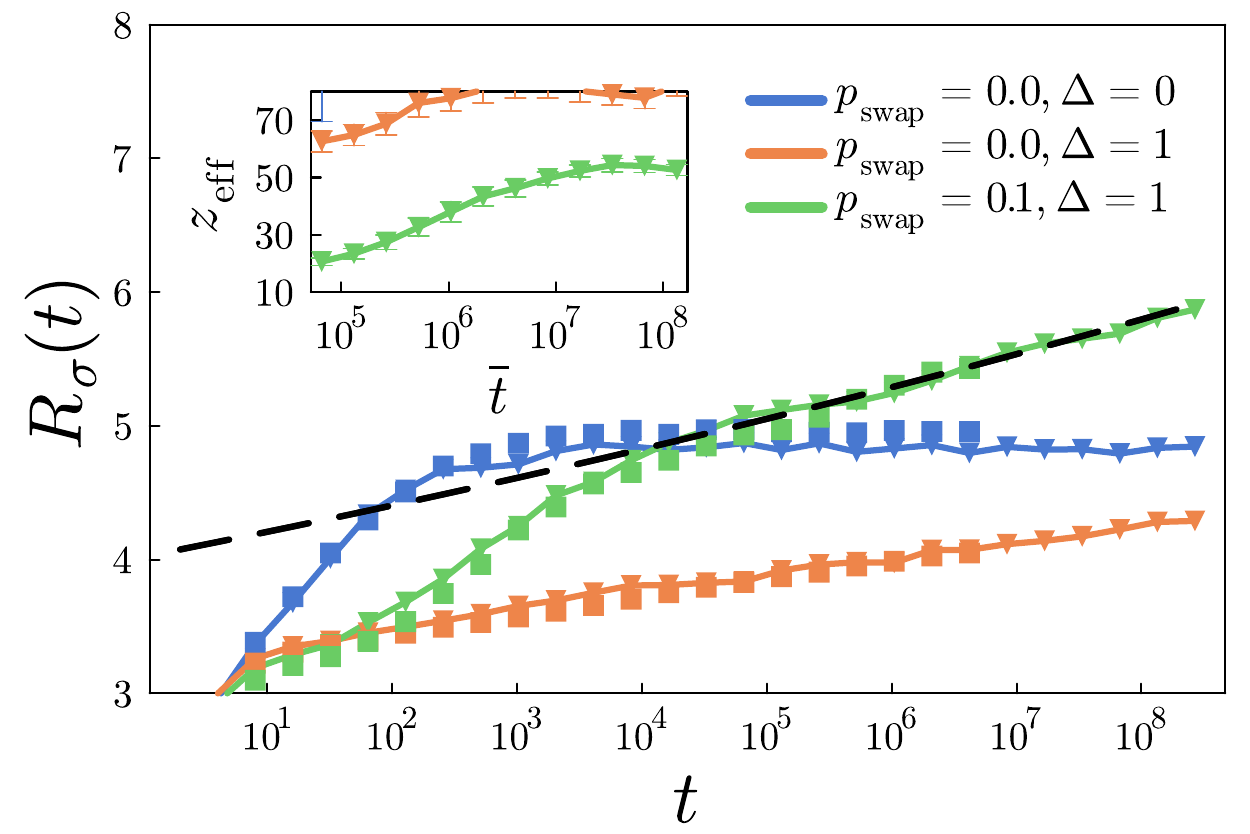}
	\vspace{-0.2cm}
	\caption{
	Spin-glass growing length in systems with $L=32$ (triangles) and $L=512$ (squares) quenched to $T = 0.1$.
	The three curves compare the single spin flip evolution of the Ising ($\Delta =0$) and soft ($\Delta =1$) models 
	to the SWAP one of the latter. 
	The inset displays the dynamical exponent $z_{\rm eff}$ as a function of $\overline{t}$,  the center of the time interval over 
	which the algebraic fit was performed (with $12$ data points, and $L=32$). 
	In dashed black is the fit with  $z_{\rm eff} = 49.5(86)$ 
	and amplitude $A = 3.94(2)$.
	}
	\label{fig:Rvt2dEAandSWAPT01}
\end{figure}
 
We henceforth focus on $T < 0.3$.
In Fig.~\ref{fig:Rvt2dEAandSWAPT01}(a) we display the spin-glass length 
for the $\Delta=0$ and $\Delta=1$ models quenched to $T=0.1$, and evolved with single spin flip and SWAP dynamics. At such 
low temperature the single spin flip dynamics of the $\Delta=0$ model freeze and 
the spin-glass length saturates to $R_\sigma \sim 5$ (blue data).  Instead, 
both single spin flips (orange) and SWAP (green) of the $\Delta=1$ model accelerate the dynamics at long times, 
with the latter becoming more efficient in the last four time decades.  We estimated running effective exponents
from fits over a moving window with $12$ data points (inset).  
Within numerical accuracy $z_{\rm eff}$  converges to 
$z_{\rm eff}^\infty \sim 49$. The dashed line is 
the algebraic law $A \, t^{1/z_{\rm eff}^\infty}$. 

However, the long-time configurations  are not in equilibrium.
First, the maximal length $R_\sigma \sim 6$ for $L=32$  is far 
from $L/2$.  Second, the $\tau_i$ variables, which are also 
dynamical under SWAP, are still evolving.	Hence, $\sigma_i^{(1)}$ and $\sigma_i^{(2)}$ may be optimized with respect to 
different ${\mathcal J}_{ij}$ and thus not really inform us  about the performance of the 
algorithm and measurement in taking one system close to equilibrium.

\subsection{B.5 The correlation of the $\tau_i$ variables}

\begin{figure}[h!]
	\centering
	\includegraphics[width=0.87\linewidth]{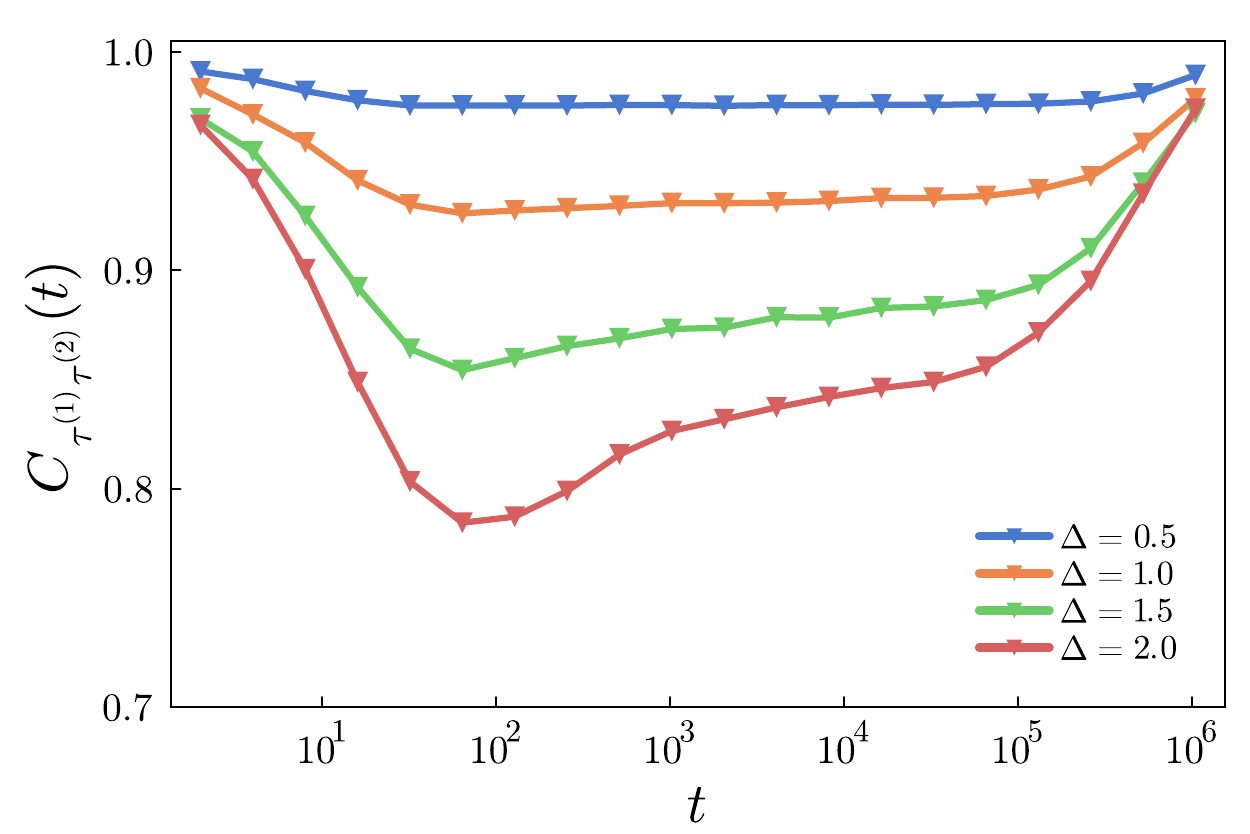}
	\caption{The correlation of the $\tau_i$ variables according to Eq.~(\ref{eq:tau-corr}). $L=16$ system annealed to zero temperature.
	}
	\label{fig:Ctautau_L161000sigma}
\end{figure}

In Fig.~\ref{fig:Ctautau_L161000sigma} we plot the time evolution of the correlation
\begin{equation}
C_{\tau^{(1)} \tau^{(2)}}(t) = 
\dfrac{ \left[ \dfrac{1}{N} \sum\limits_{i=1}^N \tau^{(1)}_i(t) \tau^{(2)}_i(t)  \right]-1}{\left(1+\dfrac{\Delta^2}{12}\right)-1}
\label{eq:tau-corr}
\end{equation}
where $\tau^{(1)}_i$ and $\tau^{(2)}_i$ are the values of the length variables in two runs of the 
same quench disordered model (same $J_{ij}$) starting from the same initial condition 
of the Ising spins and length variables and evolved with different Monte Carlo 
noises. At equal times $C_{\tau^{(1)} \tau^{(2)}}(t=0) =1$ by definition since $\tau^{(1)}_i(0) = \tau^{(2)}_i(0)$
and $[\tau^2_i]=1+\Delta^2/12$. If for $t\to\infty$ the $\tau_i^{(1)}$ and $\tau^{(2)}_i$ lengths 
fully correlate again, then 1 should be recovered in this limit as well. This is confirmed in 
Fig.~\ref{fig:Ctautau_L161000sigma} where this time-dependent correlation is shown for 
various values of $\Delta$ in a system with $L=16$ annealed following the protocol Eq. (16) from $T_0 = 0.5$ to zero temperature.

\subsection{B.6 The two-time correlation}

\begin{figure}[h!]
	\centering
	\includegraphics[width=0.87\linewidth]{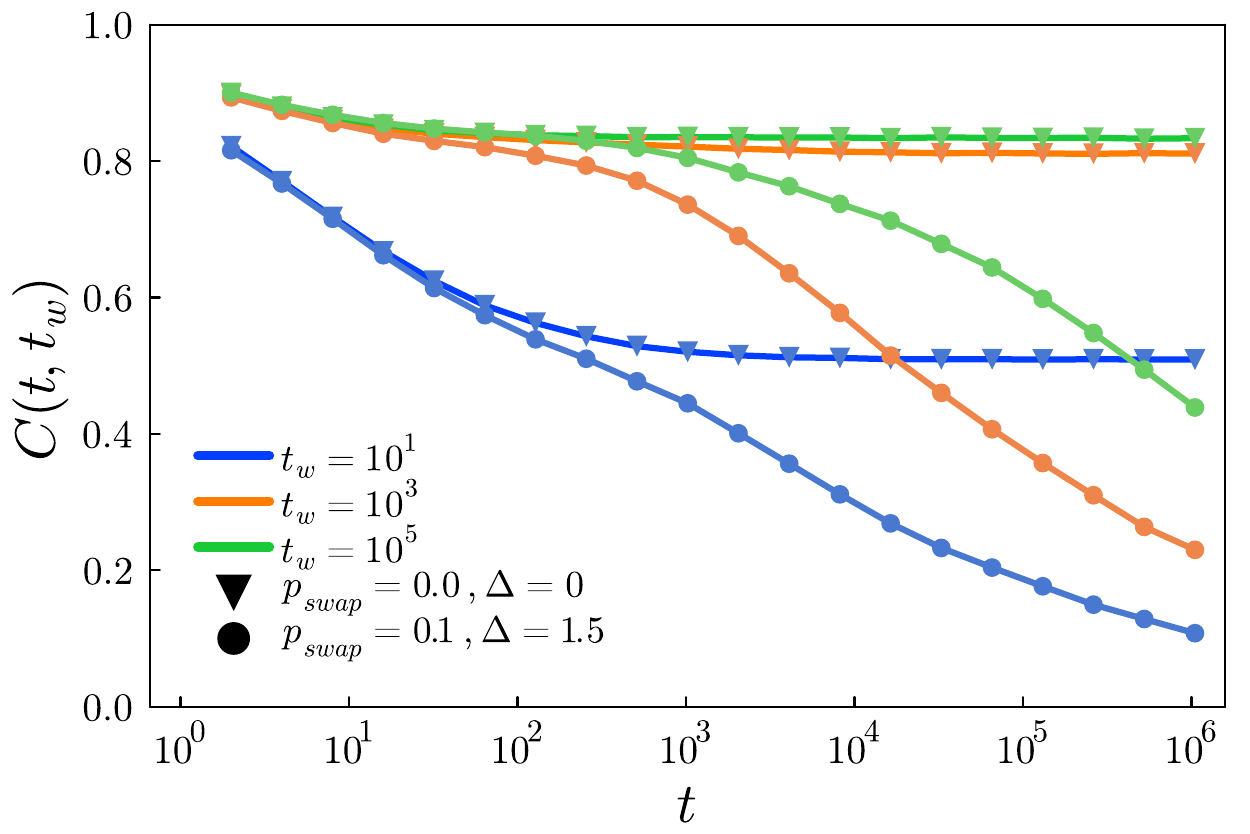}
	\caption{The two-time correlation of the $s_i$ variables, confronting the SWAP method ($p_{\textrm{swap}}=0.1$) with the $\Delta = 1.5$-model with the single-spin-flip kinetics of the 2DEA model, 
	both with sizes $L = 512$ and the waiting time values displayed in the key.}
	\label{fig:AutocorrelationSWAPvssf}
\end{figure}

Finally, we have calculated the two-time correlation function, defined as\begin{equation}
	C(t, t_w) = \dfrac{1}{N} \sum_{i=1}^N \; [\langle s_i(t) s_i(t+t_w)\rangle ]
\end{equation} 
where $s_i(t)$ = $\sigma_i(t) \tau_i(t)$ for the $\Delta$-model and $s_i(t) = \sigma_i(t)$ for the EA one. 
As can be seen in Fig.~\ref{fig:AutocorrelationSWAPvssf} the SWAP method induces a faster decay, compared to the single spin flip dynamics, in which the curves saturate rather quickly, consistent with plateau the $R_p$ found in Fig.~\ref{fig:Rvt2dEA}.

\vspace{0.5cm}

\subsection{B.7 Measures of frustration}

In Fig.~\ref{fig:snapshots-SG} we follow the time evolution of the instantaneous configuration of the system, parametrized as its projection on the 
ground state of the final couplings ${\mathcal J}^*_{ij} = {\mathcal J}_{ij}(t_{\rm max})$. More precisely, each square in the images 
represents
\begin{equation}
s_i(t) \sigma^{\rm gs}_i = \tau_i(t) \sigma_i(t) \sigma^{\rm gs}_i 
\label{eq:o-local}
\; . 
\end{equation}
The color code goes from $-2$ (light blue) to $+2$ (light red), A very negative value represents
a long spin which is anti-aligned with the 
ground state, while a very positive value represents a long spin which is aligned with the ground state. Dark squares represent short spins.
On each vertex of the (dual) lattice we place a symbol when the corresponding plaquette of the original lattice is frustrated. 
The strength of the local frustration is quantified by
\begin{equation}
f_P(t) = \prod\limits_{\langle ij\rangle \in P} {\mathcal J}_{ij}(t) = \prod\limits_{\langle ij\rangle \in P} J_{ij} \tau_i(t)  \tau_j(t) 
\; . 
\end{equation}
Concretely, on a square plaquette with site labels $1,2,3,4$, 
\begin{equation}
f_P(t) = J_{12}J_{23}J_{34}J_{41} \, \tau_1^2(t)   \tau_2^2(t)   \tau_3^2(t)   \tau_4^2 (t) 
\; . 
\end{equation}
\begin{figure}[h!]
\centerline{
\includegraphics[scale=0.3]{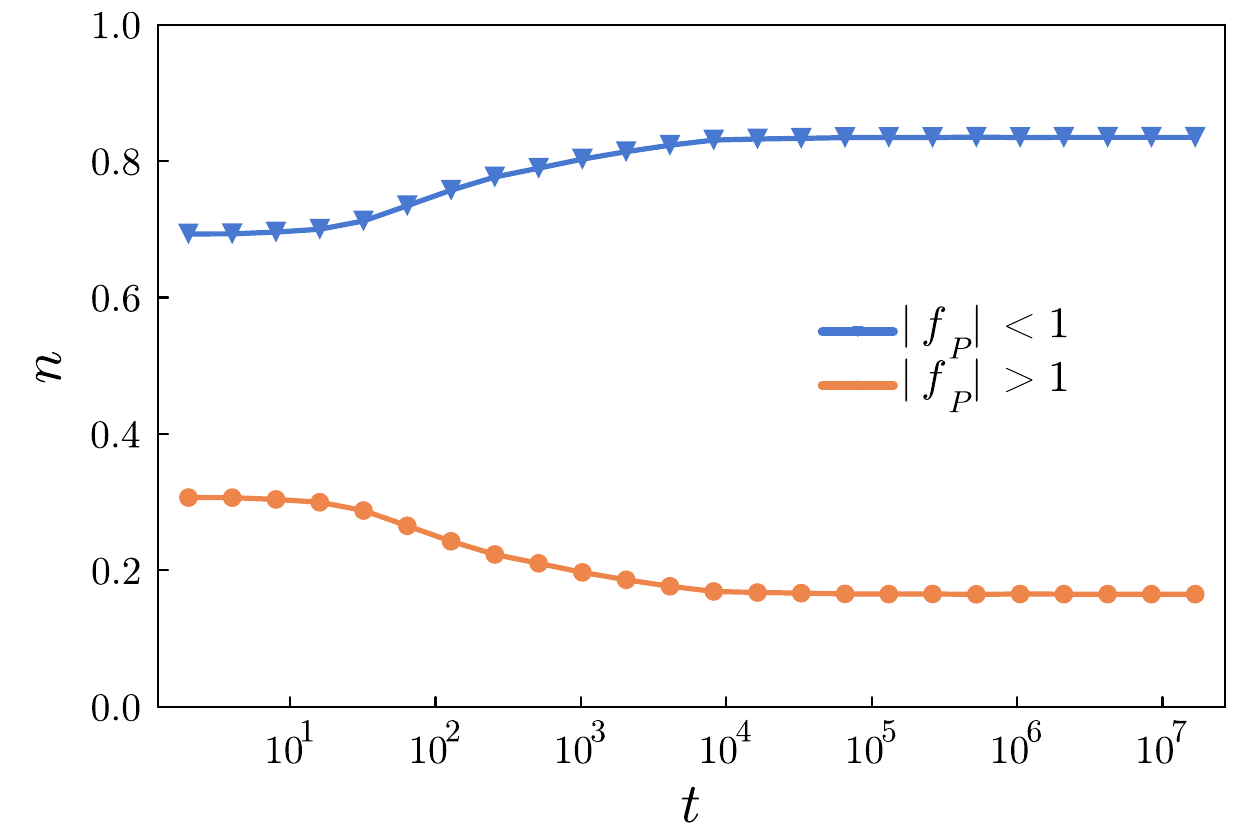}
}
\caption{
The number density of plaquettes with frustration $f_P<0$,  distinguished by the  modulus being larger or smaller than $1$.}
\label{fig:np}
\end{figure}
The sign, and whether the plaquette is frustrated or not, is decided by the factor 
$J_{12}J_{23}J_{34}J_{41}$
while the magnitude of the potential frustration is determined by $\tau_1^2(t)   \tau_2^2(t)   \tau_3^2(t)   \tau_4^2 (t)$
which depends on time. Initially one can imagine that the last four factors are independent and 
$\tau_1^2  \tau_2^2  \tau_3^2  \tau_4^2 \sim (1+\Delta^2/12)^4$. 
Going back to the snapshots, we used two kinds of symbols: 
a bullet for $|f_P(t)|$
greater than one,  a triangle for $|f_P(t)|$ smaller than 0.1, respectively. 
We called $n$ the density of each of these frustrated plaquettes, that is the number of 
the frustrated plaquettes of each kind divided by the total number of frustrated plaquettes $N_F$ which 
is constant and approximately equal to half the total number of plaquettes. 
\\
\indent
A total frustration density, defined as the normalized sum of the 
local ones, reads
\begin{equation}
f(t) = \frac{1}{N_F} \sum_P f_P(t)
\label{eq:total-f}
\end{equation}
with the sum running  over frustrated plaquettes.

\begin{figure}[h!]
\centerline{
\includegraphics[scale=0.35]{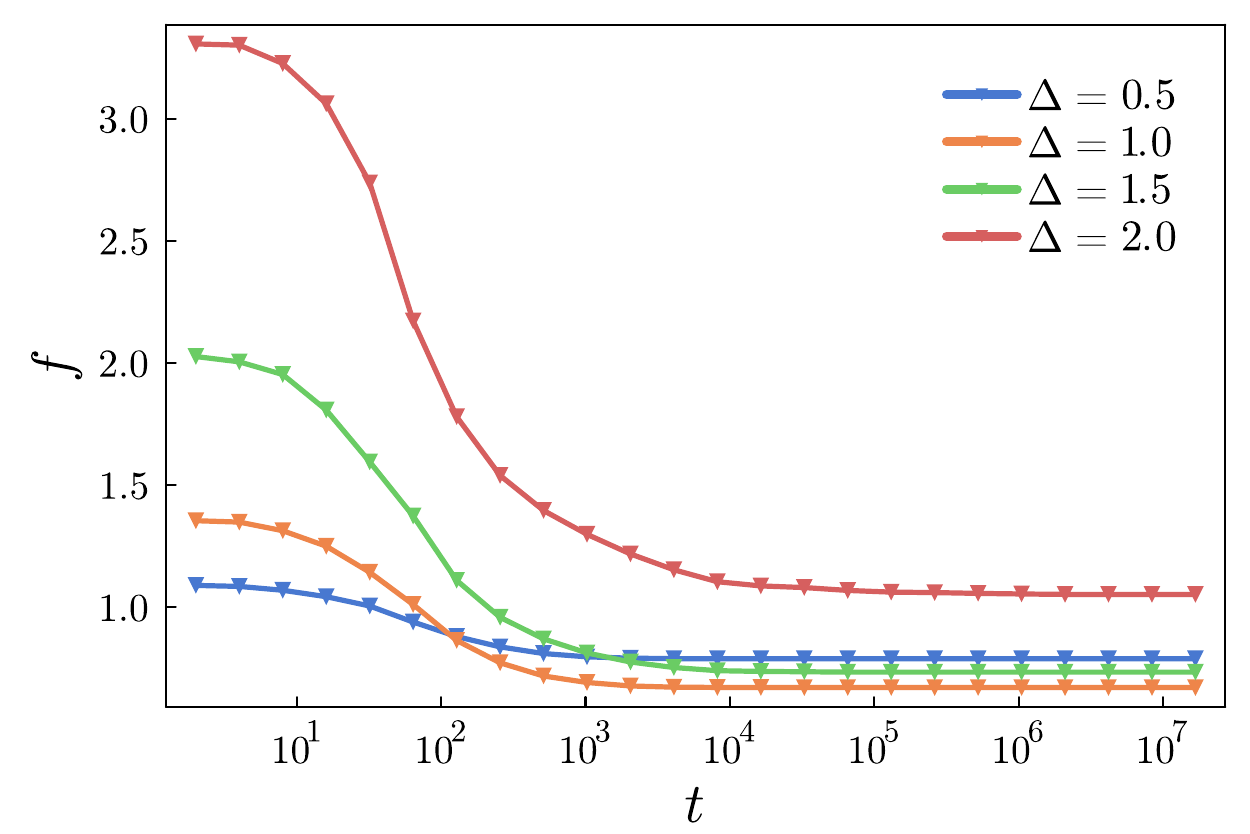}
}
\caption{
The magnitude of the total frustration $f$, defined in Eq.~(\ref{eq:total-f})
in $\Delta$-models with different values of $\Delta$ specified in the key and $L=32$ quenched to zero temperature 
and evolved with SWAP.   The initial value is $(1+\Delta^2/12)^4$.
}
\label{fig:total-f}
\end{figure}

In the course of time we see the following.
\begin{enumerate}
\item[--] The frustrated or unfrustrated nature of the plaquettes does not change in time. When there is 
a symbol (bullet or triangle) attached to them, they do not disappear. They are not created elsewhere either.
\item[--] The strength of the local frustration does vary in time, so the symbols can change, for example, from 
bullet to triangle, or {\it vice versa}. The time dependence of the densities of frustrated plaquettes of each 
kind are shown in Fig.~\ref{fig:np}. The number density of plaquettes with small frustration dominates
over the ones with large frustration asymptotically.
\item[--] 
In the course of time the symbols rearrange spatially. Rather large regions with small frustration are visible in 
the late snapshots.
\item[--]
The magnitude of the total frustration $f$ tends to diminish in time.
\item[--] Initially, the color of the images is  roughly randomly distributed over the full palette. 
In the course of time,  blue cells tend to disappear and the image becomes globally reddish. This means that 
the system orders in the direction of the selected ground state.
\item[--]
In the last images the crosses (weak frustration) are located in regions where the spins 
have short length and the bullets (strong frustration) are placed in regions where the spins have long length.
\end{enumerate}

\begin{figure}[h!]
	\centering
	\includegraphics[width=0.87\linewidth]{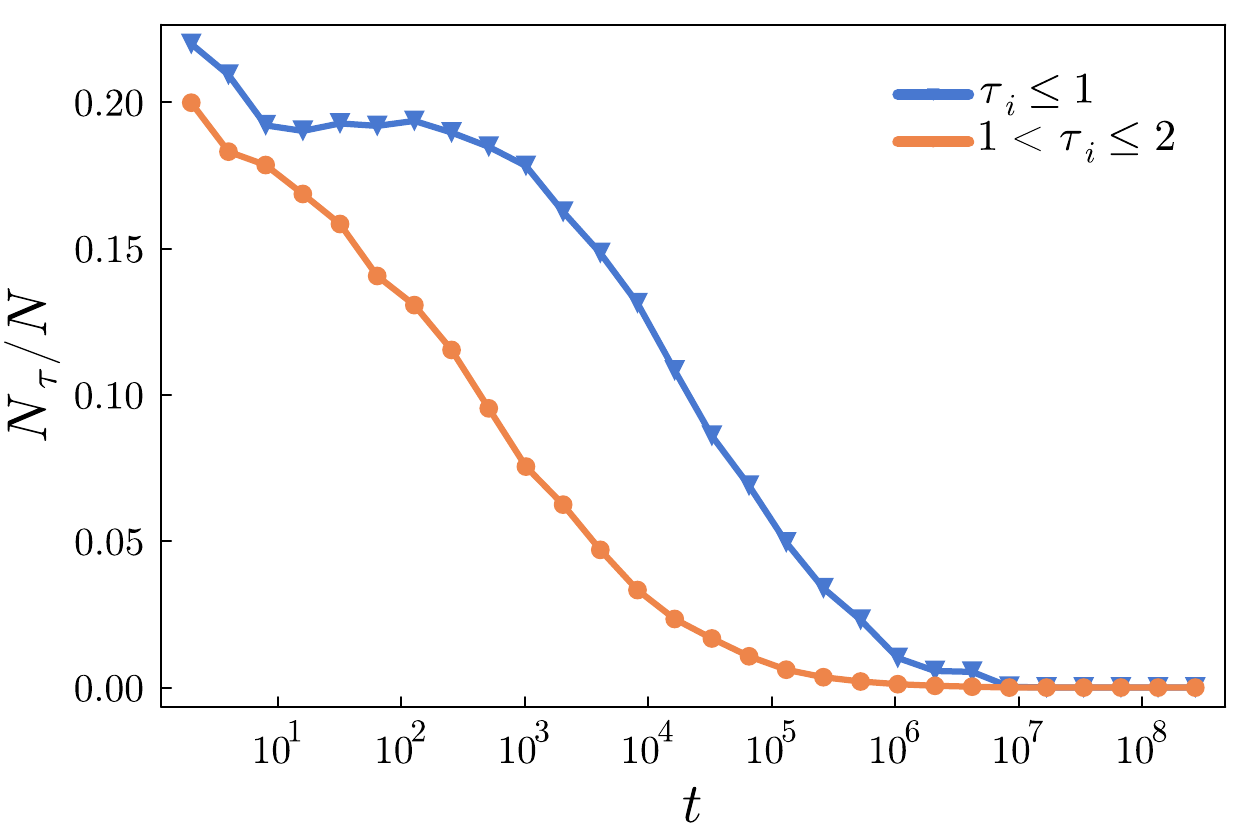}
	\caption{Time evolution for the ratio of small-length ($\tau_i \leq 1$) and large-length ($1 < \tau_i \leq 2$) spins 
	with $\sigma_i$ not aligned with the ground state 
	$\sigma^{\rm gs}_i$, i.e. $\sigma_i(t) \sigma_i^{\rm gs} < 0$.}
	\label{fig:rateoftau}
\end{figure}

\subsection{B.8 Evolution of the bond distribution}
The aforementioned change in the distribution of the frustration strength $f_P$ has direct effects on the distribution of the effective bonds $\mathcal{J}_{ij} = J_{ij} \tau_i \tau_j$. Here we show the (almost) initial case ($t=2$) and a last step ($t = 2^{24}$) of the empirical bond distribution from several runs ($N_r = 100$) after a quench at $T = 0$. 

\begin{figure}[h!]
	\centering
	\includegraphics[width=0.87\linewidth]{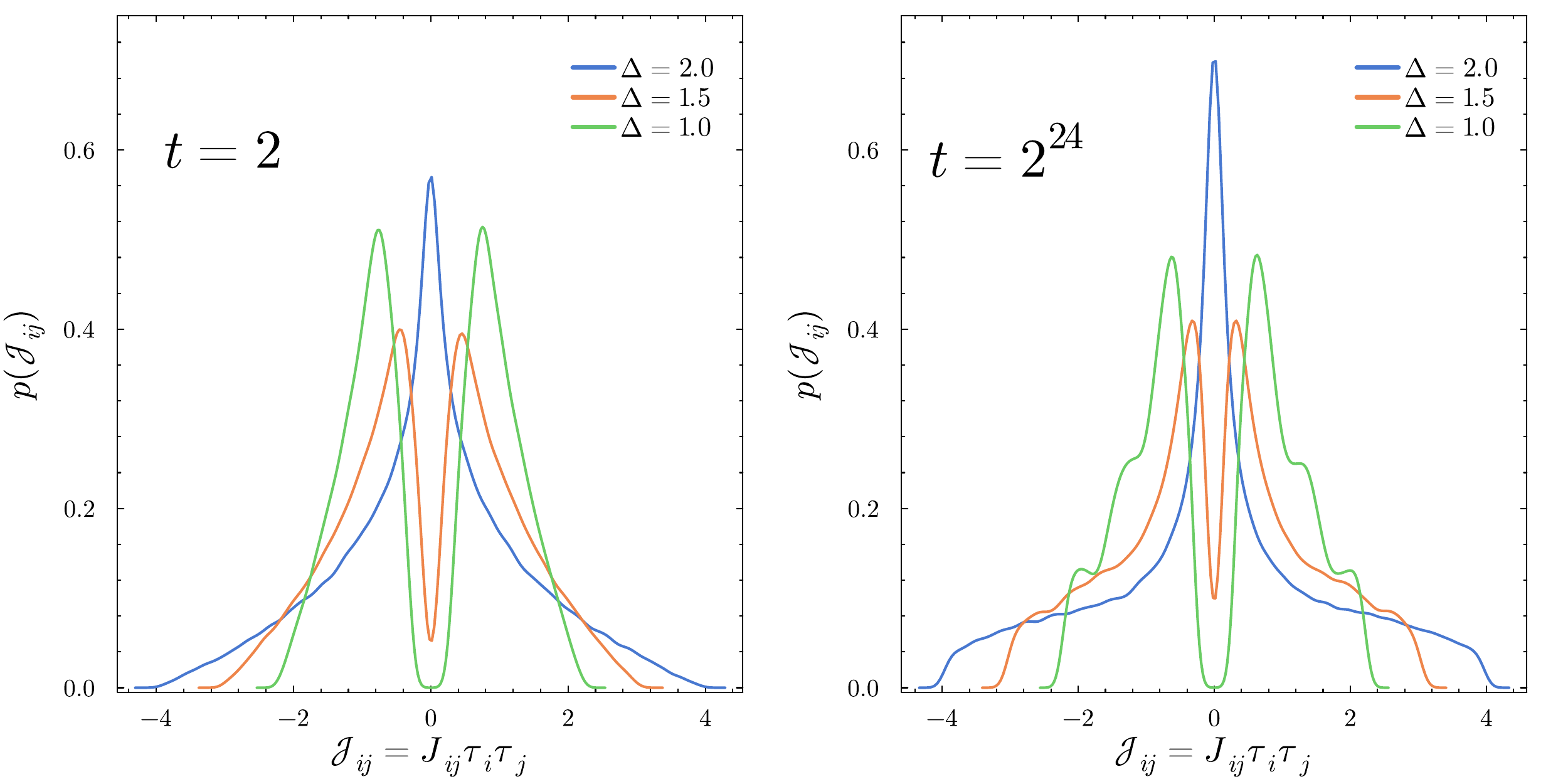}
	\caption{Distribution of the effective bonds $\mathcal{J}_{ij}$ at two instants of the simulation: (almost) initial distribution ($t = 2$), final distribution ($t = 2^{24}$) after a SWAP method was performed ($p_{\textrm{swap}} = 0.1)$. Taken from an evolution at $T = 0$ quench, with $L = 32$, and the values of $\Delta$ displayed in the key.}
	\label{fig:effective_bond_distro_MC}
\end{figure}
The distribution preserves its symmetric structure. Whereas for the (almost) initial time, $t = 2$, the distribution coincides with the expected theoretical form shown in Fig. \ref{fig:CouplingsDistro_EA}, at later times the tails widen. For $\Delta \neq 2$ new symmetric peaks arise, always lower than the original typical values, that are still present. No new peaks arise for $\Delta = 2$, however the original peak increase its probability. 
\subsection{B.9 Instantaneous configurations}
\begin{figure}[h!]
	\centering
	\includegraphics[height = 12cm]{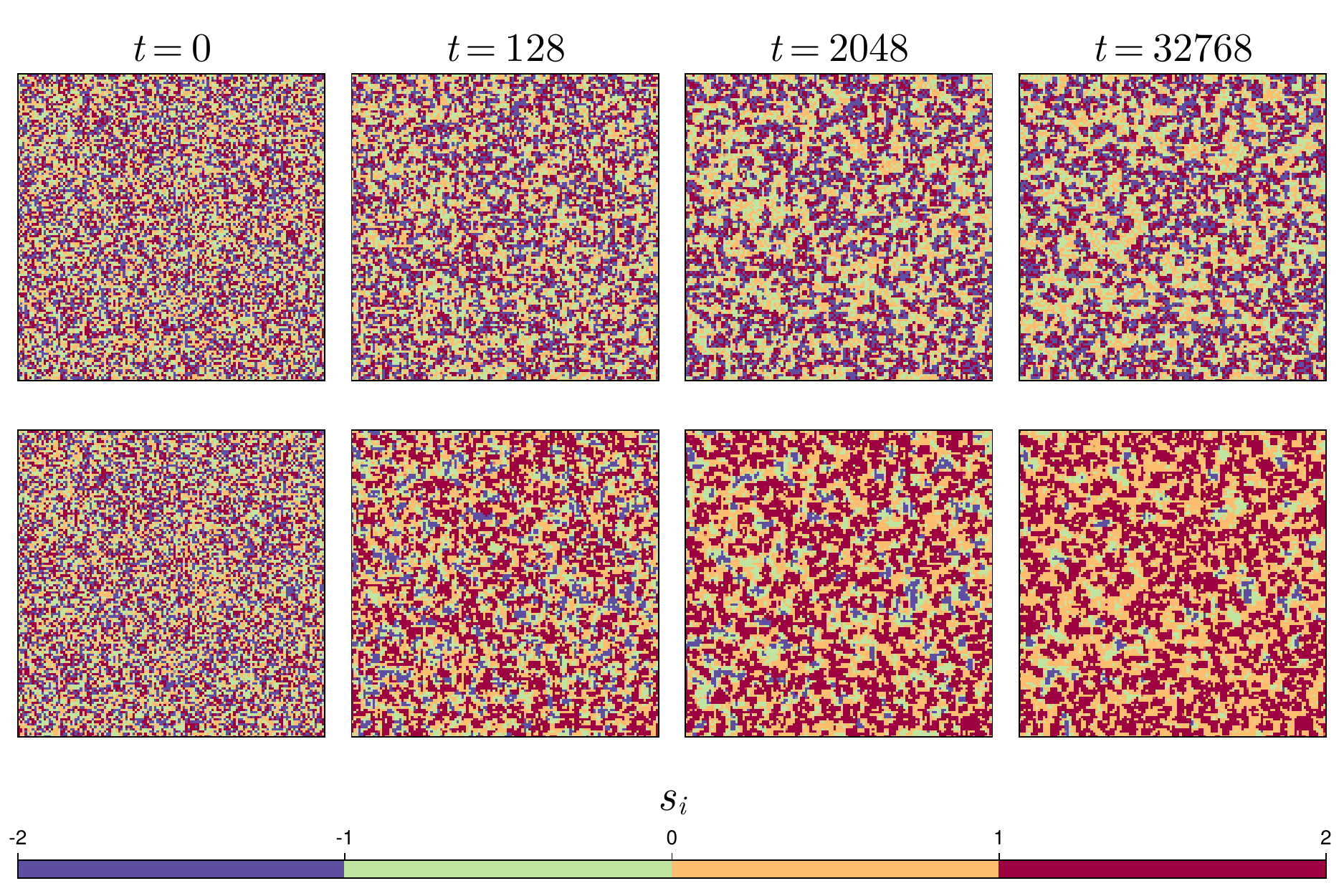}
	\caption{Instantaneous snapshots of the frustrated $\Delta$-model 
	with $\Delta = 2$, at four times indicated in the figure after a quench to $T = 0$. 
	First row: The spin $s_i$ configurations. 
	Second row: Overlap with the $\sigma$-ground state ($s_i(t) \sigma_i^{\rm gs}$).
	The color scale binned as indicated in the bar shows the lengths of the local spins.
	}
	\vspace{0.5cm}
	\label{fig:SGDomains}
\end{figure}

\begin{figure}[h!]
\includegraphics[scale=0.09]{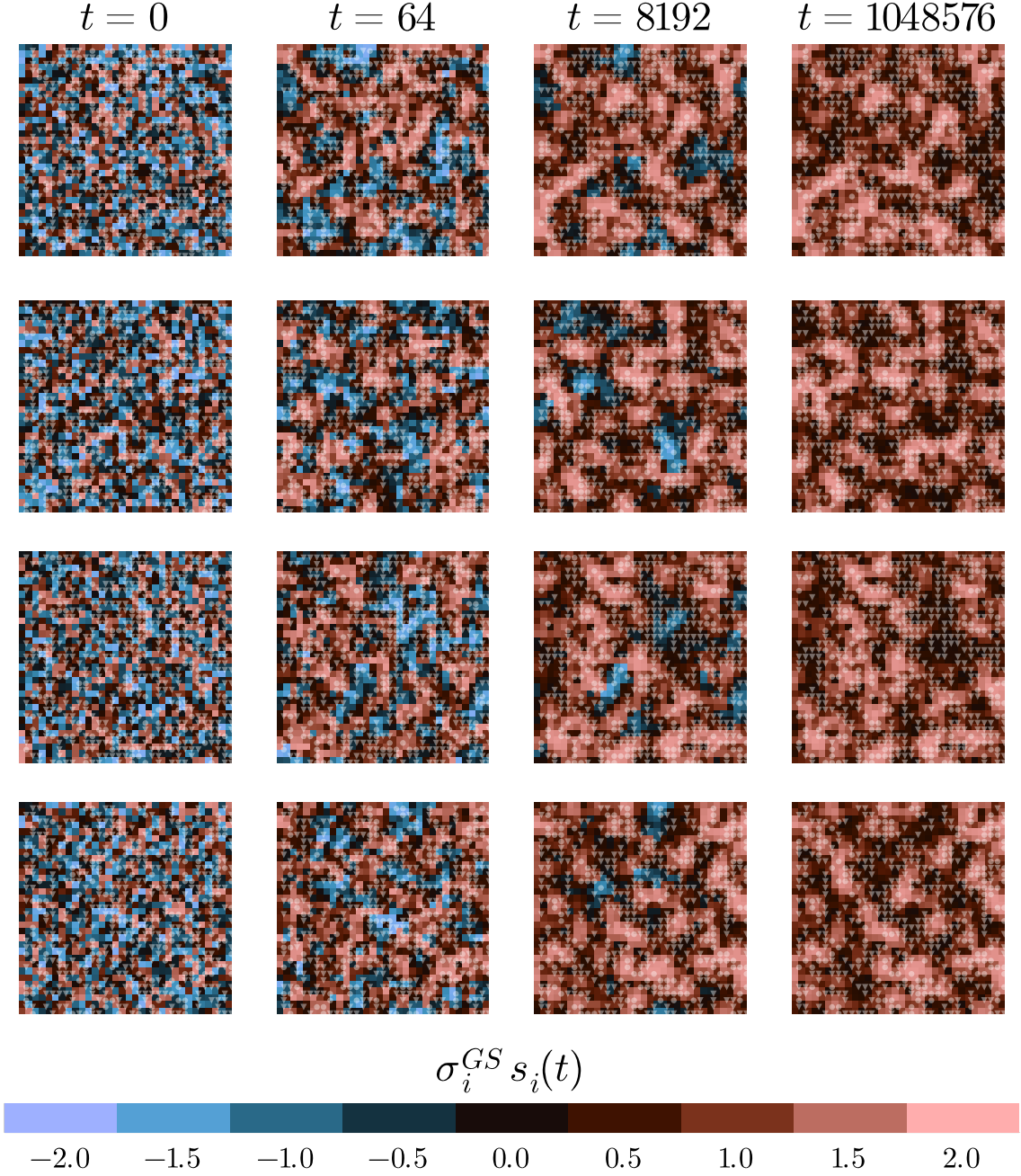}
\caption{Instantaneous configurations of the $\Delta$-model obtained 
at subsequent (though not equally spaced) times after a quench to 
$T=0$ of four initial conditions (different rows) evolved with SWAP.
Each square represents the instantaneous local overlap 
$\tau_i(t) \sigma_i(t) \sigma^{\rm gs}_i $, see Eq.~(\ref{eq:o-local}).
The color code spans the interval  [$-2$ (light blue), $2$ (light red)].
The frustrated plaquettes
are indicated with bullets and triangles according to the local frustration $f_P(t)$ being 
greater or smaller than one, respectively. 
}
\label{fig:snapshots-SG}
\end{figure}

The mechanism that renders the SWAP method efficient in the frustrated $\Delta$-Model is clarified by inspecting 
some snapshots, as we did for the ferromagnetic model. In Fig.~\ref{fig:SGDomains}, we show the actual configuration 
(first row) and the overlap with the ground state of the system with the final ${\mathcal J}_{ij}$ interactions (second row). 
(In the main text we explain how we obtain the $\sigma_i^{\rm gs}$ configuration.)
We have performed a sub-critical quench at $T = 0$, for $L=128$ using SWAP dynamics with $p_{\rm swap} = 0.1$.

The first row displays the spin configurations at four times after the quench. The images do not show much structure 
apart from a slight tendency of long spins with the same orientation to group locally. Still, and as expected, no 
 structure with long-range spin ordering develops.

One can recognize the formation of domains in the overlap of the configurations with the ground state. 
Indeed, the images shown in the second row are primarily red or yellow, that is, the system acquires a positive
overlap with the ground state all over space. 
The SWAP algorithm produces domain walls consisting mostly of short length spins (green and yellow). 
This structure lowers the energy barriers locally, making the spin-flip more feasible in these regions. The ratio of $\{\tau_i\}$ variables for which the corresponding Ising spins do not match the ground state (i.e. $\sigma_i(t)\sigma_i^{gs} < 0$) goes to zero, see Fig.~\ref{fig:rateoftau}, in the course of time. However, the decay rate depends strongly on the length of the spins. Larger spins ($1 < \tau_i \leq 2$) align faster on the ground state directions compared to smaller ones ($\tau_i \leq 1$), as can also be seen in Fig.~\ref{fig:rateoftau}, supporting the previous explanation.

\onecolumngrid

\twocolumngrid

\widetext


\end{document}